\documentclass[10pt,preprint]{sigplanconf}
 
\usepackage{abbrevs}
\usepackage{coq}
\usepackage{code}
\usepackage{qsymbols}
\usepackage{listings}
\usepackage{xcolor}
\usepackage{catchfilebetweentags}

\usepackage[color]{coqdoc}

\definecolor{kwblack}{rgb}{0,0,0}
\def\coqdockwcolor{kwred}
\renewcommand{\coqdocnotation}[1]{#1}
\renewcommand{\coqexternalref}[3]{\href{#2.html\##1}{\textcolor{black}{#3}}}
\renewcommand{\coqdef}[3]{\phantomsection\hypertarget{coq:#1}{\text{#3}}}

\newcommand\eg{\emph{e.g.}}

\newcommand\stype[3]{{\coqdocnotation{\{}}\coqdocvar{#1}{\coqdocnotation{:}}{#2}
  {\coqdocnotation{\ensuremath{~|~}}}
\coqdocvariable{#3} \coqdocvar{#1}{\coqdocnotation{\}}}}

\newcommand\cvar[1]{\coqdocvar{#1}}
\newcommand\cpair[2]{{\coqdocnotation{(}}\coqdocvar{#1}
  {\coqdocnotation{;}} \coqdocvar{#2}{\coqdocnotation{)}}}
\newcommand\cind[1]{\coqdocinductive{#1}}

\newcommand\cdef[1]{\coqdocdefinition{#1}}
\newcommand\ckw[1]{\coqdockw{#1}}
\newcommand\ccons[1]{\coqdocconstructor{#1}}

\renewcommand\coqlibrary[3]{}

\usepackage{url}

\renewcommand{\coqlibrary}[3]{}

\toappear{}

\titlebanner{In Proceedings of the ACM Dynamic Languages Symposium (DLS) 2015}        

\title{Gradual Certified Programming in Coq}
\authorinfo{{\'E}ric Tanter}
           {PLEIAD Lab, Computer Science Dept (DCC)\\ University of
             Chile, Santiago, Chile}
           {etanter@dcc.uchile.cl}
\authorinfo{Nicolas Tabareau}
           {Inria\\ Nantes, France}
           {nicolas.tabareau@inria.fr}

\begin{document}
\maketitle

\begin{abstract} 
%
Expressive static typing disciplines are a powerful way to achieve
high-quality software. However, the adoption cost of such techniques
should not be under-estimated. Just like gradual typing allows for a
smooth transition from dynamically-typed to statically-typed programs,
it seems desirable to support a gradual path to certified programming.
We explore gradual certified programming in Coq, providing the
possibility to postpone the proofs of selected properties, and to
check ``at runtime'' whether the properties actually hold.  Casts can
be integrated with the implicit coercion mechanism of Coq to support
implicit cast insertion {\em {\`a} la} gradual typing.  Additionally,
when extracting Coq functions to mainstream languages, our encoding of
casts supports lifting assumed properties into runtime checks. Much to
our surprise, it is not necessary to extend Coq in any way to support
gradual certified programming. A simple mix of type classes and axioms
makes it possible to bring gradual certified programming to Coq in a
straightforward manner.
\end{abstract}





\section{Introduction}
\label{sec:introduction}

In {\em Certified Programming with Dependent Types}, Chlipala sketches
two main approaches to certified programming~\cite{cpdt}. In the
classical program verification approach, one separately writes a program, its
specification, and the proof that the program meets its
specification. A more effective technique is to exploit rich, dependent
types to integrate programming, specification and proving into a
single phase: specifications are expressed as types, as advocated by
Sheard et al.~\cite{sheardAl:foser2010} in what they call
{\em language-based verification}. While rich types are a powerful
way to achieve high-quality software, we believe that the adoption
cost of such techniques is not to be under-estimated. Therefore, it
seems desirable to support a {\em gradual} path to certified
programming with rich types, just like gradual typing allows for a
smooth transition from dynamically-typed to statically-typed
programs~\cite{siekTaha:sfp2006}. Indeed, the idea of progressively
strengthening programs through a form of gradual checking has already
been applied to a variety of type disciplines, like
typestates~\cite{wolffAl:ecoop2011,garciaAl:toplas2014}, information
flow typing and security
types~\cite{disneyFlanagan:stop2011,fennellThiemann:csf2013},
ownership types~\cite{sergeyClarke:esop2012}, annotated type
systems~\cite{thiemannFennell:esop2014}, and
effects~\cite{banadosAl:icfp2014}. Recent developments like
property-based testing for Coq~\cite{quickchick:coq2014} and
randomized testing based on refinement types
annotations~\cite{seidelAl:esop2015} are complementary efforts to make language-based verification more practical and attractive.

In this article, we consider a gradual path to certified programming
in Coq, so that programmers can
safely postpone providing some proof terms. 
We focus mostly (but
not exclusively) on {\em subset types}, which are the canonical way to
attach a property to a value.  Subset types are of the form
\mbox{\stype{a}{\cvar{A}}{P}}, denoting the elements \cvar{a} of type
\cvar{A} for which property \cvar{P a} holds. More precisely, an
inhabitant of \stype{a}{\cvar{A}}{P} is a dependent pair \cpair{a}{p},
where \cvar{a} is a term of type \cvar{A}, and \cvar{p} is a proof
term of type \cvar{P a}. 

%
  %
Constructing a value of type
  \stype{a}{\cvar{A}}{P} requires the associated proof term of type
  \cvar{P a}. Currently, Coq has two mechanisms to delay providing
  such a proof term. First, one can use
  \ckw{Program}, a facility that allows automatic coercions to subset
  types leaving proof obligations to be fulfilled after the definition
  is completed~\cite{sozeau:types2006}. This is only a small delay
  however, because one must discharge all pending obligations before
  being able to use the defined value.
  The second mechanism is to {\em admit} the said property, which
  makes Coq accept a definition on blind faith, without any proof.
  This solution is unsatisfactory from a gradual checking point of
  view, because it is {\em unsafe}: there is no delayed checking of
  the unproven property. Therefore, a function that expects a value
  with a given property might end up producing incorrect results.
The motto of gradual checking, {\em trust but
    verify}, is therefore not supported. 

  The main contribution of this work is to provide safe
  casts\footnote{Note that we use the name ``cast'' in the standard
    way~\cite{pierce:tapl} to denote a type assertion with an
    associated runtime check---this differs from the non-traditional
    use of ``cast'' in the Coq reference manual (1.2.10) where it
    refers to a static type assertion.} for Coq, paving the way for
  gradual certified programming, and to show that this is feasible
  entirely within standard Coq, without extending the underlying
  theory and implementation.  When casting a value \cvar{a} of type
  \cvar{A} to the rich type \stype{a}{\cvar{A}}{P}, the property
  \cvar{P a} is checked as needed, forbidding unsafe projection of the
  value of type \cvar{A} from the dependent pair. Note
  that because Coq is dependently-typed (ie. types can be dependent
  arbitrarily on computations and values), there is no rigid
  compile-time/runtime distinction: therefore, cast errors can
  possibly occur both as part of standard evaluation (triggered with
  \ckw{Eval}) and as part of type checking, during type conversion.

A key feature of our development is that we
support a smooth gradual path to certified programming
that avoids imposing a global monadic discipline to handle the
possibility of cast errors.
Technically, this is achieved thanks to the (possibly controversial)
choice of representing cast failures in Coq as an
  inconsistent axiom, so
that failed casts manifest as non-canonical normal forms (\eg~a normal form of type \cind{bool} is either \ccons{true},
\ccons{false}, or a cast failure). 

Section~\ref{sec:gradual-subset-types} provides an informal tour of
gradual certified programming with subset types in Coq, through a
number of examples. We then dive into the details of the approach,
namely type classes for decidability (Section~\ref{sec:decidable}) and
an axiomatic representation of casts (Section~\ref{sec:axioms}).
Section~\ref{sec:coerce} then discusses implicit cast
insertion {\em {\`a} la} gradual typing. 
Sections~\ref{sec:hocs} and~\ref{sec:hocd} focus on higher-order casts, with
both simple and dependent function types---the latter being subtly more
challenging. 
Section~\ref{sec:extraction} describes the use of casts to protect
functions extracted to mainstream languages that do not support subset
types.
Section~\ref{sec:properties}
briefly describes the main properties of our approach, which
follow directly from being entirely developed within standard
Coq.
Section~\ref{sec:records} shows how our approach scales
beyond subset types to other dependently-typed constructions, such as
record types, and illustrates how it is possible to customize the
inference of decision procedures.
Section~\ref{sec:related-work} discusses related work and
Section~\ref{sec:conclusion} concludes.

The code presented in this paper is available as a Coq library at \url{https://github.com/tabareau/Cocasse}.

\section{Gradual Certified Programming in Action}
\label{sec:gradual-subset-types}

We start by introducing gradual certified programming with subset
types through a number of examples of increasing complexity,
culminating in a small gradually certified compiler. For now, we only
appeal to the intuition of the reader; we discuss the technical details of the
approach in Section~\ref{sec:decidable} and beyond.

\coqlibrary{Casts.simple}{Library }{Casts.simple}

\begin{coqdoccode}
\coqdocemptyline
\coqdocemptyline
\end{coqdoccode}
\subsection{First Examples}

 \noindent We now show how casts behave with examples. In this paper, we denote the first and second
projections of a pair as $._1$ and $._2$ respectively.
First consider a simple definition that is rejected by Coq:\\

\coqdockw{Definition} \coqref{Casts.simple.n not ok}{\coqdocdefinition{n\_not\_ok}} : \{\coqdocvar{n}:\coqexternalref{nat}{http://coq.inria.fr/stdlib/Coq.Init.Datatypes}{\coqdocinductive{nat}} \ensuremath{|} \coqdocvar{n} $<$ 10\} := 5. \\

\noindent This definition is rejected, because the value should be a dependent pair, not just a natural number.  Using \ckw{Program},\footnote{\ckw{Program} is a definition facility that allows automatic coercions to subset types leaving proof obligations to be fulfilled after the
  definition is completed~\cite{sozeau:types2006}, but before the definition can be used.} we are left with the obligation to prove that 5 $<$ 10, which is arguably not too hard.

We could instead use our basic cast operator---denoted ?---to promote 5 to a value of type \{\coqdocvar{n}:\coqexternalref{nat}{http://coq.inria.fr/stdlib/Coq.Init.Datatypes}{\coqdocinductive{nat}} \ensuremath{|} \coqdocvar{n} $<$ 10\}. The semantics is that, if we ever need to evaluate \coqref{Casts.simple.n good}{\coqdocdefinition{n\_good}},\footnote{Coq does not impose any fixed reduction strategy. Instead, \ckw{Eval} is parameterized by a reduction strategy, called a conversion tactic, such as \coqdoctac{cbv} (aka. \coqdoctac{compute}), \coqdoctac{lazy}, \coqdoctac{hnf}, \coqdoctac{simpl}, etc.} we will check whether 5 is less than 10:
\begin{coqdoccode}
\coqdocemptyline
\coqdocnoindent
\coqdockw{Definition} \coqdef{Casts.simple.n good}{n\_good}{\coqdocdefinition{n\_good}} : \coqexternalref{:type scope:'x7B' x ':' x '|' x 'x7D'}{http://coq.inria.fr/stdlib/Coq.Init.Specif}{\coqdocnotation{\{}}\coqdocvar{n}\coqexternalref{:type scope:'x7B' x ':' x '|' x 'x7D'}{http://coq.inria.fr/stdlib/Coq.Init.Specif}{\coqdocnotation{:}}\coqexternalref{nat}{http://coq.inria.fr/stdlib/Coq.Init.Datatypes}{\coqdocinductive{nat}} \coqexternalref{:type scope:'x7B' x ':' x '|' x 'x7D'}{http://coq.inria.fr/stdlib/Coq.Init.Specif}{\coqdocnotation{\ensuremath{|}}} \coqdocvar{n} \coqexternalref{:nat scope:x '<' x}{http://coq.inria.fr/stdlib/Coq.Init.Peano}{\coqdocnotation{$<$}} 10\coqexternalref{:type scope:'x7B' x ':' x '|' x 'x7D'}{http://coq.inria.fr/stdlib/Coq.Init.Specif}{\coqdocnotation{\}}} := ? 5.\coqdoceol
\coqdocemptyline
\coqdocnoindent
\coqdockw{Eval} \coqdoctac{compute} \coqdoctac{in} \coqref{Casts.simple.n good}{\coqdocdefinition{n\_good}}.\coqdoceol
\end{coqdoccode}
\begin{verbatim}
= (5; Le.le_n_S 5 9 (...))
: {n : nat | n < 10}
\end{verbatim}
We indeed have a dependent pair, whose first component is the number 5 and second component is the proof that 5 $<$ 10 (elided). We can naturally project the number from the pair:\begin{coqdoccode}
\coqdocemptyline
\coqdocnoindent
\coqdockw{Eval} \coqdoctac{compute} \coqdoctac{in} \coqref{Casts.simple.n good}{\coqdocdefinition{n\_good}}\coqdocnotation{$._1$}.\coqdoceol
\end{coqdoccode}
\begin{verbatim}
= 5
: nat
\end{verbatim}

  Of course, we may be mistaken and believe that 15 $<$ 10: \begin{coqdoccode}
\coqdocemptyline
\coqdocnoindent
\coqdockw{Definition} \coqdef{Casts.simple.n bad}{n\_bad}{\coqdocdefinition{n\_bad}} : \coqexternalref{:type scope:'x7B' x ':' x '|' x 'x7D'}{http://coq.inria.fr/stdlib/Coq.Init.Specif}{\coqdocnotation{\{}}\coqdocvar{n}\coqexternalref{:type scope:'x7B' x ':' x '|' x 'x7D'}{http://coq.inria.fr/stdlib/Coq.Init.Specif}{\coqdocnotation{:}}\coqexternalref{nat}{http://coq.inria.fr/stdlib/Coq.Init.Datatypes}{\coqdocinductive{nat}} \coqexternalref{:type scope:'x7B' x ':' x '|' x 'x7D'}{http://coq.inria.fr/stdlib/Coq.Init.Specif}{\coqdocnotation{\ensuremath{|}}} \coqdocvar{n} \coqexternalref{:nat scope:x '<' x}{http://coq.inria.fr/stdlib/Coq.Init.Peano}{\coqdocnotation{$<$}} 10\coqexternalref{:type scope:'x7B' x ':' x '|' x 'x7D'}{http://coq.inria.fr/stdlib/Coq.Init.Specif}{\coqdocnotation{\}}} := ? 15.\coqdoceol
\coqdocemptyline
\end{coqdoccode}
\noindent  The cast error now manifests whenever we evaluate \coqref{Casts.simple.n bad}{\coqdocdefinition{n\_bad}}:
\begin{coqdoccode}
\coqdocemptyline
\coqdocnoindent
\coqdockw{Eval} \coqdoctac{compute} \coqdoctac{in} \coqref{Casts.simple.n bad}{\coqdocdefinition{n\_bad}}.\coqdoceol
\coqdocemptyline
\end{coqdoccode}
\begin{verbatim}
= failed_cast 15 (16 <= 10)
: {n : nat | n < 10}
\end{verbatim}

\noindent Importantly, a failed cast does not manifest as an exception or error, since Coq is a purely functional programming language. Instead, as we will explain further in Section~\ref{sec:axioms}, {\tt failed\_cast} is a normal form (ie., it cannot be further reduced) of the appropriate subset type, which indicates both the casted value (15) and the violated property (16 \ensuremath{\le} 10).

Crucially, because \coqref{Casts.simple.n bad}{\coqdocdefinition{n\_bad}} evaluates to a failed cast, we cannot project the natural number, since we do not even have a proper dependent pair: \begin{coqdoccode}
\coqdocemptyline
\coqdocnoindent
\coqdockw{Eval} \coqdoctac{compute} \coqdoctac{in} \coqref{Casts.simple.n bad}{\coqdocdefinition{n\_bad}}\coqdocnotation{$._1$}.\coqdoceol
\end{coqdoccode}
\begin{verbatim}
= let (a, _) := 
   failed_cast 15 (16 <= 10) in a 
: nat
\end{verbatim}

 At this point, it is worthwhile illustrating a major difference with the use of \coqdocvar{admit}, to which we alluded in the introduction. Consider that we use \coqdocvar{admit} to lie about 15: \begin{coqdoccode}
\coqdocemptyline
\coqdocnoindent
\coqdockw{Program Definition} \coqdef{Casts.simple.n real bad}{n\_real\_bad}{\coqdocdefinition{n\_real\_bad}} : \coqexternalref{:type scope:'x7B' x ':' x '|' x 'x7D'}{http://coq.inria.fr/stdlib/Coq.Init.Specif}{\coqdocnotation{\{}}\coqdocvar{n}\coqexternalref{:type scope:'x7B' x ':' x '|' x 'x7D'}{http://coq.inria.fr/stdlib/Coq.Init.Specif}{\coqdocnotation{:}}\coqexternalref{nat}{http://coq.inria.fr/stdlib/Coq.Init.Datatypes}{\coqdocinductive{nat}} \coqexternalref{:type scope:'x7B' x ':' x '|' x 'x7D'}{http://coq.inria.fr/stdlib/Coq.Init.Specif}{\coqdocnotation{\ensuremath{|}}} \coqdocvar{n} \coqexternalref{:nat scope:x '<' x}{http://coq.inria.fr/stdlib/Coq.Init.Peano}{\coqdocnotation{$<$}} 10\coqexternalref{:type scope:'x7B' x ':' x '|' x 'x7D'}{http://coq.inria.fr/stdlib/Coq.Init.Specif}{\coqdocnotation{\}}} :=\coqdoceol
\coqdocindent{1.00em}
15.\coqdoceol
\coqdocnoindent
\coqdockw{Next} \coqdockw{Obligation}. \coqdocvar{Admitted}.\coqdoceol
\coqdocemptyline
\end{coqdoccode}
\noindent In this case, \coqref{Casts.simple.n real bad}{\coqdocdefinition{n\_real\_bad}} is an actual dependent pair, with the use of \coqdocvar{proof\_admitted} (an inhabitant of \coqdocvar{False}) in the second component: \begin{coqdoccode}
\coqdocemptyline
\coqdocnoindent
\coqdockw{Eval} \coqdoctac{compute} \coqdoctac{in} \coqref{Casts.simple.n real bad}{\coqdocdefinition{n\_real\_bad}}.\coqdoceol
\end{coqdoccode}

\begin{verbatim}
= (15; match proof_admitted return (16 <= 10) 
       ...)
: {n : nat | n < 10}
\end{verbatim}

\noindent This means that we are able to project the number from \coqref{Casts.simple.n real bad}{\coqdocdefinition{n\_real\_bad}} without revealing the lie: \begin{coqdoccode}
\coqdocemptyline
\coqdocnoindent
\coqdockw{Eval} \coqdoctac{compute} \coqdoctac{in} \coqref{Casts.simple.n real bad}{\coqdocdefinition{n\_real\_bad}}\coqdocnotation{$._1$}.\coqdoceol
\coqdocemptyline
\end{coqdoccode}
\begin{verbatim}
= 15
: nat
\end{verbatim}

\subsection{Casting Lists}

\begin{coqdoccode}
\coqdocemptyline
\end{coqdoccode}
Casting a list of elements of type \coqdocvariable{A} to a list of elements of type \{\coqdocvariable{a}: \coqdocvariable{A} \ensuremath{|} \coqdocvariable{P} \coqdocvariable{a}\} simply means mapping the cast operator ? over the list.  For instance, we can claim that the following list is a list of 3s: \begin{coqdoccode}
\coqdocemptyline
\coqdocnoindent
\coqdockw{Definition} \coqdef{Casts.simple.list of 3}{list\_of\_3}{\coqdocdefinition{list\_of\_3}}: \coqexternalref{list}{http://coq.inria.fr/stdlib/Coq.Init.Datatypes}{\coqdocinductive{list}} \coqexternalref{:type scope:'x7B' x ':' x '|' x 'x7D'}{http://coq.inria.fr/stdlib/Coq.Init.Specif}{\coqdocnotation{\{}}\coqdocvar{n}\coqexternalref{:type scope:'x7B' x ':' x '|' x 'x7D'}{http://coq.inria.fr/stdlib/Coq.Init.Specif}{\coqdocnotation{:}}\coqexternalref{nat}{http://coq.inria.fr/stdlib/Coq.Init.Datatypes}{\coqdocinductive{nat}} \coqexternalref{:type scope:'x7B' x ':' x '|' x 'x7D'}{http://coq.inria.fr/stdlib/Coq.Init.Specif}{\coqdocnotation{\ensuremath{|}}} \coqdocvar{n} \coqexternalref{:type scope:x '=' x}{http://coq.inria.fr/stdlib/Coq.Init.Logic}{\coqdocnotation{=}} 3\coqexternalref{:type scope:'x7B' x ':' x '|' x 'x7D'}{http://coq.inria.fr/stdlib/Coq.Init.Specif}{\coqdocnotation{\}}} := \coqdoceol
\coqdocindent{1.00em}
\coqexternalref{map}{http://coq.inria.fr/stdlib/Coq.Lists.List}{\coqdocdefinition{map}} ? (3 \coqexternalref{:list scope:x '::' x}{http://coq.inria.fr/stdlib/Coq.Init.Datatypes}{\coqdocnotation{::}} 2 \coqexternalref{:list scope:x '::' x}{http://coq.inria.fr/stdlib/Coq.Init.Datatypes}{\coqdocnotation{::}} 1 \coqexternalref{:list scope:x '::' x}{http://coq.inria.fr/stdlib/Coq.Init.Datatypes}{\coqdocnotation{::}} \coqexternalref{nil}{http://coq.inria.fr/stdlib/Coq.Init.Datatypes}{\coqdocconstructor{nil}}).\coqdoceol
\coqdocemptyline
\end{coqdoccode}
If we force the evaluation of \coqref{Casts.simple.list of 3}{\coqdocdefinition{list\_of\_3}}, we obtain a list of elements that are either 3 with the proof that 3 = 3, or a failed cast: \begin{coqdoccode}
\coqdocemptyline
\coqdocnoindent
\coqdockw{Eval} \coqdoctac{compute} \coqdoctac{in} \coqref{Casts.simple.list of 3}{\coqdocdefinition{list\_of\_3}}.\coqdoceol
\coqdocemptyline
\end{coqdoccode}

\begin{verbatim}
= (3; eq_refl)
    :: failed_cast 2 (2 = 3)
       :: failed_cast 1 (1 = 3) :: nil
: list {n : nat | n = 3}
\end{verbatim}

 Note the difference between a list of type \coqexternalref{list}{http://coq.inria.fr/stdlib/Coq.Init.Datatypes}{\coqdocinductive{list}} \{\coqdocvariable{a} : \coqdocvariable{A} \ensuremath{|} \coqdocvariable{P} \coqdocvariable{a}\} and a list of type \{\coqdocvar{l} : \coqexternalref{list}{http://coq.inria.fr/stdlib/Coq.Init.Datatypes}{\coqdocinductive{list}} \coqdocvariable{A} \ensuremath{|} \coqdocvariable{P} \coqdocvar{l}\}. While the former expresses that each element \coqdocvariable{a} of the list satisfies \coqdocvariable{P} \coqdocvariable{a}, the latter expresses that the list \coqdocvar{l} as a whole satisfies \coqdocvariable{P} \coqdocvar{l}.
Casting works similarly for other inductively-defined structures. \begin{coqdoccode}
\coqdocemptyline
\end{coqdoccode}

\subsection{A Gradually Certified Compiler}
\label{sec:compile}
\label{sec:grad-cert-comp}

\coqlibrary{Casts.compiler}{Library }{Casts.compiler}

\begin{coqdoccode}
\coqdocemptyline
\coqdocemptyline
\end{coqdoccode}
We now show how to apply casts to a (slightly) less artificial example. Consider a certified compiler of arithmetic expressions, adapted from Chapter 2 of CPDT~\cite{cpdt}. 

 \paragraph{Source language.}

The source language includes the following binary operations:
\begin{coqdoccode}
\coqdocemptyline
\coqdocnoindent
\coqdockw{Inductive} \coqdef{Casts.compiler.binop}{binop}{\coqdocinductive{binop}} : \coqdockw{Set} := \coqdef{Casts.compiler.Plus}{Plus}{\coqdocconstructor{Plus}} \ensuremath{|} \coqdef{Casts.compiler.Minus}{Minus}{\coqdocconstructor{Minus}} \ensuremath{|} \coqdef{Casts.compiler.Times}{Times}{\coqdocconstructor{Times}}.\coqdoceol
\coqdocemptyline
\end{coqdoccode}
Expressions are either constants or applications of a binary operation: \begin{coqdoccode}
\coqdocemptyline
\coqdocnoindent
\coqdockw{Inductive} \coqdef{Casts.compiler.exp}{exp}{\coqdocinductive{exp}} : \coqdockw{Set} := \coqdoceol
\coqdocindent{1.00em}
\ensuremath{|} \coqdef{Casts.compiler.Const}{Const}{\coqdocconstructor{Const}} : \coqexternalref{nat}{http://coq.inria.fr/stdlib/Coq.Init.Datatypes}{\coqdocinductive{nat}} \coqexternalref{:type scope:x '->' x}{http://coq.inria.fr/stdlib/Coq.Init.Logic}{\coqdocnotation{\ensuremath{\rightarrow}}} \coqref{Casts.compiler.exp}{\coqdocinductive{exp}}\coqdoceol
\coqdocindent{1.00em}
\ensuremath{|} \coqdef{Casts.compiler.Binop}{Binop}{\coqdocconstructor{Binop}} : \coqref{Casts.compiler.binop}{\coqdocinductive{binop}} \coqexternalref{:type scope:x '->' x}{http://coq.inria.fr/stdlib/Coq.Init.Logic}{\coqdocnotation{\ensuremath{\rightarrow}}} \coqref{Casts.compiler.exp}{\coqdocinductive{exp}} \coqexternalref{:type scope:x '->' x}{http://coq.inria.fr/stdlib/Coq.Init.Logic}{\coqdocnotation{\ensuremath{\rightarrow}}} \coqref{Casts.compiler.exp}{\coqdocinductive{exp}} \coqexternalref{:type scope:x '->' x}{http://coq.inria.fr/stdlib/Coq.Init.Logic}{\coqdocnotation{\ensuremath{\rightarrow}}} \coqref{Casts.compiler.exp}{\coqdocinductive{exp}}.\coqdoceol
\coqdocemptyline
\end{coqdoccode}
The semantics of binary operations is as expected: \begin{coqdoccode}
\coqdocemptyline
\coqdocnoindent
\coqdockw{Definition} \coqdef{Casts.compiler.evalBinop}{evalBinop}{\coqdocdefinition{evalBinop}} (\coqdocvar{b}: \coqref{Casts.compiler.binop}{\coqdocinductive{binop}}) : \coqexternalref{nat}{http://coq.inria.fr/stdlib/Coq.Init.Datatypes}{\coqdocinductive{nat}} \coqexternalref{:type scope:x '->' x}{http://coq.inria.fr/stdlib/Coq.Init.Logic}{\coqdocnotation{\ensuremath{\rightarrow}}} \coqexternalref{nat}{http://coq.inria.fr/stdlib/Coq.Init.Datatypes}{\coqdocinductive{nat}} \coqexternalref{:type scope:x '->' x}{http://coq.inria.fr/stdlib/Coq.Init.Logic}{\coqdocnotation{\ensuremath{\rightarrow}}} \coqexternalref{nat}{http://coq.inria.fr/stdlib/Coq.Init.Datatypes}{\coqdocinductive{nat}} :=\coqdoceol
\coqdocindent{1.00em}
\coqdockw{match} \coqdocvariable{b} \coqdockw{with}\coqdoceol
\coqdocindent{2.00em}
\ensuremath{|} \coqref{Casts.compiler.Plus}{\coqdocconstructor{Plus}} \ensuremath{\Rightarrow} \coqexternalref{plus}{http://coq.inria.fr/stdlib/Coq.Init.Peano}{\coqdocabbreviation{plus}}\coqdoceol
\coqdocindent{2.00em}
\ensuremath{|} \coqref{Casts.compiler.Minus}{\coqdocconstructor{Minus}} \ensuremath{\Rightarrow} \coqexternalref{sub}{http://coq.inria.fr/stdlib/Coq.Init.Nat}{\coqdocdefinition{sub}}\coqdoceol
\coqdocindent{2.00em}
\ensuremath{|} \coqref{Casts.compiler.Times}{\coqdocconstructor{Times}} \ensuremath{\Rightarrow} \coqexternalref{mult}{http://coq.inria.fr/stdlib/Coq.Init.Peano}{\coqdocabbreviation{mult}}\coqdoceol
\coqdocindent{1.00em}
\coqdockw{end}.\coqdoceol
\coqdocemptyline
\end{coqdoccode}
So is the semantics of evaluating expressions: \begin{coqdoccode}
\coqdocemptyline
\coqdocnoindent
\coqdockw{Fixpoint} \coqdef{Casts.compiler.evalExp}{evalExp}{\coqdocdefinition{evalExp}} (\coqdocvar{e}: \coqref{Casts.compiler.exp}{\coqdocinductive{exp}}) : \coqexternalref{nat}{http://coq.inria.fr/stdlib/Coq.Init.Datatypes}{\coqdocinductive{nat}} :=\coqdoceol
\coqdocindent{1.00em}
\coqdockw{match} \coqdocvariable{e} \coqdockw{with}\coqdoceol
\coqdocindent{2.00em}
\ensuremath{|} \coqref{Casts.compiler.Const}{\coqdocconstructor{Const}} \coqdocvar{n} \ensuremath{\Rightarrow} \coqdocvar{n}\coqdoceol
\coqdocindent{2.00em}
\ensuremath{|} \coqref{Casts.compiler.Binop}{\coqdocconstructor{Binop}} \coqdocvar{b} \coqdocvar{e1} \coqdocvar{e2} \ensuremath{\Rightarrow}\coqdoceol
\coqdocindent{3.00em}
(\coqref{Casts.compiler.evalBinop}{\coqdocdefinition{evalBinop}} \coqdocvar{b}) (\coqref{Casts.compiler.evalExp}{\coqdocdefinition{evalExp}} \coqdocvar{e1}) (\coqref{Casts.compiler.evalExp}{\coqdocdefinition{evalExp}} \coqdocvar{e2})\coqdoceol
\coqdocindent{1.00em}
\coqdockw{end}.\coqdoceol
\coqdocemptyline
\coqdocemptyline
\end{coqdoccode}
\paragraph{Stack machine.} 

 We now introduce the intermediate language of instructions for a stack machine: \begin{coqdoccode}
\coqdocemptyline
\coqdocnoindent
\coqdockw{Inductive} \coqdef{Casts.compiler.instr}{instr}{\coqdocinductive{instr}} : \coqdockw{Set} :=\coqdoceol
\coqdocindent{1.00em}
\ensuremath{|} \coqdef{Casts.compiler.iConst}{iConst}{\coqdocconstructor{iConst}} : \coqexternalref{nat}{http://coq.inria.fr/stdlib/Coq.Init.Datatypes}{\coqdocinductive{nat}} \coqexternalref{:type scope:x '->' x}{http://coq.inria.fr/stdlib/Coq.Init.Logic}{\coqdocnotation{\ensuremath{\rightarrow}}} \coqref{Casts.compiler.instr}{\coqdocinductive{instr}}\coqdoceol
\coqdocindent{1.00em}
\ensuremath{|} \coqdef{Casts.compiler.iBinop}{iBinop}{\coqdocconstructor{iBinop}} : \coqref{Casts.compiler.binop}{\coqdocinductive{binop}} \coqexternalref{:type scope:x '->' x}{http://coq.inria.fr/stdlib/Coq.Init.Logic}{\coqdocnotation{\ensuremath{\rightarrow}}} \coqref{Casts.compiler.instr}{\coqdocinductive{instr}}.\coqdoceol
\coqdocemptyline
\end{coqdoccode}
A program is a list of instructions, and a stack is a list of natural numbers: \begin{coqdoccode}
\coqdocemptyline
\coqdocnoindent
\coqdockw{Definition} \coqdef{Casts.compiler.prog}{prog}{\coqdocdefinition{prog}} := \coqexternalref{list}{http://coq.inria.fr/stdlib/Coq.Init.Datatypes}{\coqdocinductive{list}} \coqref{Casts.compiler.instr}{\coqdocinductive{instr}}.\coqdoceol
\coqdocnoindent
\coqdockw{Definition} \coqdef{Casts.compiler.stack}{stack}{\coqdocdefinition{stack}} := \coqexternalref{list}{http://coq.inria.fr/stdlib/Coq.Init.Datatypes}{\coqdocinductive{list}} \coqexternalref{nat}{http://coq.inria.fr/stdlib/Coq.Init.Datatypes}{\coqdocinductive{nat}}.\coqdoceol
\coqdocemptyline
\end{coqdoccode}
Executing an instruction on a given stack produces either a new stack or \coqexternalref{None}{http://coq.inria.fr/stdlib/Coq.Init.Datatypes}{\coqdocconstructor{None}} if the stack is in an invalid state: \begin{coqdoccode}
\coqdocemptyline
\coqdocnoindent
\coqdockw{Definition} \coqdef{Casts.compiler.runInstr}{runInstr}{\coqdocdefinition{runInstr}} (\coqdocvar{i}: \coqref{Casts.compiler.instr}{\coqdocinductive{instr}}) (\coqdocvar{s}: \coqref{Casts.compiler.stack}{\coqdocdefinition{stack}}): \coqexternalref{option}{http://coq.inria.fr/stdlib/Coq.Init.Datatypes}{\coqdocinductive{option}} \coqref{Casts.compiler.stack}{\coqdocdefinition{stack}} :=\coqdoceol
\coqdocindent{1.00em}
\coqdockw{match} \coqdocvariable{i} \coqdockw{with}\coqdoceol
\coqdocindent{2.00em}
\ensuremath{|} \coqref{Casts.compiler.iConst}{\coqdocconstructor{iConst}} \coqdocvar{n} \ensuremath{\Rightarrow} \coqexternalref{Some}{http://coq.inria.fr/stdlib/Coq.Init.Datatypes}{\coqdocconstructor{Some}} (\coqdocvar{n} \coqexternalref{:list scope:x '::' x}{http://coq.inria.fr/stdlib/Coq.Init.Datatypes}{\coqdocnotation{::}} \coqdocvariable{s})\coqdoceol
\coqdocindent{2.00em}
\ensuremath{|} \coqref{Casts.compiler.iBinop}{\coqdocconstructor{iBinop}} \coqdocvar{b} \ensuremath{\Rightarrow} \coqdoceol
\coqdocindent{3.00em}
\coqdockw{match} \coqdocvariable{s} \coqdockw{with}\coqdoceol
\coqdocindent{4.00em}
\ensuremath{|} \coqdocvar{arg1} \coqexternalref{:list scope:x '::' x}{http://coq.inria.fr/stdlib/Coq.Init.Datatypes}{\coqdocnotation{::}} \coqdocvar{arg2} \coqexternalref{:list scope:x '::' x}{http://coq.inria.fr/stdlib/Coq.Init.Datatypes}{\coqdocnotation{::}} \coqdocvar{s'} \ensuremath{\Rightarrow}\coqdoceol
\coqdocindent{5.00em}
\coqexternalref{Some}{http://coq.inria.fr/stdlib/Coq.Init.Datatypes}{\coqdocconstructor{Some}} ((\coqref{Casts.compiler.evalBinop}{\coqdocdefinition{evalBinop}} \coqdocvar{b}) \coqdocvar{arg1} \coqdocvar{arg2} \coqexternalref{:list scope:x '::' x}{http://coq.inria.fr/stdlib/Coq.Init.Datatypes}{\coqdocnotation{::}} \coqdocvar{s'})\coqdoceol
\coqdocindent{4.00em}
\ensuremath{|} \coqdocvar{\_} \ensuremath{\Rightarrow} \coqexternalref{None}{http://coq.inria.fr/stdlib/Coq.Init.Datatypes}{\coqdocconstructor{None}}\coqdoceol
\coqdocindent{3.00em}
\coqdockw{end}\coqdoceol
\coqdocindent{1.00em}
\coqdockw{end}.\coqdoceol
\coqdocemptyline
\end{coqdoccode}
\noindent Running a program simply executes each instruction, recursively: \begin{coqdoccode}
\coqdocemptyline
\coqdocnoindent
\coqdockw{Fixpoint} \coqdef{Casts.compiler.runProg}{runProg}{\coqdocdefinition{runProg}} (\coqdocvar{p}: \coqref{Casts.compiler.prog}{\coqdocdefinition{prog}}) (\coqdocvar{s}: \coqref{Casts.compiler.stack}{\coqdocdefinition{stack}}): \coqexternalref{option}{http://coq.inria.fr/stdlib/Coq.Init.Datatypes}{\coqdocinductive{option}} \coqref{Casts.compiler.stack}{\coqdocdefinition{stack}} :=\coqdoceol
\coqdocindent{1.00em}
\coqdockw{match} \coqdocvariable{p} \coqdockw{with}\coqdoceol
\coqdocindent{2.00em}
\ensuremath{|} \coqexternalref{nil}{http://coq.inria.fr/stdlib/Coq.Init.Datatypes}{\coqdocconstructor{nil}} \ensuremath{\Rightarrow} \coqexternalref{Some}{http://coq.inria.fr/stdlib/Coq.Init.Datatypes}{\coqdocconstructor{Some}} \coqdocvariable{s}\coqdoceol
\coqdocindent{2.00em}
\ensuremath{|} \coqdocvar{i} \coqexternalref{:list scope:x '::' x}{http://coq.inria.fr/stdlib/Coq.Init.Datatypes}{\coqdocnotation{::}} \coqdocvar{p'} \ensuremath{\Rightarrow} \coqdockw{match} \coqref{Casts.compiler.runInstr}{\coqdocdefinition{runInstr}} \coqdocvar{i} \coqdocvariable{s} \coqdockw{with}\coqdoceol
\coqdocindent{9.50em}
\ensuremath{|} \coqexternalref{None}{http://coq.inria.fr/stdlib/Coq.Init.Datatypes}{\coqdocconstructor{None}} \ensuremath{\Rightarrow} \coqexternalref{None}{http://coq.inria.fr/stdlib/Coq.Init.Datatypes}{\coqdocconstructor{None}}\coqdoceol
\coqdocindent{9.50em}
\ensuremath{|} \coqexternalref{Some}{http://coq.inria.fr/stdlib/Coq.Init.Datatypes}{\coqdocconstructor{Some}} \coqdocvar{s'} \ensuremath{\Rightarrow} \coqref{Casts.compiler.runProg}{\coqdocdefinition{runProg}} \coqdocvar{p'} \coqdocvar{s'}\coqdoceol
\coqdocindent{8.50em}
\coqdockw{end}\coqdoceol
\coqdocindent{1.00em}
\coqdockw{end}.\coqdoceol
\coqdocemptyline
\coqdocemptyline
\end{coqdoccode}
 \paragraph{Compiler.}

We now turn to the compiler, which is a recursive function that produces a program given an expression: \begin{coqdoccode}
\coqdocemptyline
\coqdocnoindent
\coqdockw{Fixpoint} \coqdef{Casts.compiler.compile}{compile}{\coqdocdefinition{compile}} (\coqdocvar{e}: \coqref{Casts.compiler.exp}{\coqdocinductive{exp}}) : \coqref{Casts.compiler.prog}{\coqdocdefinition{prog}} :=\coqdoceol
\coqdocindent{1.00em}
\coqdockw{match} \coqdocvariable{e} \coqdockw{with}\coqdoceol
\coqdocindent{1.00em}
\ensuremath{|} \coqref{Casts.compiler.Const}{\coqdocconstructor{Const}} \coqdocvar{n} \ensuremath{\Rightarrow} \coqref{Casts.compiler.iConst}{\coqdocconstructor{iConst}} \coqdocvar{n} \coqexternalref{:list scope:x '::' x}{http://coq.inria.fr/stdlib/Coq.Init.Datatypes}{\coqdocnotation{::}} \coqexternalref{nil}{http://coq.inria.fr/stdlib/Coq.Init.Datatypes}{\coqdocconstructor{nil}}\coqdoceol
\coqdocindent{1.00em}
\ensuremath{|} \coqref{Casts.compiler.Binop}{\coqdocconstructor{Binop}} \coqdocvar{b} \coqdocvar{e1} \coqdocvar{e2} \ensuremath{\Rightarrow}\coqdoceol
\coqdocindent{2.00em}
\coqref{Casts.compiler.compile}{\coqdocdefinition{compile}} \coqdocvar{e1} \coqexternalref{:list scope:x '++' x}{http://coq.inria.fr/stdlib/Coq.Init.Datatypes}{\coqdocnotation{++}} \coqref{Casts.compiler.compile}{\coqdocdefinition{compile}} \coqdocvar{e2} \coqexternalref{:list scope:x '++' x}{http://coq.inria.fr/stdlib/Coq.Init.Datatypes}{\coqdocnotation{++}} \coqref{Casts.compiler.iBinop}{\coqdocconstructor{iBinop}} \coqdocvar{b} \coqexternalref{:list scope:x '::' x}{http://coq.inria.fr/stdlib/Coq.Init.Datatypes}{\coqdocnotation{::}} \coqexternalref{nil}{http://coq.inria.fr/stdlib/Coq.Init.Datatypes}{\coqdocconstructor{nil}}\coqdoceol
\coqdocindent{1.00em}
\coqdockw{end}. \coqdocemptyline
\end{coqdoccode}
\noindent \emph{Hint: there is a bug!} 

 \paragraph{Correct?}
\label{sct:correct}

Of course, one would like to be sure that \coqref{Casts.compiler.compile}{\coqdocdefinition{compile}} is a {\em correct} compiler. 
The traditional way of certifying the compiler is to state and prove a correctness theorem. In CPDT, the compiler correctness is stated as follows: \begin{coqdoccode}
\coqdocemptyline
\coqdocnoindent
\coqdockw{Theorem} \coqdef{Casts.compiler.compile correct}{compile\_correct}{\coqdoclemma{compile\_correct}} : \coqdockw{\ensuremath{\forall}} (\coqdocvar{e}: \coqref{Casts.compiler.exp}{\coqdocinductive{exp}}),\coqdoceol
\coqdocindent{1.00em}
\coqref{Casts.compiler.runProg}{\coqdocdefinition{runProg}} (\coqref{Casts.compiler.compile}{\coqdocdefinition{compile}} \coqdocvariable{e}) \coqexternalref{nil}{http://coq.inria.fr/stdlib/Coq.Init.Datatypes}{\coqdocconstructor{nil}} \coqexternalref{:type scope:x '=' x}{http://coq.inria.fr/stdlib/Coq.Init.Logic}{\coqdocnotation{=}} \coqexternalref{Some}{http://coq.inria.fr/stdlib/Coq.Init.Datatypes}{\coqdocconstructor{Some}} (\coqref{Casts.compiler.evalExp}{\coqdocdefinition{evalExp}} \coqdocvariable{e} \coqexternalref{:list scope:x '::' x}{http://coq.inria.fr/stdlib/Coq.Init.Datatypes}{\coqdocnotation{::}} \coqexternalref{nil}{http://coq.inria.fr/stdlib/Coq.Init.Datatypes}{\coqdocconstructor{nil}}).\coqdoceol
\end{coqdoccode}
\vspace{2mm}\noindent Namely, executing the program returned by the compiler on an empty stack yields a well-formed stack with one element on top, which is the same value as interpreting the source program directly. 

 It turns out that the theorem cannot be proven directly by induction on expressions because of the use of \coqexternalref{nil}{http://coq.inria.fr/stdlib/Coq.Init.Datatypes}{\coqdocconstructor{nil}} in the statement of the theorem: the induction hypotheses are not useful. Instead, one has to state a generalized version of the theorem, whose proof does go by induction, and then prove \coqref{Casts.compiler.compile correct}{\coqdoclemma{compile\_correct}} as a corollary~\cite{cpdt}.

Instead of going into such a burden as soon as the compiler is defined, one may want to assert correctness and have it checked dynamically. With our framework, it is possible to simply cast the compiler to a correct compiler. To make the following exposition clearer, we first define what a correct program (for a given source expression) is: \\[2mm]  \begin{coqdoccode}
\coqdocnoindent
\coqdockw{Definition} \coqdef{Casts.compiler.correct prog}{correct\_prog}{\coqdocdefinition{correct\_prog}} (\coqdocvar{e}: \coqref{Casts.compiler.exp}{\coqdocinductive{exp}}) (\coqdocvar{p}: \coqref{Casts.compiler.prog}{\coqdocdefinition{prog}}) : \coqdockw{Prop} := \coqdoceol
\coqdocindent{1.00em}
\coqref{Casts.compiler.runProg}{\coqdocdefinition{runProg}} \coqdocvariable{p} \coqexternalref{nil}{http://coq.inria.fr/stdlib/Coq.Init.Datatypes}{\coqdocconstructor{nil}} \coqexternalref{:type scope:x '=' x}{http://coq.inria.fr/stdlib/Coq.Init.Logic}{\coqdocnotation{=}} \coqexternalref{Some}{http://coq.inria.fr/stdlib/Coq.Init.Datatypes}{\coqdocconstructor{Some}} (\coqref{Casts.compiler.evalExp}{\coqdocdefinition{evalExp}} \coqdocvariable{e} \coqexternalref{:list scope:x '::' x}{http://coq.inria.fr/stdlib/Coq.Init.Datatypes}{\coqdocnotation{::}} \coqexternalref{nil}{http://coq.inria.fr/stdlib/Coq.Init.Datatypes}{\coqdocconstructor{nil}}).\coqdoceol
\coqdocemptyline
\end{coqdoccode}

To exploit gradual certified programming to claim that
\coqref{Casts.compiler.compile}{\coqdocdefinition{compile}} is correct using a cast, we could try to use our cast
operator ?, to attempt to give \coqref{Casts.compiler.compile}{\coqdocdefinition{compile}} the type \{\coqdocvar{f}: \coqref{Casts.compiler.exp}{\coqdocinductive{exp}}
\ensuremath{\rightarrow} \coqref{Casts.compiler.prog}{\coqdocdefinition{prog}} \ensuremath{|} \coqdockw{\ensuremath{\forall}} \coqdocvariable{e}:\coqref{Casts.compiler.exp}{\coqdocinductive{exp}}, \coqref{Casts.compiler.correct prog}{\coqdocdefinition{correct\_prog}} \coqdocvariable{e} (\coqdocvar{f} \coqdocvariable{e})\}. This is however undecidable because
the property quantifies over all expressions. (In fact, such a cast is rejected by our framework, as discussed in Section~\ref{sec:decidable}.) Instead, we need to
resort to a higher-order cast operator, denoted $\forall?$, which can
lazily check that the compiler is ``apparently'' correct by checking
that it produces correct programs whenever it is used: \begin{coqdoccode}
\coqdocemptyline
\coqdocnoindent
\coqdockw{Definition} \coqdef{Casts.compiler.correct comp}{correct\_comp}{\coqdocdefinition{correct\_comp}} := \coqdoceol
\coqdocindent{1.00em}
\coqdockw{\ensuremath{\forall}} \coqdocvar{e}: \coqref{Casts.compiler.exp}{\coqdocinductive{exp}}, \coqexternalref{:type scope:'x7B' x ':' x '|' x 'x7D'}{http://coq.inria.fr/stdlib/Coq.Init.Specif}{\coqdocnotation{\{}}\coqdocvar{p}\coqexternalref{:type scope:'x7B' x ':' x '|' x 'x7D'}{http://coq.inria.fr/stdlib/Coq.Init.Specif}{\coqdocnotation{:}} \coqref{Casts.compiler.prog}{\coqdocdefinition{prog}} \coqexternalref{:type scope:'x7B' x ':' x '|' x 'x7D'}{http://coq.inria.fr/stdlib/Coq.Init.Specif}{\coqdocnotation{\ensuremath{|}}} \coqref{Casts.compiler.correct prog}{\coqdocdefinition{correct\_prog}} \coqdocvariable{e} \coqdocvar{p} \coqexternalref{:type scope:'x7B' x ':' x '|' x 'x7D'}{http://coq.inria.fr/stdlib/Coq.Init.Specif}{\coqdocnotation{\}}}.\coqdoceol
\coqdocemptyline
\coqdocnoindent
\coqdockw{Definition} \coqdef{Casts.compiler.compile ok}{compile\_ok}{\coqdocdefinition{compile\_ok}} : \coqref{Casts.compiler.correct comp}{\coqdocdefinition{correct\_comp}} := $\forall?$ \coqref{Casts.compiler.compile}{\coqdocdefinition{compile}}.\coqdoceol
\coqdocemptyline
\end{coqdoccode}
 \noindent Let us now exercise \coqref{Casts.compiler.compile ok}{\coqdocdefinition{compile\_ok}}. The following evaluation succeeds: \begin{coqdoccode}
\coqdocemptyline
\coqdocnoindent
\coqdockw{Eval} \coqdoctac{compute} \coqdoctac{in} \coqdoceol
\coqdocindent{1.00em}
\coqref{Casts.compiler.compile ok}{\coqdocdefinition{compile\_ok}} (\coqref{Casts.compiler.Binop}{\coqdocconstructor{Binop}} \coqref{Casts.compiler.Plus}{\coqdocconstructor{Plus}} (\coqref{Casts.compiler.Const}{\coqdocconstructor{Const}} 2) (\coqref{Casts.compiler.Const}{\coqdocconstructor{Const}} 2)).\coqdoceol
\coqdocemptyline
\end{coqdoccode}

\begin{verbatim}
= (iConst 2 :: iConst 2 :: iBinop Plus :: nil; 
   eq_refl)
: {p : prog | correct_prog ...}
\end{verbatim}

  \noindent However, the cast fails when using a (non-commutative!) subtraction operation: \begin{coqdoccode}
\coqdocemptyline
\coqdocnoindent
\coqdockw{Eval} \coqdoctac{compute} \coqdoctac{in} \coqdoceol
\coqdocindent{1.00em}
\coqref{Casts.compiler.compile ok}{\coqdocdefinition{compile\_ok}} (\coqref{Casts.compiler.Binop}{\coqdocconstructor{Binop}} \coqref{Casts.compiler.Minus}{\coqdocconstructor{Minus}} (\coqref{Casts.compiler.Const}{\coqdocconstructor{Const}} 2) (\coqref{Casts.compiler.Const}{\coqdocconstructor{Const}} 1)).\coqdoceol
\coqdocemptyline
\end{coqdoccode}

\begin{verbatim}
= failed_cast (iConst 2 :: iConst 1 
                        :: iBinop Minus :: nil)
            (Some (0 :: nil) = Some (1 :: nil))
: {p : prog | correct_prog ...}
\end{verbatim}

 Indeed, the compiler incorrectly compiles application nodes, compiling sub-expressions in the wrong order! The last argument of {\tt failed\_cast}---the invalid property---is explicit about what went wrong: the compiler produced a program that returns 0, while the interpreter returned 1.

 Finally, suppose we write a \coqref{Casts.compiler.runc}{\coqdocdefinition{runc}} function that requires a correct compiler as argument: \begin{coqdoccode}
\coqdocemptyline
\coqdocnoindent
\coqdockw{Definition} \coqdef{Casts.compiler.runc}{runc}{\coqdocdefinition{runc}} (\coqdocvar{c}: \coqref{Casts.compiler.correct comp}{\coqdocdefinition{correct\_comp}}) (\coqdocvar{e}: \coqref{Casts.compiler.exp}{\coqdocinductive{exp}}) :=\coqdoceol
\coqdocindent{1.00em}
\coqref{Casts.compiler.runProg}{\coqdocdefinition{runProg}} \coqdocnotation{(}\coqdocvariable{c} \coqdocvariable{e}\coqdocnotation{$)._1$} \coqexternalref{nil}{http://coq.inria.fr/stdlib/Coq.Init.Datatypes}{\coqdocconstructor{nil}}.\coqdoceol
\coqdocemptyline
\end{coqdoccode}
We can use the cast framework to pass \coqref{Casts.compiler.compile}{\coqdocdefinition{compile}} as argument, but in case the compiler behaves badly, \coqref{Casts.compiler.runc}{\coqdocdefinition{runc}} fails because it cannot apply the projection $._1$ to a failed cast: \begin{coqdoccode}
\coqdocemptyline
\coqdocnoindent
\coqdockw{Eval} \coqdoctac{compute} \coqdoctac{in} \coqref{Casts.compiler.runc}{\coqdocdefinition{runc}} $(\forall?$ \coqref{Casts.compiler.compile}{\coqdocdefinition{compile}}) \coqdoceol
\coqdocindent{9.00em}
(\coqref{Casts.compiler.Binop}{\coqdocconstructor{Binop}} \coqref{Casts.compiler.Minus}{\coqdocconstructor{Minus}} (\coqref{Casts.compiler.Const}{\coqdocconstructor{Const}} 2) (\coqref{Casts.compiler.Const}{\coqdocconstructor{Const}} 1)).\coqdoceol
\coqdocemptyline
\end{coqdoccode}
\begin{verbatim}
= ...
  (let (a, _) := failed_cast 
    (Some (0 :: nil) = Some (1 :: nil)) ...
: option stack
\end{verbatim}

 \noindent Again, note that if we had used \coqdocvar{admit} to lie about \coqref{Casts.compiler.compile}{\coqdocdefinition{compile}}, then \coqref{Casts.compiler.runc}{\coqdocdefinition{runc}} would not have detected the violation of the property, and would have therefore returned an incorrect result. \begin{coqdoccode}
\end{coqdoccode}

\coqlibrary{Casts.cast1}{Library }{Casts.cast1}

\begin{coqdoccode}
\coqdocemptyline
\coqdocemptyline
\end{coqdoccode}

\section{Casts and Decidability}

\label{sec:decidable}

What exactly does it mean to {\em cast} a value \coqdocvariable{a} of type \coqdocvariable{A} to a value of the rich type \{\coqdocvariable{a} : \coqdocvariable{A} \ensuremath{|} \coqdocvariable{P} \coqdocvariable{a}\}?
There are two challenges to be addressed. First, because we are talking about {\em safe} casts, it must be possible
to check, for a given \coqdocvariable{a}, whether \coqdocvariable{P} \coqdocvariable{a} holds. This means that \coqdocvariable{P} \coqdocvariable{a} must be {\em decidable}. 
Second, because it may be the case that \coqdocvariable{P} \coqdocvariable{a} does not hold, we must consider how to represent such a ``cast error'', considering that Coq does not have any built-in exception mechanism. For decidability, we exploit the type class mechanism of Coq, as explained in this section. For failed casts, we exploit axioms (Section~\ref{sec:axioms}).

\subsection{Decidable Properties}

The \coqref{Casts.cast1.Decidable}{\coqdocrecord{$\mathsf{Decidable}$}} type class, which is used in the Coq/HoTT
library\footnote{\url{https://github.com/HoTT/HoTT}}, is a way to
characterize properties that are decidable. To establish that a
property is decidable, one must provide an explicit proof that it 
either holds or not: 
 \begin{coqdoccode}
\coqdocemptyline
\coqdocnoindent
\coqdockw{Class} \coqdef{Casts.cast1.Decidable}{$\mathsf{Decidable}$}{\coqdocrecord{$\mathsf{Decidable}$}} (\coqdocvar{P} : \coqdockw{Prop}) := \coqdef{Casts.cast1.dec}{dec}{\coqdocprojection{dec}} : \coqdocvariable{P} \coqexternalref{:type scope:x '+' x}{http://coq.inria.fr/stdlib/Coq.Init.Datatypes}{\coqdocnotation{+}} \coqexternalref{:type scope:'x7E' x}{http://coq.inria.fr/stdlib/Coq.Init.Logic}{\coqdocnotation{$\neg$}} \coqdocvariable{P} .\coqdoceol
\coqdocemptyline
\end{coqdoccode}
Note that the disjunction is encoded using a sum type (+, which is in \coqdockw{Type}) instead of a propositional disjunction (\ensuremath{\lor}, which is in \coqdockw{Prop}) in order to support projecting the underlying proof term and use it computationally as a decision procedure for the property.\footnote{An equivalent decision procedure mechanism is implemented in the Ssreflect library~\cite{ssreflect}, using boolean reflection. We discuss the differences between the two approaches in Appendix~\ref{sec:refl}. It must be noticed already that the differences are minor and our cast mechanism works perfectly well with both ways of formalizing decidability.}
\begin{coqdoccode}
\coqdocemptyline
\coqdocemptyline
\end{coqdoccode}
The Coq type class system can automatically infer the decision procedure of a complex property, using type class resolution, when a cast is performed. For that, the appropriate generic decidability instances must be provided first, but those instances are implemented only once and are already part of the \coqref{Casts.cast1.Decidable}{\coqdocrecord{$\mathsf{Decidable}$}} library or can be added as needed.
For example, the following instance definition (definition omitted) allows Coq to infer decidability---and build the associated decision procedure---for a conjunction of two decidable properties by evaluating the decision procedure for each property:\begin{coqdoccode}
\coqdocemptyline
\coqdocnoindent
\coqdockw{Instance} \coqdef{Casts.cast1.Decidable and}{Decidable\_and}{\coqdocinstance{Decidable\_and}} (\coqdocvar{P} \coqdocvar{Q}: \coqdockw{Prop}) (\coqdocvar{HP} : \coqref{Casts.cast1.Decidable}{\coqdocclass{$\mathsf{Decidable}$}} \coqdocvariable{P})\coqdoceol
\coqdocindent{4.50em}
(\coqdocvar{HQ} : \coqref{Casts.cast1.Decidable}{\coqdocclass{$\mathsf{Decidable}$}} \coqdocvariable{Q}) :  \coqref{Casts.cast1.Decidable}{\coqdocclass{$\mathsf{Decidable}$}} (\coqdocvariable{P} \coqexternalref{:type scope:x '/x5C' x}{http://coq.inria.fr/stdlib/Coq.Init.Logic}{\coqdocnotation{\ensuremath{\land}}} \coqdocvariable{Q}).\coqdoceol
 \coqdocemptyline
\end{coqdoccode}
\noindent Also, whenever a proposition has been proven, it is obviously decidable (\coqexternalref{inl}{http://coq.inria.fr/stdlib/Coq.Init.Datatypes}{\coqdocconstructor{inl}} is the left injection on a sum type): \begin{coqdoccode}
\coqdocemptyline
\coqdocnoindent
\coqdockw{Instance} \coqdef{Casts.cast1.Decidable proven}{Decidable\_proven}{\coqdocinstance{Decidable\_proven}} (\coqdocvar{P} : \coqdockw{Prop}) (\coqdocvar{ev} :  \coqdocvariable{P}):\coqdoceol
\coqdocindent{1.00em}
\coqref{Casts.cast1.Decidable}{\coqdocclass{$\mathsf{Decidable}$}} \coqdocvariable{P} := \coqexternalref{inl}{http://coq.inria.fr/stdlib/Coq.Init.Datatypes}{\coqdocconstructor{inl}} \coqdocvar{ev}.\coqdoceol
\coqdocemptyline
\end{coqdoccode}
\noindent This instance allows programmers to mix proven and decidable properties, for instance by inferring that \coqdocvariable{P} \ensuremath{\land} \coqdocvariable{Q} is decidable if \coqdocvariable{P} is decidable and \coqdocvariable{Q} is proven.

Another interesting instance is the one exploits the fact that every
property that is equivalent to a decidable property is decidable: \begin{coqdoccode}
\coqdocemptyline
\coqdocnoindent
\coqdockw{Definition} \coqdef{Casts.cast1.Decidable equivalent}{Decidable\_equivalent}{\coqdocdefinition{Decidable\_equivalent}} \{\coqdocvar{P} \coqdocvar{P'} : \coqdockw{Prop}\}\coqdoceol
\coqdocindent{2.50em}
(\coqdocvar{HPP'} : \coqdocvariable{P'} \coqexternalref{:type scope:x '<->' x}{http://coq.inria.fr/stdlib/Coq.Init.Logic}{\coqdocnotation{\ensuremath{\leftrightarrow}}} \coqdocvariable{P}) `\{\coqref{Casts.cast1.Decidable}{\coqdocclass{$\mathsf{Decidable}$}} \coqdocvariable{P'}\} : \coqref{Casts.cast1.Decidable}{\coqdocclass{$\mathsf{Decidable}$}} \coqdocvariable{P}.\coqdoceol
\coqdocemptyline
\end{coqdoccode}
We will exploit this instance in Section~\ref{sec:records} to synthesize more efficient decision procedures. 

 If type class resolution cannot find an instance of the \coqref{Casts.cast1.Decidable}{\coqdocrecord{$\mathsf{Decidable}$}} class for a given property, then casting to a subset type with that property fails statically. This happens if we try to cast \coqref{Casts.cast1.compile}{\coqdocdefinition{compile}} directly to a function subset type with a universally-quantified property, as discussed in Section~\ref{sec:gradual-subset-types}. \begin{coqdoccode}
\coqdocemptyline
\coqdocemptyline
\end{coqdoccode}
\subsection{Leveraging Type Class Resolution}

\label{sec:leveraging}

 Depending on the structure of the property to be established, we can
get decidability entirely for free.  In fact, in the compiler example (Section~\ref{sct:correct}), the decidability of \coqref{Casts.cast1.correct prog}{\coqdocdefinition{correct\_prog}} was
automatically inferred!  We now explain how this automation was achieved.

The \coqref{Casts.cast1.correct prog}{\coqdocdefinition{correct\_prog}} property is about equality of the results of
running programs, which are \coqexternalref{option}{http://coq.inria.fr/stdlib/Coq.Init.Datatypes}{\coqdocinductive{option}} \coqref{Casts.cast1.stack}{\coqdocdefinition{stack}}s, or more explicitly,
\coqexternalref{option}{http://coq.inria.fr/stdlib/Coq.Init.Datatypes}{\coqdocinductive{option}} \coqexternalref{list}{http://coq.inria.fr/stdlib/Coq.Init.Datatypes}{\coqdocinductive{list}} \coqexternalref{nat}{http://coq.inria.fr/stdlib/Coq.Init.Datatypes}{\coqdocinductive{nat}}s. The \coqref{Casts.cast1.Decidable}{\coqdocrecord{$\mathsf{Decidable}$}} type class already allows, with
its instances, to automatically obtain complex correct decision
procedures based on composition of atomic ones
(Sect.~\ref{sec:decidable}).  For \coqref{Casts.cast1.correct prog}{\coqdocdefinition{correct\_prog}} to enjoy this full
automation, the \coqref{Casts.cast1.Decidable}{\coqdocrecord{$\mathsf{Decidable}$}} library needs to include instances that
allow equality of lists and options to be inferred. More precisely, we
provide a type class for decidable equality,
\coqref{Casts.cast1.EqDecidable}{\coqdocrecord{$\mathsf{Decidable}_=$}}:\footnote{A similar type class is also used in the
Coq/HoTT library under the name \cind{DecidablePaths}.} \begin{coqdoccode}
\coqdocemptyline
\coqdocnoindent
\coqdockw{Class} \coqdef{Casts.cast1.EqDecidable}{$\mathsf{Decidable}_=$}{\coqdocrecord{$\mathsf{Decidable}_=$}} (\coqdocvar{A} : \coqdockw{Type}) := \coqdoceol
\coqdocindent{1.00em}
\{ \coqdef{Casts.cast1.eq dec}{eq\_dec}{\coqdocprojection{eq\_dec}} : \coqdockw{\ensuremath{\forall}} \coqdocvar{a} \coqdocvar{b} : \coqdocvariable{A}, \coqdocclass{$\mathsf{Decidable}$} (\coqdocvariable{a} \coqexternalref{:type scope:x '=' x}{http://coq.inria.fr/stdlib/Coq.Init.Logic}{\coqdocnotation{=}} \coqdocvariable{b}) \}.\coqdoceol
\coqdocemptyline
\end{coqdoccode}
Based on this decidable equality class, we can define once and for
all how to derive the decidability of the equality between lists of
\coqdocvariable{A} or options of \coqdocvariable{A} provided that equality is decidable for \coqdocvariable{A}: \begin{coqdoccode}
\coqdocemptyline
\coqdocnoindent
\coqdockw{Instance} \coqdef{Casts.cast1.Decidable eq list}{Decidable\_eq\_list}{\coqdocinstance{Decidable\_eq\_list}} : \coqdockw{\ensuremath{\forall}} \coqdocvar{A} (\coqdocvar{HA}: \coqref{Casts.cast1.EqDecidable}{\coqdocclass{$\mathsf{Decidable}_=$}} \coqdocvariable{A}) \coqdoceol
\coqdocindent{1.00em}
(\coqdocvar{l} \coqdocvar{l'}: \coqexternalref{list}{http://coq.inria.fr/stdlib/Coq.Init.Datatypes}{\coqdocinductive{list}} \coqdocvariable{A}), \coqdocclass{$\mathsf{Decidable}$} (\coqdocvariable{l} \coqexternalref{:type scope:x '=' x}{http://coq.inria.fr/stdlib/Coq.Init.Logic}{\coqdocnotation{=}} \coqdocvariable{l'}).\coqdoceol
\coqdocemptyline
\coqdocnoindent
\coqdockw{Instance} \coqdef{Casts.cast1.Decidable eq option}{Decidable\_eq\_option}{\coqdocinstance{Decidable\_eq\_option}} : \coqdockw{\ensuremath{\forall}} \coqdocvar{A} (\coqdocvar{HA}: \coqref{Casts.cast1.EqDecidable}{\coqdocclass{$\mathsf{Decidable}_=$}} \coqdocvariable{A})\coqdoceol
\coqdocindent{1.00em}
(\coqdocvar{o} \coqdocvar{o'}: \coqexternalref{option}{http://coq.inria.fr/stdlib/Coq.Init.Datatypes}{\coqdocinductive{option}} \coqdocvariable{A}), \coqdocclass{$\mathsf{Decidable}$} (\coqdocvariable{o} \coqexternalref{:type scope:x '=' x}{http://coq.inria.fr/stdlib/Coq.Init.Logic}{\coqdocnotation{=}} \coqdocvariable{o'}).\coqdoceol
\coqdocemptyline
\end{coqdoccode}
By also declaring the corresponding \coqref{Casts.cast1.EqDecidable}{\coqdocrecord{$\mathsf{Decidable}_=$}} instances for the \coqexternalref{list}{http://coq.inria.fr/stdlib/Coq.Init.Datatypes}{\coqdocinductive{list}} and \coqexternalref{option}{http://coq.inria.fr/stdlib/Coq.Init.Datatypes}{\coqdocinductive{option}} type constructors, the type class resolution mechanism of Coq is able to automatically build the correct decision procedures for properties that state equality between arbitrary nestings of these type constructors, such as \coqref{Casts.cast1.correct prog}{\coqdocdefinition{correct\_prog}}. A well-furnished decidability library allows developers to seamlessly enjoy the benefits of gradual certified programming. 

 We come back to decidability in Section~\ref{sec:records}, when describing casts on rich records, in order to show how one can specialize the decision procedure to use in specific cases, for instance to obtain a  procedure that is more efficient than the default one.

\section{Casts and Axioms}

\label{sec:axioms}

Intuitively, the basic cast operator ? should be defined as a function \coqref{Casts.cast1.cast}{\coqdocdefinition{cast}} of type \coqdocvariable{A} \ensuremath{\rightarrow} \{\coqdocvariable{a} : \coqdocvariable{A} \ensuremath{|} \coqdocvariable{P} \coqdocvariable{a}\} (assuming that \coqdocvariable{P} \coqdocvariable{a} is decidable). To perform such a cast implies exploiting the decidability of \coqdocvariable{P} \coqdocvariable{a}: checking (and hence evaluating) whether \coqdocvariable{P} \coqdocvariable{a} holds or not. If it holds true, the cast succeeds. The \coqref{Casts.cast1.cast}{\coqdocdefinition{cast}} function can simply return the dependent pair with the value \coqdocvariable{a} and the proof. If \coqdocvariable{P} \coqdocvariable{a} does not hold, the cast fails. How should such errors manifest?

\subsection{The Monadic Approach}

The traditional way to support errors in a purely functional setting is to adopt a monadic style. For instance, we could define \coqref{Casts.cast1.cast}{\coqdocdefinition{cast}} to return \coqexternalref{option}{http://coq.inria.fr/stdlib/Coq.Init.Datatypes}{\coqdocinductive{option}} \{\coqdocvariable{a}:\coqdocvariable{A} \ensuremath{|} \coqdocvariable{P} \coqdocvariable{a}\} instead of just \{\coqdocvariable{a}:\coqdocvariable{A} \ensuremath{|} \coqdocvariable{P} \coqdocvariable{a}\}. Then, a cast failure would simply manifest as \coqdocconstructor{None}. This is all well and understood, but has serious consequences from a software engineering point of view: it forces all code that (potentially) uses casts to also be written in monadic style. Because the philosophy of gradual typing entails that casts may be added (or removed) anywhere as the software evolves, it means that the entire development has to be defensively written in monadic style. 
For instance, consider the definition of \coqref{Casts.cast1.runc}{\coqdocdefinition{runc}} in Section~\ref{sec:grad-cert-comp}: \begin{coqdoccode}
\coqdocemptyline
\coqdocnoindent
\coqdockw{Definition} \coqdef{Casts.cast1.runc}{runc}{\coqdocdefinition{runc}} (\coqdocvar{c}: \coqdocdefinition{correct\_comp}) (\coqdocvar{e}: \coqdocinductive{exp}) :=\coqdoceol
\coqdocindent{1.00em}
\coqdocdefinition{runProg} \coqdocnotation{(}\coqdocvariable{c} \coqdocvariable{e}\coqdocnotation{$)._1$} \coqexternalref{nil}{http://coq.inria.fr/stdlib/Coq.Init.Datatypes}{\coqdocconstructor{nil}}.\coqdoceol
\coqdocemptyline
\coqdocemptyline
\end{coqdoccode}
If it were possible to check eagerly that \coqref{Casts.cast1.compile}{\coqdocdefinition{compile}} is correct, the monadic cast would produce a value of type \coqexternalref{option}{http://coq.inria.fr/stdlib/Coq.Init.Datatypes}{\coqdocinductive{option}} \coqdocdefinition{correct\_comp}, and the client calling \coqref{Casts.cast1.runc}{\coqdocdefinition{runc}} (? \coqref{Casts.cast1.compile}{\coqdocdefinition{compile}}) would simply have to locally deal with the potential of failure. However, since \coqdocdefinition{correct\_comp} is undecidable, the only solution is to delay casts, which means that the casted compiler would now have type
$\forall$ \coqdocvariable{e} : \coqdocinductive{exp}, \coqexternalref{option}{http://coq.inria.fr/stdlib/Coq.Init.Datatypes}{\coqdocinductive{option}} \{\coqdocvar{p} : \coqdocvar{prog} \ensuremath{|} \coqref{Casts.cast1.correct prog}{\coqdocdefinition{correct\_prog}} \coqdocvariable{e} \coqdocvar{p}\}. This in turn implies that {\em all users} of the compiler (such as \coqref{Casts.cast1.runc}{\coqdocdefinition{runc}}) have to be prepared to deal with optional values. The argument type of \coqref{Casts.cast1.runc}{\coqdocdefinition{runc}} would have to be changed, and its body as well because (\coqdocvariable{c} \coqdocvariable{e}) would now return an \coqexternalref{option}{http://coq.inria.fr/stdlib/Coq.Init.Datatypes}{\coqdocinductive{option}} \coqref{Casts.cast1.correct prog}{\coqdocdefinition{correct\_prog}}, not a \coqref{Casts.cast1.correct prog}{\coqdocdefinition{correct\_prog}}. 
This non-local impact of deciding to statically establish guarantees or defer them to runtime is contrary to the smooth transition path that gradual typing is meant to support. 

 After all, every practical functional programming language does some compromise with purity\footnote{Even Haskell has impure features such as {\tt undefined}, {\tt unsafeCoerce} and {\tt unsafePerformIO}, for pragmatic reasons.}, supporting side effects like references and exceptions directly in the language, rather than through an explicit monadic encoding. The upside of sacrificing purity is that these side effecting operations can be used ``transparently'', without having to adopt a rigid discipline like monads, which---despite various improvements such as~\cite{schrijversOliveira:icfp2011}---is still not free from software engineering challenges. So, what does it mean to embed cast errors in such a transparent manner in Coq?

\subsection{The Axiomatic Approach}

We introduce a novel use of axioms, not to represent what is assumed to be true, but to represent errors. This allows us to provide the \coqref{Casts.cast1.cast}{\coqdocdefinition{cast}} operator as a function of type \coqdocvariable{A} \ensuremath{\rightarrow} \{\coqdocvariable{a} : \coqdocvariable{A} \ensuremath{|} \coqdocvariable{P} \coqdocvariable{a}\}. 
Specifically, we introduce one axiom, \coqref{Casts.cast1.failed cast}{\coqdocaxiom{failed\_cast}}, which states that for any indexed property on elements of type \coqdocvariable{A}, we can build a value of type \{\coqdocvariable{a} : \coqdocvariable{A} \ensuremath{|} \coqdocvariable{P} \coqdocvariable{a}\}:
\footnote{We declare the two first arguments of \coqref{Casts.cast1.failed cast}{\coqdocaxiom{failed\_cast}} as implicit (between \{\}), and only leave the value \coqdocvariable{a} and the \coqdocvariable{msg} argument as explicit. The argument \coqdocvariable{msg} is apparently redundant, since it is just defined as \coqdocvariable{P} \coqdocvariable{a} in \coqref{Casts.cast1.cast}{\coqdocdefinition{cast}}; however, declaring it as an explicit argument together with \coqdocvariable{a} allows for clear and concise error messages when cast fails, reporting the violated property for a given value, as illustrated in Section~\ref{sec:gradual-subset-types}.}

\begin{coqdoccode}
\coqdocemptyline
\coqdocnoindent
\coqdockw{Axiom} \coqdef{Casts.cast1.failed cast}{failed\_cast}{\coqdocaxiom{failed\_cast}} : \coqdockw{\ensuremath{\forall}} \{\coqdocvar{A}:\coqdockw{Type}\} \{\coqdocvar{P} : \coqdocvariable{A} \coqexternalref{:type scope:x '->' x}{http://coq.inria.fr/stdlib/Coq.Init.Logic}{\coqdocnotation{\ensuremath{\rightarrow}}} \coqdockw{Prop}\}\coqdoceol
\coqdocindent{1.00em}
(\coqdocvar{a}:\coqdocvariable{A}) (\coqdocvar{msg}: \coqdockw{Prop}), \coqexternalref{:type scope:'x7B' x ':' x '|' x 'x7D'}{http://coq.inria.fr/stdlib/Coq.Init.Specif}{\coqdocnotation{\{}}\coqdocvar{a} \coqexternalref{:type scope:'x7B' x ':' x '|' x 'x7D'}{http://coq.inria.fr/stdlib/Coq.Init.Specif}{\coqdocnotation{:}} \coqdocvariable{A} \coqexternalref{:type scope:'x7B' x ':' x '|' x 'x7D'}{http://coq.inria.fr/stdlib/Coq.Init.Specif}{\coqdocnotation{\ensuremath{|}}} \coqdocvariable{P} \coqdocvariable{a}\coqexternalref{:type scope:'x7B' x ':' x '|' x 'x7D'}{http://coq.inria.fr/stdlib/Coq.Init.Specif}{\coqdocnotation{\}}}.\coqdoceol
\coqdocemptyline
\coqdocemptyline
\coqdocemptyline
\end{coqdoccode}

Obviously, \coqref{Casts.cast1.failed cast}{\coqdocaxiom{failed\_cast}} is a lie. This lie is used in the definition of the \coqref{Casts.cast1.cast}{\coqdocdefinition{cast}} operator, in  case the decision procedure indicates that the property does not hold:  \begin{coqdoccode}
\coqdocemptyline
\coqdocnoindent
\coqdockw{Definition} \coqdef{Casts.cast1.cast}{cast}{\coqdocdefinition{cast}} (\coqdocvar{A}:\coqdockw{Type}) (\coqdocvar{P} : \coqdocvariable{A} \coqexternalref{:type scope:x '->' x}{http://coq.inria.fr/stdlib/Coq.Init.Logic}{\coqdocnotation{\ensuremath{\rightarrow}}} \coqdockw{Prop}) \coqdoceol
\coqdocindent{1.00em}
(\coqdocvar{dec} : \coqdockw{\ensuremath{\forall}} \coqdocvar{a}, \coqdocclass{$\mathsf{Decidable}$} (\coqdocvariable{P} \coqdocvariable{a})) : \coqdocvariable{A} \coqexternalref{:type scope:x '->' x}{http://coq.inria.fr/stdlib/Coq.Init.Logic}{\coqdocnotation{\ensuremath{\rightarrow}}} \coqexternalref{:type scope:'x7B' x ':' x '|' x 'x7D'}{http://coq.inria.fr/stdlib/Coq.Init.Specif}{\coqdocnotation{\{}}\coqdocvar{a} \coqexternalref{:type scope:'x7B' x ':' x '|' x 'x7D'}{http://coq.inria.fr/stdlib/Coq.Init.Specif}{\coqdocnotation{:}} \coqdocvariable{A} \coqexternalref{:type scope:'x7B' x ':' x '|' x 'x7D'}{http://coq.inria.fr/stdlib/Coq.Init.Specif}{\coqdocnotation{\ensuremath{|}}} \coqdocvariable{P} \coqdocvar{a}\coqexternalref{:type scope:'x7B' x ':' x '|' x 'x7D'}{http://coq.inria.fr/stdlib/Coq.Init.Specif}{\coqdocnotation{\}}} :=\coqdoceol
\coqdocnoindent
\coqdockw{fun} \coqdocvar{a}: \coqdocvariable{A} \ensuremath{\Rightarrow} \coqdoceol
\coqdocindent{1.00em}
\coqdockw{match} \coqdocvariable{dec} \coqdocvariable{a} \coqdockw{with}\coqdoceol
\coqdocindent{2.00em}
\ensuremath{|} \coqexternalref{inl}{http://coq.inria.fr/stdlib/Coq.Init.Datatypes}{\coqdocconstructor{inl}} \coqdocvar{p} \ensuremath{\Rightarrow} \coqdocnotation{(}\coqdocvariable{a} \coqdocnotation{;} \coqdocvar{p}\coqdocnotation{)}\coqdoceol
\coqdocindent{2.00em}
\ensuremath{|} \coqexternalref{inr}{http://coq.inria.fr/stdlib/Coq.Init.Datatypes}{\coqdocconstructor{inr}} \coqdocvar{\_} \ensuremath{\Rightarrow} \coqref{Casts.cast1.failed cast}{\coqdocaxiom{failed\_cast}} \coqdocvariable{a} (\coqdocvariable{P} \coqdocvariable{a})\coqdoceol
\coqdocindent{1.00em}
\coqdockw{end}.\coqdoceol
\coqdocemptyline
\coqdocemptyline
\end{coqdoccode}
The \coqref{Casts.cast1.cast}{\coqdocdefinition{cast}} operator applies the decision procedure to the given value and, depending on the outcome, returns either the dependent pair with the obtained proof, or a \coqref{Casts.cast1.failed cast}{\coqdocaxiom{failed\_cast}}. Considering the definition of \coqref{Casts.cast1.cast}{\coqdocdefinition{cast}}, we see that a cast fails if and only if the property \coqdocvariable{P} \coqdocvariable{a} does not hold according to the decision procedure.

A subtlety in the definition of \coqref{Casts.cast1.cast}{\coqdocdefinition{cast}} is that the casted value must not be exposed as a dependent pair if the decision procedure fails. An alternative definition could always return (\coqdocvariable{a} ; \coqdocvariable{x}) where \coqdocvariable{x} is some error axiom if the cast failed.  Our definition has the advantage of reporting a cast failure as soon as the casted value is used (even though the property attached to it is not).\footnote{Appendix B briefly discusses the interplay of evaluation regimes and the representation of cast failures as non-canonical normal forms.}

We introduce the ? notation for \coqref{Casts.cast1.cast}{\coqdocdefinition{cast}}, asking Coq to infer the property and the evidence of its decidability from the context:
\begin{coqdoccode}
\coqdocemptyline
\coqdocnoindent
\coqdockw{Notation} \coqdef{Casts.cast1.::'?'}{"}{"}?" := (\coqref{Casts.cast1.cast}{\coqdocdefinition{cast}} \coqdocvar{\_} \coqdocvar{\_} \coqdocvar{\_}).\coqdoceol
\coqdocemptyline
\end{coqdoccode}
\subsection{Heresy!}

Using an axiom to represent failed casts is (slightly!) heretical from a theoretical viewpoint. As a matter of fact, one can use a cast to inhabit \coqexternalref{False}{http://coq.inria.fr/stdlib/Coq.Init.Logic}{\coqdocinductive{False}}, for instance by pretending that 0 comes with a proof of \coqexternalref{False}{http://coq.inria.fr/stdlib/Coq.Init.Logic}{\coqdocinductive{False}} and then projecting the second component:\begin{coqdoccode}
\coqdocemptyline
\coqdocnoindent
\coqdockw{Definition} \coqdef{Casts.cast1.unsound}{unsound}{\coqdocdefinition{unsound}} : \coqexternalref{False}{http://coq.inria.fr/stdlib/Coq.Init.Logic}{\coqdocinductive{False}} := \coqdocnotation{(?} 0\coqdocnotation{$)._2$}.\coqdoceol
\coqdocemptyline
\coqdocemptyline
\end{coqdoccode}
\noindent In this sense, the monadic approach is preferrable, as it preserves consistency. However, the axiomatic approach is an interesting alternative to using plain axioms and admitted definitions in Coq---which are, after all, the only pragmatic solutions available to a Coq practitioner who does not want to wrestle with a given proof immediately. Axiomatic casts are superior in many ways: 
\begin{itemize}
\item As discussed above, we cannot project the value component of a subset type with a failed cast (recall that using \coqdocvar{admit} provides no such guarantee). 
\item When things go right (i.e. the cast succeeds), there is no axiom or admitted definition that will block type conversion and evaluation. 
\item Statically establishing a property or using a cast does not affect the types involved, so the programmer can seamlessly navigate the gradual checking spectrum without having to perform non-local refactorings.
\end{itemize}

All in all, both the monadic and axiomatic approaches to gradual verification are feasible, and are likely to please different crowds. In fact, we have implemented both approaches in the Cocasse library. In this paper we focus on the axiomatic approach, because of its disruptive potential and software engineering benefits. We believe this approach will be more appealing to pragmatic practitioners who are willing to compromise consistency to some extent in order to enjoy a smooth gradual verification environment. Also, as we discuss in Section~\ref{sec:conclusion}, there are alternatives to be explored to make the axiomatic approach less heretical.
\begin{coqdoccode}
\coqdocemptyline
\coqdocemptyline
\coqdocemptyline
\end{coqdoccode}
\section{Implicit Casts}

\label{sec:coerce}
\coqlibrary{Casts.coercions}{Library }{Casts.coercions}

\begin{coqdoccode}
\coqdocemptyline
\coqdocemptyline
\end{coqdoccode}
The major technical challenge addressed in this work is to provide casts for subset types within Coq. These casts have to be \emph{explicitly} placed by programmers, much like in the seminal work of Abadi \emph{et al.} on integrating static and dynamic typing~\cite{abadiAl:toplas1991}, or in the gradual information flow type system proposed by Disney and Flanagan~\cite{disneyFlanagan:stop2011}. Gradual typing is however generally associated with a mechanism of \emph{implicit} cast insertion: the source language, which may not even feature explicit casts, is translated to an internal language with explicit casts~\cite{siekTaha:sfp2006}.

It is possible to achieve implicit cast insertion in Coq by exploiting the {\em implicit coercion} mechanism.\footnote{\url{https://coq.inria.fr/distrib/current/refman/Reference-Manual021.html}}
\begin{coqdoccode}
\coqdocemptyline
\coqdocemptyline
\end{coqdoccode}
\paragraph{Implicit coercions in a nutshell.} Let us first briefly illustrate implicit coercions in Coq. Assume a decidable indexed property on \coqexternalref{nat}{http://coq.inria.fr/stdlib/Coq.Init.Datatypes}{\coqdocinductive{nat}}, which is used to define a type \coqref{Casts.coercions.rich nat}{\coqdocdefinition{rich\_nat}}: \begin{coqdoccode}
\coqdocemptyline
\coqdocindent{1.00em}
\coqdockw{Variable} \coqdef{Casts.coercions.imp coerce.P}{P}{\coqdocvariable{P}} : \coqexternalref{nat}{http://coq.inria.fr/stdlib/Coq.Init.Datatypes}{\coqdocinductive{nat}} \coqexternalref{:type scope:x '->' x}{http://coq.inria.fr/stdlib/Coq.Init.Logic}{\coqdocnotation{\ensuremath{\rightarrow}}} \coqdockw{Prop}.\coqdoceol
\coqdocindent{1.00em}
\coqdockw{Variable} \coqdef{Casts.coercions.imp coerce.P dec}{P\_dec}{\coqdocvariable{P\_dec}} : \coqdockw{\ensuremath{\forall}} \coqdocvar{n}:\coqexternalref{nat}{http://coq.inria.fr/stdlib/Coq.Init.Datatypes}{\coqdocinductive{nat}}, \coqdocclass{$\mathsf{Decidable}$} (\coqdocvariable{P} \coqdocvariable{n}).\coqdoceol
\coqdocindent{1.00em}
\coqdockw{Definition} \coqdef{Casts.coercions.rich nat}{rich\_nat}{\coqdocdefinition{rich\_nat}} := \coqexternalref{:type scope:'x7B' x ':' x '|' x 'x7D'}{http://coq.inria.fr/stdlib/Coq.Init.Specif}{\coqdocnotation{\{}}\coqdocvar{n}\coqexternalref{:type scope:'x7B' x ':' x '|' x 'x7D'}{http://coq.inria.fr/stdlib/Coq.Init.Specif}{\coqdocnotation{:}} \coqexternalref{nat}{http://coq.inria.fr/stdlib/Coq.Init.Datatypes}{\coqdocinductive{nat}} \coqexternalref{:type scope:'x7B' x ':' x '|' x 'x7D'}{http://coq.inria.fr/stdlib/Coq.Init.Specif}{\coqdocnotation{\ensuremath{|}}} \coqdocvariable{P} \coqdocvar{n}\coqexternalref{:type scope:'x7B' x ':' x '|' x 'x7D'}{http://coq.inria.fr/stdlib/Coq.Init.Specif}{\coqdocnotation{\}}}.\coqdoceol
\coqdocemptyline
\end{coqdoccode}
To define an implicit coercion from values of type \coqref{Casts.coercions.rich nat}{\coqdocdefinition{rich\_nat}} to \coqexternalref{nat}{http://coq.inria.fr/stdlib/Coq.Init.Datatypes}{\coqdocinductive{nat}}, we first define a function with the appropriate type, and then declare it as an implicit \coqdockw{Coercion}: 
\begin{coqdoccode}
\coqdocindent{1.00em}
\coqdockw{Definition} \coqdef{Casts.coercions.rnat to nat}{rnat\_to\_nat}{\coqdocdefinition{rnat\_to\_nat}} : \coqref{Casts.coercions.rich nat}{\coqdocdefinition{rich\_nat}} \coqexternalref{:type scope:x '->' x}{http://coq.inria.fr/stdlib/Coq.Init.Logic}{\coqdocnotation{\ensuremath{\rightarrow}}} \coqexternalref{nat}{http://coq.inria.fr/stdlib/Coq.Init.Datatypes}{\coqdocinductive{nat}} :=\coqdoceol
\coqdocindent{2.00em}
\coqdockw{fun} \coqdocvar{n} \ensuremath{\Rightarrow} \coqdocvariable{n}\coqdocnotation{$._1$}.\coqdoceol
\coqdocindent{1.00em}
\coqdockw{Coercion} \coqref{Casts.coercions.rnat to nat}{\coqdocdefinition{rnat\_to\_nat}} : \coqdocvar{rich\_nat} $\rightarrowtail$ \coqdocvar{nat}.\coqdoceol
\coqdocemptyline
\end{coqdoccode}
\noindent We can now pass a \coqref{Casts.coercions.rich nat}{\coqdocdefinition{rich\_nat}} to a function that expects a \coqexternalref{nat}{http://coq.inria.fr/stdlib/Coq.Init.Datatypes}{\coqdocinductive{nat}}, without having to explicitly apply the coercion function: \begin{coqdoccode}
\coqdocemptyline
\coqdocindent{1.00em}
\coqdockw{Variable} \coqdef{Casts.coercions.imp coerce.f}{f}{\coqdocvariable{f}} : \coqexternalref{nat}{http://coq.inria.fr/stdlib/Coq.Init.Datatypes}{\coqdocinductive{nat}} \coqexternalref{:type scope:x '->' x}{http://coq.inria.fr/stdlib/Coq.Init.Logic}{\coqdocnotation{\ensuremath{\rightarrow}}} \coqexternalref{nat}{http://coq.inria.fr/stdlib/Coq.Init.Datatypes}{\coqdocinductive{nat}}.\coqdoceol
\coqdocindent{1.00em}
\coqdockw{Variable} \coqdef{Casts.coercions.imp coerce.s}{s}{\coqdocvariable{s}} : \coqref{Casts.coercions.rich nat}{\coqdocdefinition{rich\_nat}}.\coqdoceol
\coqdocindent{1.00em}
\coqdockw{Check} \coqdocvariable{f} \coqdocvariable{s}.\coqdoceol
\coqdocemptyline
\end{coqdoccode}
\paragraph{Implicit cast insertion.} In order to implicitly insert casts, it is enough to define a
  standard implicit coercion based on a function that introduces
  casts. For instance, we define an implicit coercion (cast insertion)
  from \coqexternalref{nat}{http://coq.inria.fr/stdlib/Coq.Init.Datatypes}{\coqdocinductive{nat}} to \coqref{Casts.coercions.rich nat}{\coqdocdefinition{rich\_nat}}: \begin{coqdoccode}
\coqdocemptyline
\coqdocindent{1.00em}
\coqdockw{Definition} \coqdef{Casts.coercions.nat to rnat}{nat\_to\_rnat}{\coqdocdefinition{nat\_to\_rnat}} : \coqexternalref{nat}{http://coq.inria.fr/stdlib/Coq.Init.Datatypes}{\coqdocinductive{nat}} \coqexternalref{:type scope:x '->' x}{http://coq.inria.fr/stdlib/Coq.Init.Logic}{\coqdocnotation{\ensuremath{\rightarrow}}} \coqref{Casts.coercions.rich nat}{\coqdocdefinition{rich\_nat}} := ?.\coqdoceol
\coqdocindent{1.00em}
\coqdockw{Coercion} \coqref{Casts.coercions.nat to rnat}{\coqdocdefinition{nat\_to\_rnat}} : \coqdocvar{nat} $\rightarrowtail$ \coqdocvar{rich\_nat}.\coqdoceol
\coqdocemptyline
\end{coqdoccode}
Calling a function that expects a \coqref{Casts.coercions.rich nat}{\coqdocdefinition{rich\_nat}} with a \coqexternalref{nat}{http://coq.inria.fr/stdlib/Coq.Init.Datatypes}{\coqdocinductive{nat}} argument is now type-correct. Under the hood, the implicit coercion takes care of inserting the cast: \begin{coqdoccode}
\coqdocemptyline
\coqdocindent{1.00em}
\coqdockw{Variable} \coqdef{Casts.coercions.imp coerce.g}{g}{\coqdocvariable{g}} : \coqref{Casts.coercions.rich nat}{\coqdocdefinition{rich\_nat}} \coqexternalref{:type scope:x '->' x}{http://coq.inria.fr/stdlib/Coq.Init.Logic}{\coqdocnotation{\ensuremath{\rightarrow}}} \coqexternalref{nat}{http://coq.inria.fr/stdlib/Coq.Init.Datatypes}{\coqdocinductive{nat}}.\coqdoceol
\coqdocindent{1.00em}
\coqdockw{Variable} \coqdef{Casts.coercions.imp coerce.n}{n}{\coqdocvariable{n}}: \coqexternalref{nat}{http://coq.inria.fr/stdlib/Coq.Init.Datatypes}{\coqdocinductive{nat}}.\coqdoceol
\coqdocindent{1.00em}
\coqdockw{Check} \coqdocvariable{g} \coqdocvariable{n}.\coqdoceol
\coqdocemptyline
\coqdocemptyline
\coqdocemptyline
\end{coqdoccode}
Compared to standard gradual typing, the limitation of this approach is that Coq does not support universal coercions, so one needs to explicitly define the specific coercions that are permitted. This is arguably less convenient than a general implicit cast insertion mechanism,  but it is also more controlled. Because types are so central to Coq programming, it is unclear whether general implicit cast insertion would be useful and not an endless source of confusion. Actually, even in gradually-typed languages with much less powerful type systems, it has been argued that a mechanism to control implicit cast insertion is important~\cite{allendeAl:oopsla2014}. We believe that the implicit coercion mechanism of Coq combined with casts might be a good tradeoff in practice. \begin{coqdoccode}
\end{coqdoccode}

\section{Higher-Order Casts, Simply}

\label{sec:hocs}

We now consider cast operators for functions. As expected, function casts are enforced lazily similarly to higher-order contracts~\cite{findlerFelleisen:icfp2002}.
We first focus on non-dependent function types of the form \coqdocvariable{A}\ensuremath{\rightarrow}\coqdocvariable{B}. One could want to either strengthen the range of the function, claiming that the return type is \{\coqdocvariable{b} : \coqdocvariable{B} \ensuremath{|} \coqdocvariable{P} \coqdocvariable{b}\}, or vice-versa, to hide the fact that a function expects rich arguments of type \{\coqdocvariable{a} : \coqdocvariable{A} \ensuremath{|} \coqdocvariable{P} \coqdocvariable{a}\}.

\subsection{Strengthening the Range}

The \coqref{Casts.cast1.cast fun range}{\coqdocdefinition{cast\_fun\_range}} operator below takes a function of type \coqdocvariable{A}\ensuremath{\rightarrow}\coqdocvariable{B} and returns a function of type \coqdocvariable{A} \ensuremath{\rightarrow} \{\coqdocvariable{b} : \coqdocvariable{B} \ensuremath{|} \coqdocvariable{P} \coqdocvariable{b}\}. It simply casts the return value to the expected subset type:
\begin{coqdoccode}
\coqdocemptyline
\coqdocnoindent
\coqdockw{Definition} \coqdef{Casts.cast1.cast fun range}{cast\_fun\_range}{\coqdocdefinition{cast\_fun\_range}} (\coqdocvar{A} \coqdocvar{B} : \coqdockw{Type}) (\coqdocvar{P} : \coqdocvariable{B} \coqexternalref{:type scope:x '->' x}{http://coq.inria.fr/stdlib/Coq.Init.Logic}{\coqdocnotation{\ensuremath{\rightarrow}}} \coqdockw{Prop}) \coqdoceol
\coqdocindent{1.00em}
(\coqdocvar{dec} : \coqdockw{\ensuremath{\forall}} \coqdocvar{b}, \coqdocclass{$\mathsf{Decidable}$} (\coqdocvariable{P} \coqdocvariable{b})) :\coqdoceol
\coqdocindent{2.00em}
\coqexternalref{:type scope:x '->' x}{http://coq.inria.fr/stdlib/Coq.Init.Logic}{\coqdocnotation{(}}\coqdocvariable{A} \coqexternalref{:type scope:x '->' x}{http://coq.inria.fr/stdlib/Coq.Init.Logic}{\coqdocnotation{\ensuremath{\rightarrow}}} \coqdocvariable{B}\coqexternalref{:type scope:x '->' x}{http://coq.inria.fr/stdlib/Coq.Init.Logic}{\coqdocnotation{)}} \coqexternalref{:type scope:x '->' x}{http://coq.inria.fr/stdlib/Coq.Init.Logic}{\coqdocnotation{\ensuremath{\rightarrow}}} \coqdocvariable{A} \coqexternalref{:type scope:x '->' x}{http://coq.inria.fr/stdlib/Coq.Init.Logic}{\coqdocnotation{\ensuremath{\rightarrow}}} \coqexternalref{:type scope:'x7B' x ':' x '|' x 'x7D'}{http://coq.inria.fr/stdlib/Coq.Init.Specif}{\coqdocnotation{\{}}\coqdocvar{b} \coqexternalref{:type scope:'x7B' x ':' x '|' x 'x7D'}{http://coq.inria.fr/stdlib/Coq.Init.Specif}{\coqdocnotation{:}} \coqdocvariable{B} \coqexternalref{:type scope:'x7B' x ':' x '|' x 'x7D'}{http://coq.inria.fr/stdlib/Coq.Init.Specif}{\coqdocnotation{\ensuremath{|}}} \coqdocvariable{P} \coqdocvar{b}\coqexternalref{:type scope:'x7B' x ':' x '|' x 'x7D'}{http://coq.inria.fr/stdlib/Coq.Init.Specif}{\coqdocnotation{\}}} :=\coqdoceol
\coqdocindent{1.00em}
\coqdockw{fun} \coqdocvar{f} \coqdocvar{a} \ensuremath{\Rightarrow} ? (\coqdocvariable{f} \coqdocvariable{a}).\coqdoceol
\coqdocnoindent
\coqdockw{Notation} \coqdef{Casts.cast1.::'?>'}{"}{"}$\rightarrow?$" := (\coqref{Casts.cast1.cast fun range}{\coqdocdefinition{cast\_fun\_range}} \coqdocvar{\_} \coqdocvar{\_} \coqdocvar{\_} \coqdocvar{\_}).\coqdoceol
\coqdocemptyline
\end{coqdoccode}
\paragraph{Example.} We can cast a \coqexternalref{nat}{http://coq.inria.fr/stdlib/Coq.Init.Datatypes}{\coqdocinductive{nat}} \ensuremath{\rightarrow} \coqexternalref{nat}{http://coq.inria.fr/stdlib/Coq.Init.Datatypes}{\coqdocinductive{nat}} function such as \coqexternalref{S}{http://coq.inria.fr/stdlib/Coq.Init.Datatypes}{\coqdocconstructor{S}} (successor) to a function type that ensures the returned value is less than 10: \begin{coqdoccode}
\coqdocemptyline
\coqdocnoindent
\coqdockw{Definition} \coqdef{Casts.cast1.top succ}{top\_succ}{\coqdocdefinition{top\_succ}} : \coqexternalref{nat}{http://coq.inria.fr/stdlib/Coq.Init.Datatypes}{\coqdocinductive{nat}} \coqexternalref{:type scope:x '->' x}{http://coq.inria.fr/stdlib/Coq.Init.Logic}{\coqdocnotation{\ensuremath{\rightarrow}}} \coqexternalref{:type scope:'x7B' x ':' x '|' x 'x7D'}{http://coq.inria.fr/stdlib/Coq.Init.Specif}{\coqdocnotation{\{}}\coqdocvar{n}\coqexternalref{:type scope:'x7B' x ':' x '|' x 'x7D'}{http://coq.inria.fr/stdlib/Coq.Init.Specif}{\coqdocnotation{:}}\coqexternalref{nat}{http://coq.inria.fr/stdlib/Coq.Init.Datatypes}{\coqdocinductive{nat}} \coqexternalref{:type scope:'x7B' x ':' x '|' x 'x7D'}{http://coq.inria.fr/stdlib/Coq.Init.Specif}{\coqdocnotation{\ensuremath{|}}} \coqdocvar{n} \coqexternalref{:nat scope:x '<' x}{http://coq.inria.fr/stdlib/Coq.Init.Peano}{\coqdocnotation{$<$}} 10\coqexternalref{:type scope:'x7B' x ':' x '|' x 'x7D'}{http://coq.inria.fr/stdlib/Coq.Init.Specif}{\coqdocnotation{\}}} := $\rightarrow?$ \coqexternalref{S}{http://coq.inria.fr/stdlib/Coq.Init.Datatypes}{\coqdocconstructor{S}}.\coqdoceol
\coqdocemptyline
\end{coqdoccode}
\noindent Then, as expected: \begin{coqdoccode}
\coqdocemptyline
\coqdocnoindent
\coqdockw{Eval} \coqdoctac{compute} \coqdoctac{in} \coqref{Casts.cast1.top succ}{\coqdocdefinition{top\_succ}} 6.\coqdoceol
\coqdocemptyline
\end{coqdoccode}
\begin{verbatim}
= (7; Le.le_n_S 7 9 ...)
: {n : nat | n < 10}
\end{verbatim}

 \noindent And: \begin{coqdoccode}
\coqdocemptyline
\coqdocnoindent
\coqdockw{Eval} \coqdoctac{compute} \coqdoctac{in} \coqref{Casts.cast1.top succ}{\coqdocdefinition{top\_succ}} 9.\coqdoceol
\coqdocemptyline
\end{coqdoccode}
\begin{verbatim}
= failed_cast 10 (11 <= 10)
: {n : nat | n < 10}
\end{verbatim}

\subsection{Weakening the Domain}

\label{sec:weaksimple}

 Similarly, \coqref{Casts.cast1.cast fun dom}{\coqdocdefinition{cast\_fun\_dom}} turns a function of type \{\coqdocvariable{a} : \coqdocvariable{A} \ensuremath{|} \coqdocvariable{P} \coqdocvariable{a}\} \ensuremath{\rightarrow} \coqdocvariable{B}, which expects a value of a subset type, into a standard function of type \coqdocvariable{A}\ensuremath{\rightarrow}\coqdocvariable{B}, by casting the argument to the expected subset type:
\begin{coqdoccode}
\coqdocemptyline
\coqdocnoindent
\coqdockw{Definition} \coqdef{Casts.cast1.cast fun dom}{cast\_fun\_dom}{\coqdocdefinition{cast\_fun\_dom}} (\coqdocvar{A} \coqdocvar{B} : \coqdockw{Type}) (\coqdocvar{P}: \coqdocvariable{A} \coqexternalref{:type scope:x '->' x}{http://coq.inria.fr/stdlib/Coq.Init.Logic}{\coqdocnotation{\ensuremath{\rightarrow}}} \coqdockw{Prop}) \coqdoceol
\coqdocindent{1.00em}
(\coqdocvar{dec}: \coqdockw{\ensuremath{\forall}} \coqdocvar{a}, \coqdocclass{$\mathsf{Decidable}$} (\coqdocvariable{P} \coqdocvariable{a})) :\coqdoceol
\coqdocindent{2.00em}
\coqexternalref{:type scope:x '->' x}{http://coq.inria.fr/stdlib/Coq.Init.Logic}{\coqdocnotation{(}}\coqexternalref{:type scope:'x7B' x ':' x '|' x 'x7D'}{http://coq.inria.fr/stdlib/Coq.Init.Specif}{\coqdocnotation{\{}}\coqdocvar{a} \coqexternalref{:type scope:'x7B' x ':' x '|' x 'x7D'}{http://coq.inria.fr/stdlib/Coq.Init.Specif}{\coqdocnotation{:}} \coqdocvariable{A} \coqexternalref{:type scope:'x7B' x ':' x '|' x 'x7D'}{http://coq.inria.fr/stdlib/Coq.Init.Specif}{\coqdocnotation{\ensuremath{|}}} \coqdocvariable{P} \coqdocvar{a}\coqexternalref{:type scope:'x7B' x ':' x '|' x 'x7D'}{http://coq.inria.fr/stdlib/Coq.Init.Specif}{\coqdocnotation{\}}} \coqexternalref{:type scope:x '->' x}{http://coq.inria.fr/stdlib/Coq.Init.Logic}{\coqdocnotation{\ensuremath{\rightarrow}}} \coqdocvariable{B}\coqexternalref{:type scope:x '->' x}{http://coq.inria.fr/stdlib/Coq.Init.Logic}{\coqdocnotation{)}}  \coqexternalref{:type scope:x '->' x}{http://coq.inria.fr/stdlib/Coq.Init.Logic}{\coqdocnotation{\ensuremath{\rightarrow}}} \coqdocvariable{A} \coqexternalref{:type scope:x '->' x}{http://coq.inria.fr/stdlib/Coq.Init.Logic}{\coqdocnotation{\ensuremath{\rightarrow}}} \coqdocvariable{B} :=\coqdoceol
\coqdocindent{1.00em}
\coqdockw{fun} \coqdocvar{f} \coqdocvar{a} \ensuremath{\Rightarrow} \coqdocvariable{f} (? \coqdocvariable{a}).\coqdoceol
\coqdocnoindent
\coqdockw{Notation} \coqdef{Casts.cast1.::'<?'}{"}{"}$?\hspace{-1mm}\rightarrow$" := (\coqref{Casts.cast1.cast fun dom}{\coqdocdefinition{cast\_fun\_dom}} \coqdocvar{\_} \coqdocvar{\_} \coqdocvar{\_} \coqdocvar{\_}).\coqdoceol
\coqdocemptyline
\end{coqdoccode}
\paragraph{Example.} The standard division function on natural numbers in Coq, \coqexternalref{div}{http://coq.inria.fr/stdlib/Coq.Init.Nat}{\coqdocdefinition{div}}, is total and pure, but incorrect: when the divisor is 0, the result is 0. We can use subset types to define a pure and correct version, \coqref{Casts.cast1.divide}{\coqdocdefinition{divide}}, which is total on a restricted domain, by requiring its second argument to be strictly positive: \begin{coqdoccode}
\coqdocemptyline
\coqdocnoindent
\coqdockw{Definition} \coqdef{Casts.cast1.divide}{divide}{\coqdocdefinition{divide}}: \coqexternalref{nat}{http://coq.inria.fr/stdlib/Coq.Init.Datatypes}{\coqdocinductive{nat}} \coqexternalref{:type scope:x '->' x}{http://coq.inria.fr/stdlib/Coq.Init.Logic}{\coqdocnotation{\ensuremath{\rightarrow}}} \coqexternalref{:type scope:'x7B' x ':' x '|' x 'x7D'}{http://coq.inria.fr/stdlib/Coq.Init.Specif}{\coqdocnotation{\{}}\coqdocvar{n}\coqexternalref{:type scope:'x7B' x ':' x '|' x 'x7D'}{http://coq.inria.fr/stdlib/Coq.Init.Specif}{\coqdocnotation{:}} \coqexternalref{nat}{http://coq.inria.fr/stdlib/Coq.Init.Datatypes}{\coqdocinductive{nat}} \coqexternalref{:type scope:'x7B' x ':' x '|' x 'x7D'}{http://coq.inria.fr/stdlib/Coq.Init.Specif}{\coqdocnotation{\ensuremath{|}}} \coqdocvar{n} \coqexternalref{:nat scope:x '>' x}{http://coq.inria.fr/stdlib/Coq.Init.Peano}{\coqdocnotation{$>$}} 0\coqexternalref{:type scope:'x7B' x ':' x '|' x 'x7D'}{http://coq.inria.fr/stdlib/Coq.Init.Specif}{\coqdocnotation{\}}} \coqexternalref{:type scope:x '->' x}{http://coq.inria.fr/stdlib/Coq.Init.Logic}{\coqdocnotation{\ensuremath{\rightarrow}}} \coqexternalref{nat}{http://coq.inria.fr/stdlib/Coq.Init.Datatypes}{\coqdocinductive{nat}} := \coqdoceol
\coqdocindent{1.00em}
\coqdockw{fun} \coqdocvar{a} \coqdocvar{b} \ensuremath{\Rightarrow} \coqexternalref{div}{http://coq.inria.fr/stdlib/Coq.Init.Nat}{\coqdocdefinition{div}} \coqdocvariable{a} \coqdocvariable{b}\coqdocnotation{$._1$}.\coqdoceol
\coqdocemptyline
\end{coqdoccode}
Using this function now forces the programmer to provide a proof that the second argument is strictly positive. This can be achieved with the standard cast operator ?. Alternatively, we can cast \coqref{Casts.cast1.divide}{\coqdocdefinition{divide}} into a function that accepts plain \coqexternalref{nat}{http://coq.inria.fr/stdlib/Coq.Init.Datatypes}{\coqdocinductive{nat}}s, but internally casts the second argument to ensure it is strictly positive: \begin{coqdoccode}
\coqdocemptyline
\coqdocnoindent
\coqdockw{Definition} \coqdef{Casts.cast1.divide'}{divide'}{\coqdocdefinition{divide'}}: \coqexternalref{nat}{http://coq.inria.fr/stdlib/Coq.Init.Datatypes}{\coqdocinductive{nat}} \coqexternalref{:type scope:x '->' x}{http://coq.inria.fr/stdlib/Coq.Init.Logic}{\coqdocnotation{\ensuremath{\rightarrow}}} \coqexternalref{nat}{http://coq.inria.fr/stdlib/Coq.Init.Datatypes}{\coqdocinductive{nat}} \coqexternalref{:type scope:x '->' x}{http://coq.inria.fr/stdlib/Coq.Init.Logic}{\coqdocnotation{\ensuremath{\rightarrow}}} \coqexternalref{nat}{http://coq.inria.fr/stdlib/Coq.Init.Datatypes}{\coqdocinductive{nat}} := \coqdoceol
\coqdocindent{1.00em}
\coqdockw{fun} \coqdocvar{a} \ensuremath{\Rightarrow} \coqref{Casts.cast1.::'<?'}{\coqdocnotation{$?\hspace{-1mm}\rightarrow$}} (\coqref{Casts.cast1.divide}{\coqdocdefinition{divide}} \coqdocvariable{a}).\coqdoceol
\coqdocemptyline
\end{coqdoccode}
As expected, applying \coqref{Casts.cast1.divide'}{\coqdocdefinition{divide'}} with 0 as second argument produces a cast failure. \begin{coqdoccode}
\coqdocemptyline
\coqdocnoindent
\coqdockw{Eval} \coqdoctac{compute} \coqdoctac{in} \coqref{Casts.cast1.divide'}{\coqdocdefinition{divide'}} 1 0.\coqdoceol
\coqdocemptyline
\end{coqdoccode}
\begin{verbatim}
= match (let (a, _) := failed_cast 0 (1 <= 0) ...
: nat
\end{verbatim}

 Arguably, it is more correct for division by zero to manifest as a failure than to silently returning 0. We will also see in Section~\ref{sec:extraction} that weakening the domain of a function is helpful when extracting it to a target language that does not support subset types, because the assumptions expressed in the richly-typed world translate into runtime checks. 

\section{Higher-Order Casts, Dependently}

\label{sec:hocd}

The higher-order cast operators defined above are not applicable when
the target function type is dependently-typed. Recall that in Coq, a dependently-typed function has a type of the form \mbox{\coqdockw{\ensuremath{\forall}} \coqdocvariable{a}: \coqdocvariable{A}, \coqdocvariable{B} \coqdocvariable{a}}, meaning that the type of the result (\coqdocvariable{B} \coqdocvariable{a}) can depend on the value of the argument \coqdocvariable{a}.

For instance, in Section~\ref{sec:compile}, we cast
\coqref{Casts.cast1.compile}{\coqdocdefinition{compile}} to the dependent function type \coqdocdefinition{correct\_comp}, which is an alias for the type \coqdockw{\ensuremath{\forall}} \coqdocvariable{e}: \coqdocinductive{exp}, \{\coqdocvar{p}: \coqdocvar{prog} \ensuremath{|} \coqref{Casts.cast1.correct prog}{\coqdocdefinition{correct\_prog}} \coqdocvariable{e} \coqdocvar{p}\}. An alternative would have been to downcast \coqref{Casts.cast1.runc}{\coqdocdefinition{runc}}, which expects a correct compiler, to a looser function type that accepts any compiler (similarly to what we have done above with \coqref{Casts.cast1.divide}{\coqdocdefinition{divide}}). We now discuss both forms of casts; as it turns out, weakening the domain of a dependently-typed function is a bit of a challenge.

\subsection{Strengthening the Range}

Strengthening a function type so that it returns a rich dependent type is not more complex than with a simply-typed function; it just brings the possibility that the claimed property on the returned value also depends on the argument:
\begin{coqdoccode}
\coqdocemptyline
\coqdocnoindent
\coqdockw{Definition} \coqdef{Casts.cast1.cast forall range}{cast\_forall\_range}{\coqdocdefinition{cast\_forall\_range}} (\coqdocvar{A}: \coqdockw{Type}) (\coqdocvar{B}: \coqdocvariable{A} \coqexternalref{:type scope:x '->' x}{http://coq.inria.fr/stdlib/Coq.Init.Logic}{\coqdocnotation{\ensuremath{\rightarrow}}} \coqdockw{Type}) \coqdoceol
\coqdocindent{1.00em}
(\coqdocvar{P} : \coqdockw{\ensuremath{\forall}} \coqdocvar{a}:\coqdocvariable{A}, \coqdocvariable{B} \coqdocvariable{a} \coqexternalref{:type scope:x '->' x}{http://coq.inria.fr/stdlib/Coq.Init.Logic}{\coqdocnotation{\ensuremath{\rightarrow}}} \coqdockw{Prop}) \coqdoceol
\coqdocindent{1.00em}
(\coqdocvar{dec} : \coqdockw{\ensuremath{\forall}} \coqdocvar{a} (\coqdocvar{b} : \coqdocvariable{B} \coqdocvariable{a}), \coqdocclass{$\mathsf{Decidable}$} (\coqdocvariable{P} \coqdocvariable{a} \coqdocvariable{b})) :\coqdoceol
\coqdocindent{2.00em}
\coqexternalref{:type scope:x '->' x}{http://coq.inria.fr/stdlib/Coq.Init.Logic}{\coqdocnotation{(}}\coqdockw{\ensuremath{\forall}} \coqdocvar{a}: \coqdocvariable{A}, \coqdocvariable{B} \coqdocvariable{a}\coqexternalref{:type scope:x '->' x}{http://coq.inria.fr/stdlib/Coq.Init.Logic}{\coqdocnotation{)}} \coqexternalref{:type scope:x '->' x}{http://coq.inria.fr/stdlib/Coq.Init.Logic}{\coqdocnotation{\ensuremath{\rightarrow}}} \coqdockw{\ensuremath{\forall}} \coqdocvar{a}: \coqdocvariable{A}, \coqexternalref{:type scope:'x7B' x ':' x '|' x 'x7D'}{http://coq.inria.fr/stdlib/Coq.Init.Specif}{\coqdocnotation{\{}}\coqdocvar{b} \coqexternalref{:type scope:'x7B' x ':' x '|' x 'x7D'}{http://coq.inria.fr/stdlib/Coq.Init.Specif}{\coqdocnotation{:}} \coqdocvariable{B} \coqdocvariable{a} \coqexternalref{:type scope:'x7B' x ':' x '|' x 'x7D'}{http://coq.inria.fr/stdlib/Coq.Init.Specif}{\coqdocnotation{\ensuremath{|}}} \coqdocvariable{P} \coqdocvariable{a} \coqdocvar{b}\coqexternalref{:type scope:'x7B' x ':' x '|' x 'x7D'}{http://coq.inria.fr/stdlib/Coq.Init.Specif}{\coqdocnotation{\}}} :=\coqdoceol
\coqdocindent{1.00em}
\coqdockw{fun} \coqdocvar{f} \coqdocvar{a} \ensuremath{\Rightarrow} ? (\coqdocvariable{f} \coqdocvariable{a}).\coqdoceol
\coqdocnoindent
\coqdockw{Notation} \coqdef{Casts.cast1.::'??>'}{"}{"}$\forall?$" := (\coqref{Casts.cast1.cast forall range}{\coqdocdefinition{cast\_forall\_range}} \coqdocvar{\_} \coqdocvar{\_} \coqdocvar{\_} \coqdocvar{\_}).\coqdoceol
\coqdocemptyline
\end{coqdoccode}
\paragraph{Examples.} We can cast a \coqexternalref{nat}{http://coq.inria.fr/stdlib/Coq.Init.Datatypes}{\coqdocinductive{nat}} \ensuremath{\rightarrow} \coqexternalref{nat}{http://coq.inria.fr/stdlib/Coq.Init.Datatypes}{\coqdocinductive{nat}} function to a dependently-typed function that guarantees that it always returns a value that is greater than or equal to its argument: \begin{coqdoccode}
\coqdocemptyline
\coqdocnoindent
\coqdockw{Definition} \coqdef{Casts.cast1.f inc}{f\_inc}{\coqdocdefinition{f\_inc}} : \coqdoceol
\coqdocindent{1.00em}
\coqexternalref{:type scope:x '->' x}{http://coq.inria.fr/stdlib/Coq.Init.Logic}{\coqdocnotation{(}}\coqexternalref{nat}{http://coq.inria.fr/stdlib/Coq.Init.Datatypes}{\coqdocinductive{nat}} \coqexternalref{:type scope:x '->' x}{http://coq.inria.fr/stdlib/Coq.Init.Logic}{\coqdocnotation{\ensuremath{\rightarrow}}} \coqexternalref{nat}{http://coq.inria.fr/stdlib/Coq.Init.Datatypes}{\coqdocinductive{nat}}\coqexternalref{:type scope:x '->' x}{http://coq.inria.fr/stdlib/Coq.Init.Logic}{\coqdocnotation{)}} \coqexternalref{:type scope:x '->' x}{http://coq.inria.fr/stdlib/Coq.Init.Logic}{\coqdocnotation{\ensuremath{\rightarrow}}} \coqdockw{\ensuremath{\forall}} \coqdocvar{n} : \coqexternalref{nat}{http://coq.inria.fr/stdlib/Coq.Init.Datatypes}{\coqdocinductive{nat}}, \coqexternalref{:type scope:'x7B' x ':' x '|' x 'x7D'}{http://coq.inria.fr/stdlib/Coq.Init.Specif}{\coqdocnotation{\{}}\coqdocvar{m}\coqexternalref{:type scope:'x7B' x ':' x '|' x 'x7D'}{http://coq.inria.fr/stdlib/Coq.Init.Specif}{\coqdocnotation{:}}\coqexternalref{nat}{http://coq.inria.fr/stdlib/Coq.Init.Datatypes}{\coqdocinductive{nat}} \coqexternalref{:type scope:'x7B' x ':' x '|' x 'x7D'}{http://coq.inria.fr/stdlib/Coq.Init.Specif}{\coqdocnotation{\ensuremath{|}}} \coqexternalref{:type scope:'x7B' x ':' x '|' x 'x7D'}{http://coq.inria.fr/stdlib/Coq.Init.Specif}{\coqdocnotation{(}}\coqdocvariable{n} \coqexternalref{:nat scope:x '<=' x}{http://coq.inria.fr/stdlib/Coq.Init.Peano}{\coqdocnotation{\ensuremath{\le}}} \coqdocvar{m}\coqexternalref{:type scope:'x7B' x ':' x '|' x 'x7D'}{http://coq.inria.fr/stdlib/Coq.Init.Specif}{\coqdocnotation{)\}}} := $\forall?$.\coqdoceol
\coqdocemptyline
\end{coqdoccode}
\noindent Then, as expected: \begin{coqdoccode}
\coqdocemptyline
\coqdocnoindent
\coqdockw{Eval} \coqdoctac{compute} \coqdoctac{in} \coqref{Casts.cast1.f inc}{\coqdocdefinition{f\_inc}} \coqexternalref{S}{http://coq.inria.fr/stdlib/Coq.Init.Datatypes}{\coqdocconstructor{S}} 3.\coqdoceol
\coqdocemptyline
\end{coqdoccode}
\begin{verbatim}
= (4; Le.le_n_S 2 3 ...)
: {m : nat | 3 <= m}
\end{verbatim}

 \noindent And: \begin{coqdoccode}
\coqdocemptyline
\coqdocnoindent
\coqdockw{Eval} \coqdoctac{compute} \coqdoctac{in} \coqref{Casts.cast1.f inc}{\coqdocdefinition{f\_inc}} (\coqdockw{fun} \coqdocvar{\_} \ensuremath{\Rightarrow} \coqexternalref{O}{http://coq.inria.fr/stdlib/Coq.Init.Datatypes}{\coqdocconstructor{O}}) 3.\coqdoceol
\coqdocemptyline
\end{coqdoccode}
\begin{verbatim}
= failed_cast 0 (3 <= 0)
: {m : nat | 3 <= m}
\end{verbatim}

 The above example casts a simply-typed function to a dependently-typed function, also illustrating the binary property \coqdocvariable{P} \coqdocvariable{a} \coqdocvariable{b} in the range. In the following example, the casted function is dependently-typed. Consider the inductive type of length-indexed lists of \coqexternalref{nat}{http://coq.inria.fr/stdlib/Coq.Init.Datatypes}{\coqdocinductive{nat}}, and the dependently-typed constructor \coqref{Casts.cast1.build list}{\coqdocdefinition{build\_list}}: \label{ilist}
\begin{coqdoccode}
\coqdocemptyline
\coqdocnoindent
\coqdockw{Inductive} \coqdef{Casts.cast1.ilist}{ilist}{\coqdocinductive{ilist}} : \coqexternalref{nat}{http://coq.inria.fr/stdlib/Coq.Init.Datatypes}{\coqdocinductive{nat}} \coqexternalref{:type scope:x '->' x}{http://coq.inria.fr/stdlib/Coq.Init.Logic}{\coqdocnotation{\ensuremath{\rightarrow}}} \coqdockw{Set} :=\coqdoceol
\coqdocindent{0.50em}
\ensuremath{|} \coqdef{Casts.cast1.Nil}{Nil}{\coqdocconstructor{Nil}} : \coqref{Casts.cast1.ilist}{\coqdocinductive{ilist}} \coqexternalref{O}{http://coq.inria.fr/stdlib/Coq.Init.Datatypes}{\coqdocconstructor{O}}\coqdoceol
\coqdocindent{0.50em}
\ensuremath{|} \coqdef{Casts.cast1.Cons}{Cons}{\coqdocconstructor{Cons}} : \coqdockw{\ensuremath{\forall}} \coqdocvar{n}, \coqexternalref{nat}{http://coq.inria.fr/stdlib/Coq.Init.Datatypes}{\coqdocinductive{nat}} \coqexternalref{:type scope:x '->' x}{http://coq.inria.fr/stdlib/Coq.Init.Logic}{\coqdocnotation{\ensuremath{\rightarrow}}} \coqref{Casts.cast1.ilist}{\coqdocinductive{ilist}} \coqdocvariable{n} \coqexternalref{:type scope:x '->' x}{http://coq.inria.fr/stdlib/Coq.Init.Logic}{\coqdocnotation{\ensuremath{\rightarrow}}} \coqref{Casts.cast1.ilist}{\coqdocinductive{ilist}} (\coqexternalref{S}{http://coq.inria.fr/stdlib/Coq.Init.Datatypes}{\coqdocconstructor{S}} \coqdocvariable{n}).\coqdoceol
\coqdocemptyline
\coqdocnoindent
\coqdockw{Fixpoint} \coqdef{Casts.cast1.build list}{build\_list}{\coqdocdefinition{build\_list}} (\coqdocvar{n}: \coqexternalref{nat}{http://coq.inria.fr/stdlib/Coq.Init.Datatypes}{\coqdocinductive{nat}}) : \coqref{Casts.cast1.ilist}{\coqdocinductive{ilist}} \coqdocvariable{n} :=\coqdoceol
\coqdocindent{0.50em}
\coqdockw{match} \coqdocvariable{n} \coqdockw{with}\coqdoceol
\coqdocindent{1.50em}
\ensuremath{|} \coqexternalref{O}{http://coq.inria.fr/stdlib/Coq.Init.Datatypes}{\coqdocconstructor{O}} \ensuremath{\Rightarrow} \coqref{Casts.cast1.Nil}{\coqdocconstructor{Nil}}\coqdoceol
\coqdocindent{1.50em}
\ensuremath{|} \coqexternalref{S}{http://coq.inria.fr/stdlib/Coq.Init.Datatypes}{\coqdocconstructor{S}} \coqdocvar{m} \ensuremath{\Rightarrow} \coqref{Casts.cast1.Cons}{\coqdocconstructor{Cons}} \coqdocvar{\_} \coqexternalref{O}{http://coq.inria.fr/stdlib/Coq.Init.Datatypes}{\coqdocconstructor{O}} (\coqref{Casts.cast1.build list}{\coqdocdefinition{build\_list}} \coqdocvar{m})\coqdoceol
\coqdocindent{0.50em}
\coqdockw{end}.\coqdoceol
\coqdocemptyline
\end{coqdoccode}
We can cast \coqref{Casts.cast1.build list}{\coqdocdefinition{build\_list}} (of type \coqdockw{\ensuremath{\forall}} \coqdocvariable{n}:\coqexternalref{nat}{http://coq.inria.fr/stdlib/Coq.Init.Datatypes}{\coqdocinductive{nat}}, \coqref{Casts.cast1.ilist}{\coqdocinductive{ilist}} \coqdocvariable{n}) to a function type that additionally guarantees that the produced list is not empty. \begin{coqdoccode}
\coqdocemptyline
\coqdocnoindent
\coqdockw{Definition} \coqdef{Casts.cast1.non empty build}{non\_empty\_build}{\coqdocdefinition{non\_empty\_build}}:\coqdoceol
\coqdocindent{2.00em}
\coqdockw{\ensuremath{\forall}} \coqdocvar{n}:\coqexternalref{nat}{http://coq.inria.fr/stdlib/Coq.Init.Datatypes}{\coqdocinductive{nat}},  \coqexternalref{:type scope:'x7B' x ':' x '|' x 'x7D'}{http://coq.inria.fr/stdlib/Coq.Init.Specif}{\coqdocnotation{\{}}\coqdocvar{\_}\coqexternalref{:type scope:'x7B' x ':' x '|' x 'x7D'}{http://coq.inria.fr/stdlib/Coq.Init.Specif}{\coqdocnotation{:}} \coqref{Casts.cast1.ilist}{\coqdocinductive{ilist}} \coqdocvariable{n} \coqexternalref{:type scope:'x7B' x ':' x '|' x 'x7D'}{http://coq.inria.fr/stdlib/Coq.Init.Specif}{\coqdocnotation{\ensuremath{|}}} \coqdocvariable{n} \coqexternalref{:nat scope:x '>' x}{http://coq.inria.fr/stdlib/Coq.Init.Peano}{\coqdocnotation{$>$}} 0 \coqexternalref{:type scope:'x7B' x ':' x '|' x 'x7D'}{http://coq.inria.fr/stdlib/Coq.Init.Specif}{\coqdocnotation{\}}} := $\forall?$ \coqref{Casts.cast1.build list}{\coqdocdefinition{build\_list}}.\coqdoceol
\coqdocemptyline
\end{coqdoccode}
\noindent Then, as expected: \begin{coqdoccode}
\coqdocemptyline
\coqdocnoindent
\coqdockw{Eval} \coqdoctac{compute} \coqdoctac{in} \coqref{Casts.cast1.non empty build}{\coqdocdefinition{non\_empty\_build}} 2.\coqdoceol
\coqdocemptyline
\end{coqdoccode}
\begin{verbatim}
= (Cons 1 0 (Cons 0 0 Nil); ...)
: {_ : ilist 2 | 2 > 0}
\end{verbatim}

 \noindent And: \begin{coqdoccode}
\coqdocemptyline
\coqdocnoindent
\coqdockw{Eval} \coqdoctac{compute} \coqdoctac{in} \coqref{Casts.cast1.non empty build}{\coqdocdefinition{non\_empty\_build}} 0.\coqdoceol
\coqdocemptyline
\end{coqdoccode}
\begin{verbatim}
= failed_cast Nil (1 <= 0)
: {_ : ilist 0 | 0 > 0}
\end{verbatim}

\subsection{Weakening the Domain}

\label{sec:weakdep}

Consider a function that expects an argument of a subset type \{\coqdocvariable{a} : \coqdocvariable{A} \ensuremath{|} \coqdocvariable{P} \coqdocvariable{a}\}, and whose return type depends on the value component of the dependent pair. Such a function has type \coqdockw{\ensuremath{\forall}} \coqdocvariable{x}: \{\coqdocvariable{a} : \coqdocvariable{A} \ensuremath{|} \coqdocvariable{P} \coqdocvariable{a}\}, \coqdocvariable{B} \coqdocvariable{x}$._1$. Weakening the domain in this case means casting this function to the dependent type \coqdockw{\ensuremath{\forall}} \coqdocvariable{a} : \coqdocvariable{A}, \coqdocvariable{B} \coqdocvariable{a}.

Notably, defining such a cast operator leads to an interesting insight regarding casts in a dependently-typed language. Because \coqref{Casts.cast1.cast}{\coqdocdefinition{cast}} hides a lie about a value, when casting the argument of a dependently-typed function, the lie percolates at the type level due to the dependency. Consider the intuitive definition of \coqref{Casts.cast1.cast forall dom}{\coqdocdefinition{cast\_forall\_dom}}, which simply applies \coqref{Casts.cast1.cast}{\coqdocdefinition{cast}} to the argument:\\

\coqdockw{Definition} \coqref{Casts.cast1.cast forall dom}{\coqdocdefinition{cast\_forall\_dom}} (\coqdocvariable{A}: \coqdockw{Type}) (\coqdocvariable{P}: \coqdocvariable{A} \ensuremath{\rightarrow} \coqdockw{Prop})

\qquad\qquad (\coqdocvariable{B}: \coqdocvariable{A} \ensuremath{\rightarrow} \coqdockw{Type}) (\coqref{Casts.cast1.dec}{\coqdocprojection{dec}}: \coqdockw{\ensuremath{\forall}} \coqdocvariable{a}, \coqref{Casts.cast1.Decidable}{\coqdocrecord{$\mathsf{Decidable}$}} (\coqdocvariable{P} \coqdocvariable{a})) :

\qquad(\coqdockw{\ensuremath{\forall}} \coqdocvariable{x}: \{\coqdocvariable{a} : \coqdocvariable{A} \ensuremath{|} \coqdocvariable{P} \coqdocvariable{a}\}, \coqdocvariable{B} \coqdocvariable{x}$._1)$  \ensuremath{\rightarrow} (\coqdockw{\ensuremath{\forall}} \coqdocvariable{a} : \coqdocvariable{A}, \coqdocvariable{B} \coqdocvariable{a}) :=

\qquad\coqdockw{fun} \coqdocvariable{f} \coqdocvariable{a} \ensuremath{\Rightarrow} \coqdocvariable{f} (? \coqdocvariable{a}).

\vspace{2mm}
\noindent Coq (rightfully) complains that:
\begin{verbatim}
The term "f (? a)" has type "B (? a).1" 
while it is expected to have type "B a".
\end{verbatim}

Indeed, the return type of the casted function can depend on the argument, yet we are lying about the argument by claiming that it has the subset type \{\coqdocvariable{a} : \coqdocvariable{A} \ensuremath{|} \coqdocvariable{P} \coqdocvariable{a}\}. Therefore, in all honesty, the only thing we know about \coqdocvariable{f} (? \coqdocvariable{a}) is that it has type \coqdocvariable{B} \coqdocvariable{a} \emph{only if the cast succeeds}---in which case (? \coqdocvariable{a}$)._1$ = \coqdocvariable{a}. But the cast may fail, in which case ? \coqdocvariable{a} is not a dependent pair and (? \coqdocvariable{a}$)._1$ cannot be reduced: it is a cast error at the type level.

What can we do about this? We know that cast errors can occur, but we do not want to pollute all types with that uncertainty. Following the axiomatic approach to casts, we can introduce a second axiom, \coqref{Casts.cast1.failed cast proj1}{\coqdocaxiom{failed\_cast\_proj1}}, to  {\em hide} the fact that cast errors can occur at the type level. Note that we do not want to pose the equality (? \coqdocvariable{a}$)._1$ = \coqdocvariable{a} as an axiom, otherwise we would be relying on the axiom even though the cast succeeds. The axiom is required only to {\em pretend} that the first projection of a failed cast is actually the casted value\footnote{The key word in the sentence is \emph{pretend}: the new axiom does {\em not} allow one to actually project a value out of a failed cast; it only serves to hide the potential for cast failure from the types.}:
\begin{coqdoccode}
\coqdocemptyline
\coqdocnoindent
\coqdockw{Axiom} \coqdef{Casts.cast1.failed cast proj1}{failed\_cast\_proj1}{\coqdocaxiom{failed\_cast\_proj1}} : \coqdoceol
\coqdocindent{1.00em}
\coqdockw{\ensuremath{\forall}} \{\coqdocvar{A}:\coqdockw{Type}\} \{\coqdocvar{P} : \coqdocvariable{A} \coqexternalref{:type scope:x '->' x}{http://coq.inria.fr/stdlib/Coq.Init.Logic}{\coqdocnotation{\ensuremath{\rightarrow}}} \coqdockw{Prop}\} \{\coqdocvar{a}: \coqdocvariable{A}\} (\coqdocvar{msg}:\coqdockw{Prop}),\coqdoceol
\coqdocindent{2.00em}
\coqdocnotation{(}\coqref{Casts.cast1.failed cast}{\coqdocaxiom{failed\_cast}} (\coqdocvar{P}:=\coqdocvariable{P}) \coqdocvariable{a} \coqdocvariable{msg}\coqdocnotation{$)._1$} \coqexternalref{:type scope:x '=' x}{http://coq.inria.fr/stdlib/Coq.Init.Logic}{\coqdocnotation{=}} \coqdocvariable{a}.\coqdoceol
\coqdocemptyline
\end{coqdoccode}
\noindent Using this axiom allows us to define an operator to hide casts from types, \coqref{Casts.cast1.hide cast proj1}{\coqdocdefinition{hide\_cast\_proj1}} (notation [?]), as follows:\footnote{This time, we use tactics to define \cdef{hide\_cast\_proj1}, instead of giving the functional term explicitly as we did for \cdef{cast}. The reason is that because of the dependency, a simple pattern matching does not suffice and extra type annotations have to be added to \ckw{match} in order to help Coq typecheck the dependent pattern matching.} \begin{coqdoccode}
\coqdocemptyline
\coqdocnoindent
\coqdockw{Definition} \coqdef{Casts.cast1.hide cast proj1}{hide\_cast\_proj1}{\coqdocdefinition{hide\_cast\_proj1}} (\coqdocvar{A}: \coqdockw{Type}) (\coqdocvar{P}: \coqdocvariable{A} \coqexternalref{:type scope:x '->' x}{http://coq.inria.fr/stdlib/Coq.Init.Logic}{\coqdocnotation{\ensuremath{\rightarrow}}} \coqdockw{Prop})\coqdoceol
\coqdocindent{1.00em}
(\coqdocvar{B}: \coqdocvariable{A} \coqexternalref{:type scope:x '->' x}{http://coq.inria.fr/stdlib/Coq.Init.Logic}{\coqdocnotation{\ensuremath{\rightarrow}}} \coqdockw{Type}) (\coqdocvar{dec}: \coqdockw{\ensuremath{\forall}} \coqdocvar{a}, \coqdocclass{$\mathsf{Decidable}$} (\coqdocvariable{P} \coqdocvariable{a})) (\coqdocvar{a}:\coqdocvariable{A}):\coqdoceol
\coqdocindent{1.00em}
\coqdocvariable{B} \coqdocnotation{(?} \coqdocvariable{a}\coqdocnotation{$)._1$} \coqexternalref{:type scope:x '->' x}{http://coq.inria.fr/stdlib/Coq.Init.Logic}{\coqdocnotation{\ensuremath{\rightarrow}}} \coqdocvariable{B} \coqdocvariable{a}.\coqdoceol
\coqdocnoindent
\coqdockw{Proof}.\coqdoceol
\coqdocindent{1.00em}
\coqdoctac{unfold} \coqref{Casts.cast1.cast}{\coqdocdefinition{cast}}. \coqdoctac{case} (\coqdocvar{dec} \coqdocvar{a}); \coqdoctac{intro} \coqdocvar{p}.\coqdoceol
\coqdocindent{1.00em}
- \coqdoctac{exact} (\coqdockw{fun} \coqdocvar{b} \ensuremath{\Rightarrow} \coqdocvariable{b}).\coqdoceol
\coqdocindent{1.00em}
- \coqdoctac{exact} (\coqdockw{fun} \coqdocvar{b} \ensuremath{\Rightarrow} \coqexternalref{eq rect}{http://coq.inria.fr/stdlib/Coq.Init.Logic}{\coqdocdefinition{eq\_rect}} \coqdocvar{\_} \coqdocvar{\_} \coqdocvariable{b} \coqdocvar{\_}\coqdoceol
\coqdocindent{12.50em}
(\coqref{Casts.cast1.failed cast proj1}{\coqdocaxiom{failed\_cast\_proj1}} (\coqdocvar{P} \coqdocvar{a}))).\coqdoceol
\coqdocnoindent
\coqdockw{Defined}.\coqdoceol
\coqdocemptyline
\coqdocnoindent
\coqdockw{Notation} \coqdef{Casts.cast1.::'[?]'}{"}{"}[?]" := (\coqref{Casts.cast1.hide cast proj1}{\coqdocdefinition{hide\_cast\_proj1}} \coqdocvar{\_} \coqdocvar{\_} \coqdocvar{\_} \coqdocvar{\_} \coqdocvar{\_}).\coqdoceol
\coqdocemptyline
\end{coqdoccode}
\noindent 
The equality coming from \coqref{Casts.cast1.failed cast proj1}{\coqdocaxiom{failed\_cast\_proj1}} is necessary 
to transform the term \coqdocvariable{b} of type \coqdocvariable{B} (\coqref{Casts.cast1.failed cast}{\coqdocaxiom{failed\_cast}} \coqdocvar{\_} \coqdocvariable{P} \coqdocvariable{a} \coqdocvariable{msg}$)._1$ to a
term of type \coqdocvariable{B} \coqdocvariable{a}. This is done using the elimination rule \coqexternalref{eq rect}{http://coq.inria.fr/stdlib/Coq.Init.Logic}{\coqdocdefinition{eq\_rect}}
of the equality type. Here again, we can see that a
\coqref{Casts.cast1.failed cast proj1}{\coqdocaxiom{failed\_cast\_proj1}} error will only occur if the property \coqdocvariable{P} \coqdocvariable{a} does
not hold. 

 We can now define \coqref{Casts.cast1.cast forall dom}{\coqdocdefinition{cast\_forall\_dom}} as expected, by adding the hiding of the cast in the return type: \begin{coqdoccode}
\coqdocemptyline
\coqdocnoindent
\coqdockw{Definition} \coqdef{Casts.cast1.cast forall dom}{cast\_forall\_dom}{\coqdocdefinition{cast\_forall\_dom}} (\coqdocvar{A}: \coqdockw{Type}) (\coqdocvar{P}: \coqdocvariable{A} \coqexternalref{:type scope:x '->' x}{http://coq.inria.fr/stdlib/Coq.Init.Logic}{\coqdocnotation{\ensuremath{\rightarrow}}} \coqdockw{Prop}) \coqdoceol
\coqdocindent{5.50em}
(\coqdocvar{B}: \coqdocvariable{A} \coqexternalref{:type scope:x '->' x}{http://coq.inria.fr/stdlib/Coq.Init.Logic}{\coqdocnotation{\ensuremath{\rightarrow}}} \coqdockw{Type}) (\coqdocvar{dec}: \coqdockw{\ensuremath{\forall}} \coqdocvar{a}, \coqdocclass{$\mathsf{Decidable}$} (\coqdocvariable{P} \coqdocvariable{a})) :\coqdoceol
\coqdocindent{1.50em}
\coqexternalref{:type scope:x '->' x}{http://coq.inria.fr/stdlib/Coq.Init.Logic}{\coqdocnotation{(}}\coqdockw{\ensuremath{\forall}} \coqdocvar{x}: \coqexternalref{:type scope:'x7B' x ':' x '|' x 'x7D'}{http://coq.inria.fr/stdlib/Coq.Init.Specif}{\coqdocnotation{\{}}\coqdocvar{a} \coqexternalref{:type scope:'x7B' x ':' x '|' x 'x7D'}{http://coq.inria.fr/stdlib/Coq.Init.Specif}{\coqdocnotation{:}} \coqdocvariable{A} \coqexternalref{:type scope:'x7B' x ':' x '|' x 'x7D'}{http://coq.inria.fr/stdlib/Coq.Init.Specif}{\coqdocnotation{\ensuremath{|}}} \coqdocvariable{P} \coqdocvar{a}\coqexternalref{:type scope:'x7B' x ':' x '|' x 'x7D'}{http://coq.inria.fr/stdlib/Coq.Init.Specif}{\coqdocnotation{\}}}, \coqdocvariable{B} \coqdocvariable{x}\coqdocnotation{$._1$}\coqexternalref{:type scope:x '->' x}{http://coq.inria.fr/stdlib/Coq.Init.Logic}{\coqdocnotation{)}}  \coqexternalref{:type scope:x '->' x}{http://coq.inria.fr/stdlib/Coq.Init.Logic}{\coqdocnotation{\ensuremath{\rightarrow}}} \coqexternalref{:type scope:x '->' x}{http://coq.inria.fr/stdlib/Coq.Init.Logic}{\coqdocnotation{(}}\coqdockw{\ensuremath{\forall}} \coqdocvar{a} : \coqdocvariable{A}, \coqdocvariable{B} \coqdocvariable{a}\coqexternalref{:type scope:x '->' x}{http://coq.inria.fr/stdlib/Coq.Init.Logic}{\coqdocnotation{)}} :=\coqdoceol
\coqdocindent{1.00em}
\coqdockw{fun} \coqdocvar{f} \coqdocvar{a} \ensuremath{\Rightarrow} \coqref{Casts.cast1.::'[?]'}{\coqdocnotation{[?]}} (\coqdocvariable{f} (? \coqdocvariable{a})).\coqdoceol
\coqdocnoindent
\coqdockw{Notation} \coqdef{Casts.cast1.::'<??'}{"}{"}$?\forall$" := (\coqref{Casts.cast1.cast forall dom}{\coqdocdefinition{cast\_forall\_dom}} \coqdocvar{\_} \coqdocvar{\_} \coqdocvar{\_} \coqdocvar{\_}).\coqdoceol
\coqdocemptyline
\end{coqdoccode}
\paragraph{Example.} Recall the length-indexed lists of Sect.~\ref{ilist}. Consider the following dependently-typed function with a rich domain type, which specifies that given a strictly positive \coqexternalref{nat}{http://coq.inria.fr/stdlib/Coq.Init.Datatypes}{\coqdocinductive{nat}}, it returns an \coqref{Casts.cast1.ilist}{\coqdocinductive{ilist}} of that length:
\begin{coqdoccode}
\coqdocemptyline
\coqdocnoindent
\coqdockw{Definition} \coqdef{Casts.cast1.build pos}{build\_pos}{\coqdocdefinition{build\_pos}} : \coqexternalref{:type scope:'xE2x88x80' x '..' x ',' x}{http://coq.inria.fr/stdlib/Coq.Unicode.Utf8\_core}{\coqdocnotation{$\forall$}} \coqdocvar{x}: \coqexternalref{:type scope:'x7B' x ':' x '|' x 'x7D'}{http://coq.inria.fr/stdlib/Coq.Init.Specif}{\coqdocnotation{\{}}\coqdocvar{n}\coqexternalref{:type scope:'x7B' x ':' x '|' x 'x7D'}{http://coq.inria.fr/stdlib/Coq.Init.Specif}{\coqdocnotation{:}} \coqexternalref{nat}{http://coq.inria.fr/stdlib/Coq.Init.Datatypes}{\coqdocinductive{nat}} \coqexternalref{:type scope:'x7B' x ':' x '|' x 'x7D'}{http://coq.inria.fr/stdlib/Coq.Init.Specif}{\coqdocnotation{\ensuremath{|}}} \coqdocvar{n} \coqexternalref{:nat scope:x '>' x}{http://coq.inria.fr/stdlib/Coq.Init.Peano}{\coqdocnotation{$>$}} 0 \coqexternalref{:type scope:'x7B' x ':' x '|' x 'x7D'}{http://coq.inria.fr/stdlib/Coq.Init.Specif}{\coqdocnotation{\}}}\coqexternalref{:type scope:'xE2x88x80' x '..' x ',' x}{http://coq.inria.fr/stdlib/Coq.Unicode.Utf8\_core}{\coqdocnotation{,}} \coqref{Casts.cast1.ilist}{\coqdocinductive{ilist}} (\coqdocvariable{x}\coqdocnotation{$._1$}) :=\coqdoceol
\coqdocindent{0.50em}
\coqdockw{fun} \coqdocvar{n} \ensuremath{\Rightarrow} \coqref{Casts.cast1.build list}{\coqdocdefinition{build\_list}} (\coqdocvariable{n}\coqdocnotation{$._1$}).\coqdoceol
\coqdocemptyline
\end{coqdoccode}
We can use $?\forall$ to safely hide the requirement that \coqdocvariable{n} $>$ 0: \begin{coqdoccode}
\coqdocemptyline
\coqdocnoindent
\coqdockw{Definition} \coqdef{Casts.cast1.build pos'}{build\_pos'}{\coqdocdefinition{build\_pos'}} : \coqdockw{\ensuremath{\forall}} \coqdocvar{n}: \coqexternalref{nat}{http://coq.inria.fr/stdlib/Coq.Init.Datatypes}{\coqdocinductive{nat}}, \coqref{Casts.cast1.ilist}{\coqdocinductive{ilist}} \coqdocvariable{n} := \coqref{Casts.cast1.::'<??'}{\coqdocnotation{$?\forall$}} \coqref{Casts.cast1.build pos}{\coqdocdefinition{build\_pos}}.\coqdoceol
\coqdocemptyline
\end{coqdoccode}
\noindent Then, as expected: \begin{coqdoccode}
\coqdocemptyline
\coqdocnoindent
\coqdockw{Eval} \coqdoctac{compute} \coqdoctac{in} \coqref{Casts.cast1.build pos'}{\coqdocdefinition{build\_pos'}} 2.\coqdoceol
\coqdocemptyline
\end{coqdoccode}
\begin{verbatim}
= Cons 1 0 (Cons 0 0 Nil)
: ilist 2
\end{verbatim}

 \noindent\label{sec:wdd} And we can now see \coqref{Casts.cast1.failed cast proj1}{\coqdocaxiom{failed\_cast\_proj1}} appearing: \begin{coqdoccode}
\coqdocemptyline
\coqdocemptyline
\coqdocnoindent
\coqdockw{Eval} \coqdoctac{compute} \coqdoctac{in} \coqref{Casts.cast1.build pos'}{\coqdocdefinition{build\_pos'}} 0.\coqdoceol
\coqdocemptyline
\coqdocemptyline
\end{coqdoccode}

\begin{verbatim}
= eq_rect ... 
 ((fix build_list (n : nat) : ilist n := ...)
  (let (a, _) := failed_cast 0 (1 <= 0) in a)) 
 0 (failed_cast_proj1 (1 <= 0))
: ilist 0
\end{verbatim}
\begin{coqdoccode}
\coqdocemptyline
\coqdocemptyline
\end{coqdoccode}

\section{Extraction}
\label{sec:extraction}

An interesting feature of Coq in terms of bridging certified
programming with practical developments is the possibility to {\em
  extract} definitions to mainstream languages. The standard
distribution of Coq supports extraction to Ocaml, Haskell, and Scheme;
and there exists several experimental projects for extracting Coq to
other languages like Scala and Erlang.

Coq establishes a strong distinction between programs (in
\coqdockw{Type}), which have computational content, and proofs (in
\coqdockw{Prop}), which are devoid of computational meaning and are
therefore erased during extraction. This allows for extracted programs
to be efficient and not carry around the burden of unnecessary proof
terms. However, this erasure of proofs also means that subset types
are extracted to plain types, without any safeguards. It also means
that the use of admitted properties is simply and unsafely erased!

To address these issues, we can exploit our cast framework.  By
establishing a bridge between properties and computation, casts are
extracted as runtime checks, and cast failures manifest as runtime
exceptions---which is exactly how standard casts work in mainstream
programming languages.  This ensures that the assumptions made by
certified components extracted to a mainstream language are
dynamically enforced.

\coqlibrary{Casts.extraction}{Library }{Casts.extraction}

\begin{coqdoccode}
\coqdocemptyline
\coqdocemptyline
\end{coqdoccode}

\paragraph{Example.} Recall from Section~\ref{sec:weaksimple} the \coqref{Casts.extraction.divide}{\coqdocdefinition{divide}} function  of type
\coqexternalref{nat}{http://coq.inria.fr/stdlib/Coq.Init.Datatypes}{\coqdocinductive{nat}} \ensuremath{\rightarrow} \{\coqdocvar{n}: \coqexternalref{nat}{http://coq.inria.fr/stdlib/Coq.Init.Datatypes}{\coqdocinductive{nat}} \ensuremath{|} \coqdocvar{n} $>$ 0\} \ensuremath{\rightarrow} \coqexternalref{nat}{http://coq.inria.fr/stdlib/Coq.Init.Datatypes}{\coqdocinductive{nat}}. 
To define \coqref{Casts.extraction.divide}{\coqdocdefinition{divide}}, the programmer works under
the assumption that the second argument is strictly positive. However,
this guarantee is lost when extracting the function to a mainstream
programming language, because the extracted function has the plain type
\coqexternalref{nat}{http://coq.inria.fr/stdlib/Coq.Init.Datatypes}{\coqdocinductive{nat}}\ensuremath{\rightarrow}\coqexternalref{nat}{http://coq.inria.fr/stdlib/Coq.Init.Datatypes}{\coqdocinductive{nat}}\ensuremath{\rightarrow}\coqexternalref{nat}{http://coq.inria.fr/stdlib/Coq.Init.Datatypes}{\coqdocinductive{nat}}:
\begin{coqdoccode}
\coqdocemptyline
\coqdocnoindent
\coqdockw{Definition} \coqdef{Casts.extraction.divide}{divide}{\coqdocdefinition{divide}}: \coqexternalref{nat}{http://coq.inria.fr/stdlib/Coq.Init.Datatypes}{\coqdocinductive{nat}} \coqexternalref{:type scope:x '->' x}{http://coq.inria.fr/stdlib/Coq.Init.Logic}{\coqdocnotation{\ensuremath{\rightarrow}}} \coqexternalref{:type scope:'x7B' x ':' x '|' x 'x7D'}{http://coq.inria.fr/stdlib/Coq.Init.Specif}{\coqdocnotation{\{}}\coqdocvar{n}\coqexternalref{:type scope:'x7B' x ':' x '|' x 'x7D'}{http://coq.inria.fr/stdlib/Coq.Init.Specif}{\coqdocnotation{:}} \coqexternalref{nat}{http://coq.inria.fr/stdlib/Coq.Init.Datatypes}{\coqdocinductive{nat}} \coqexternalref{:type scope:'x7B' x ':' x '|' x 'x7D'}{http://coq.inria.fr/stdlib/Coq.Init.Specif}{\coqdocnotation{\ensuremath{|}}} \coqdocvar{n} \coqexternalref{:nat scope:x '>' x}{http://coq.inria.fr/stdlib/Coq.Init.Peano}{\coqdocnotation{$>$}} 0\coqexternalref{:type scope:'x7B' x ':' x '|' x 'x7D'}{http://coq.inria.fr/stdlib/Coq.Init.Specif}{\coqdocnotation{\}}} \coqexternalref{:type scope:x '->' x}{http://coq.inria.fr/stdlib/Coq.Init.Logic}{\coqdocnotation{\ensuremath{\rightarrow}}} \coqexternalref{nat}{http://coq.inria.fr/stdlib/Coq.Init.Datatypes}{\coqdocinductive{nat}} := \coqdoceol
\coqdocindent{1.00em}
\coqdockw{fun} \coqdocvar{a} \coqdocvar{b} \ensuremath{\Rightarrow} \coqexternalref{div}{http://coq.inria.fr/stdlib/Coq.Init.Nat}{\coqdocdefinition{div}} \coqdocvariable{a} \coqdocvariable{b}\coqdocnotation{$._1$}.\coqdoceol
\coqdocemptyline
\coqdocnoindent
\coqdockw{Extraction} \coqdocvar{Language} \coqdocvar{Ocaml}.\coqdoceol
\coqdocnoindent
\coqdockw{Extraction} \coqref{Casts.extraction.divide}{\coqdocdefinition{divide}}.\coqdoceol
\coqdocemptyline
\end{coqdoccode}

\begin{verbatim}
let divide a b = div a b
\end{verbatim}

 The dependent pair corresponding to the subset type
has been erased, and \coqref{Casts.extraction.divide}{\coqdocdefinition{divide}} does not check that the second argument
is positive (we extract \coqexternalref{nat}{http://coq.inria.fr/stdlib/Coq.Init.Datatypes}{\coqdocinductive{nat}} to OCaml's {\tt int}):

\begin{verbatim}
# divide 1 0;;
- : int = 0
\end{verbatim}

 If we instead first cast \coqref{Casts.extraction.divide}{\coqdocdefinition{divide}} to the \coqref{Casts.extraction.divide'}{\coqdocdefinition{divide'}} function with plain type \coqexternalref{nat}{http://coq.inria.fr/stdlib/Coq.Init.Datatypes}{\coqdocinductive{nat}}\ensuremath{\rightarrow}\coqexternalref{nat}{http://coq.inria.fr/stdlib/Coq.Init.Datatypes}{\coqdocinductive{nat}}\ensuremath{\rightarrow}\coqexternalref{nat}{http://coq.inria.fr/stdlib/Coq.Init.Datatypes}{\coqdocinductive{nat}}, and then extract \coqref{Casts.extraction.divide'}{\coqdocdefinition{divide'}}: \begin{coqdoccode}
\coqdocemptyline
\coqdocnoindent
\coqdockw{Definition} \coqdef{Casts.extraction.divide'}{divide'}{\coqdocdefinition{divide'}}: \coqexternalref{nat}{http://coq.inria.fr/stdlib/Coq.Init.Datatypes}{\coqdocinductive{nat}} \coqexternalref{:type scope:x '->' x}{http://coq.inria.fr/stdlib/Coq.Init.Logic}{\coqdocnotation{\ensuremath{\rightarrow}}} \coqexternalref{nat}{http://coq.inria.fr/stdlib/Coq.Init.Datatypes}{\coqdocinductive{nat}} \coqexternalref{:type scope:x '->' x}{http://coq.inria.fr/stdlib/Coq.Init.Logic}{\coqdocnotation{\ensuremath{\rightarrow}}} \coqexternalref{nat}{http://coq.inria.fr/stdlib/Coq.Init.Datatypes}{\coqdocinductive{nat}} := \coqdoceol
\coqdocindent{1.00em}
\coqdockw{fun} \coqdocvar{a} \ensuremath{\Rightarrow} \coqdocnotation{$?\hspace{-1mm}\rightarrow$} (\coqref{Casts.extraction.divide}{\coqdocdefinition{divide}} \coqdocvariable{a}).\coqdoceol
\coqdocemptyline
\coqdocnoindent
\coqdockw{Extraction} \coqref{Casts.extraction.divide'}{\coqdocdefinition{divide'}}.\coqdoceol
\coqdocemptyline
\end{coqdoccode}

\begin{verbatim}
let divide' a =
  cast_fun_dom (decidable_le_nat 1) (divide a)
\end{verbatim}

 The inserted cast translates to a runtime check in the extracted code, whose failure results in a runtime cast error: 
\begin{verbatim}
# divide' 1 0;;
Exception: Failure "Cast has failed".
\end{verbatim}

 \paragraph{Extracting axioms as exceptions.}
By default, the use of an axiom translates to a runtime exception in Ocaml. In order to make the error message more informative, we explicitly instruct Coq to extract uses of \coqdocvar{failed\_cast} as follows:\footnote{To be more helpful in the error reporting, we do provide a string representation of the casted value by using a showable type class, similar to {\tt Show} in Haskell (see code in the distribution). However, we cannot provide the information of the violated property, because there is currently no way to obtain the string representation of an arbitrary \coqdockw{Prop} within Coq.} \begin{coqdoccode}
\coqdocemptyline
\coqdocnoindent
\coqdockw{Extract} \coqdocvar{Constant} \coqdocvar{failed\_cast} \ensuremath{\Rightarrow}\coqdoceol
\coqdocindent{1.00em}
"failwith ""Cast has failed""".\coqdoceol
\coqdocemptyline
\end{coqdoccode}

Appendix~\ref{sec:eval}, which discusses evaluation regimes, includes discussion about some subtleties that arise when extracting to an eager language like Scheme or Ocaml.

Finally, note that the second axiom we introduced in Section~\ref{sec:weakdep}, \coqdocvar{failed\_cast\_proj1}, does not need to be extracted at all: it is used to convert two types that are equal after extraction (because they only differ in propositional content). \begin{coqdoccode}
\coqdocemptyline
\end{coqdoccode}

\section{Properties}
\label{sec:properties}

The development of gradual checking of subset types we have presented is
entirely internalized in Coq: we have neither extended the underlying
theory nor modified the implementation. The only peculiarities are the use of the \coqref{Casts.cast1.failed
  cast}{\coqdocaxiom{failed\_cast}}
and 
\coqref{Casts.cast1.failed cast
  proj1}{\coqdocaxiom{failed\_cast\_proj1}}
axioms.
As a consequence, a number of
properties come ``for free''.

\paragraph{Canonicity.} Coq without axioms enjoys a canonicity
property, which states that all normal forms correspond to canonical
forms. For instance, all normal forms of type {\tt bool} are either
{\tt true} or {\tt false}.  

\paragraph{Cast errors.} Canonicity is only violated by the use of
axioms. Here, this means that the only non-canonical normal
forms are terms with \coqref{Casts.cast1.failed cast}{\coqdocaxiom{failed\_cast}} (or \coqref{Casts.cast1.failed cast type}{\coqdocaxiom{failed\_cast\_proj1}})
inside. More precisely, a cast failure in Coq is any term $t$ such
that $t = E[\texttt{failed\_cast}~v~p]$,
where $v$ is the casted value and $p$ is a false property (ditto for 
\coqdocaxiom{failed\_cast\_proj1}).
In Coq, for cast errors that manifest at the value
level, the evaluation context $E$ is determined by the evaluation
regime specified when calling \ckw{Eval}. For cast errors that
manifest at the type level, $E$ follows the reduction strategy for
type conversion, which is coarsely a head normal-form evaluation with
optimization for constants.

\paragraph{Soundness via extraction.} The canonicity of Coq and the
definition of cast errors, together with the assumption that program
extraction in Coq is correct (and axioms are extracted as runtime
errors), entail the typical type soundness property for
gradually-typed programs, i.e. programs with safe runtime
casts~\cite{siekTaha:sfp2006,igarashi01fj}: the only stuck terms at
runtime are cast errors.\footnote{Note that if the target language is
  impure, then it is possible to break the safety of program
  extraction altogether (eg. by passing an impure Ocaml function as input to
  a Coq-extracted function). This general issue is independent of
  casting and beyond the scope of this work. Ensuring safe
  interoperability between a purely functional dependently-typed
  language like Coq and a language with impure features is a
  challenging research venue.}

\paragraph{Termination of casts.} Because decision procedures are
defined within Coq, casts are guaranteed to terminate. This is in
contrast to some approaches, like hybrid type checking in 
Sage~\cite{knowlesFlanagan:toplas2010,sage}, in which decision
procedures are defined within a language for which termination is not
guaranteed.

\paragraph{Simplification at extraction.}
Because propositions are erased at extraction, the
\coqref{Casts.cast1.failed cast
  proj1}{\coqdocaxiom{failed\_cast\_proj1}} axiom is never extracted in
the target language and thus cannot fail. This means that in the
extracted program, \coqref{Casts.cast1.hide
  cast proj1}{\coqdocdefinition{hide\_cast\_proj1}} is always extracted to the
identity function, and errors can only manifest through the \coqref{Casts.cast1.failed cast}{\coqdocaxiom{failed\_cast}} axiom.


\section{Casting More Dependent Types: Records}
\label{sec:records}
\coqlibrary{Casts.records}{Library }{Casts.records}

\begin{coqdoccode}
\coqdocemptyline
\coqdocemptyline
\end{coqdoccode}
Until now, we have developed the axiomatic approach to gradual verification in Coq with subset types, because they are the canonical way to attach a property to a value.
However, the approach is not specific to subset types and accomodates other dependently-typed structures commonly used by Coq developers, such as {\em record types}. To stress that our approach is not restricted to subset types, we now illustrate how to deal with dependent records. We also use this example as a case study in customizing the synthesis of correct decision procedures through the \coqdocclass{$\mathsf{Decidable}$} type class.

\paragraph{Rationals.} Consider the prototypical example provided in the Coq reference manual\footnote{\url{https://coq.inria.fr/refman/Reference-Manual004.html\#sec61}}: a record type for rational numbers, which embeds the property that the divisor is not zero, and that the fraction is irreducible. The type of rational numbers, with their properties, is defined as: 
\begin{coqdoccode}
\coqdocemptyline
\coqdocnoindent
\coqdockw{Record} \coqdef{Casts.records.Rat}{Rat}{\coqdocrecord{Rat}} : \coqdockw{Set} := \coqdef{Casts.records.mkRat}{mkRat}{\coqdocconstructor{mkRat}}\coqdoceol
\coqdocindent{1.50em}
\{\coqdef{Casts.records.sign}{sign}{\coqdocprojection{sign}} : \coqexternalref{bool}{http://coq.inria.fr/stdlib/Coq.Init.Datatypes}{\coqdocinductive{bool}};\coqdoceol
\coqdocindent{2.00em}
\coqdef{Casts.records.top}{top}{\coqdocprojection{top}} : \coqexternalref{nat}{http://coq.inria.fr/stdlib/Coq.Init.Datatypes}{\coqdocinductive{nat}};\coqdoceol
\coqdocindent{2.00em}
\coqdef{Casts.records.bottom}{bottom}{\coqdocprojection{bottom}} : \coqexternalref{nat}{http://coq.inria.fr/stdlib/Coq.Init.Datatypes}{\coqdocinductive{nat}};\coqdoceol
\coqdocindent{2.00em}
\coqdef{Casts.records.Rat bottom cond}{Rat\_bottom\_cond}{\coqdocprojection{Rat\_bottom\_cond}} : 0 \coqexternalref{:type scope:x '<>' x}{http://coq.inria.fr/stdlib/Coq.Init.Logic}{\coqdocnotation{\ensuremath{\not=}}} \coqref{Casts.records.bottom}{\coqdocmethod{bottom}};\coqdoceol
\coqdocindent{2.00em}
\coqdef{Casts.records.Rat irred cond}{Rat\_irred\_cond}{\coqdocprojection{Rat\_irred\_cond}} : \coqdockw{\ensuremath{\forall}} \coqdocvar{x} \coqdocvar{y} \coqdocvar{z},\coqdoceol
\coqdocindent{4.00em}
\coqdocvariable{y} \coqexternalref{:nat scope:x '*' x}{http://coq.inria.fr/stdlib/Coq.Init.Nat}{\coqdocnotation{\ensuremath{\times}}} \coqdocvariable{x} \coqexternalref{:type scope:x '=' x}{http://coq.inria.fr/stdlib/Coq.Init.Logic}{\coqdocnotation{=}} \coqref{Casts.records.top}{\coqdocmethod{top}} \coqexternalref{:type scope:x '/x5C' x}{http://coq.inria.fr/stdlib/Coq.Init.Logic}{\coqdocnotation{\ensuremath{\land}}} \coqdocvariable{z} \coqexternalref{:nat scope:x '*' x}{http://coq.inria.fr/stdlib/Coq.Init.Nat}{\coqdocnotation{\ensuremath{\times}}} \coqdocvariable{x} \coqexternalref{:type scope:x '=' x}{http://coq.inria.fr/stdlib/Coq.Init.Logic}{\coqdocnotation{=}} \coqref{Casts.records.bottom}{\coqdocmethod{bottom}} \coqexternalref{:type scope:x '->' x}{http://coq.inria.fr/stdlib/Coq.Init.Logic}{\coqdocnotation{\ensuremath{\rightarrow}}} 1 \coqexternalref{:type scope:x '=' x}{http://coq.inria.fr/stdlib/Coq.Init.Logic}{\coqdocnotation{=}} \coqdocvariable{x}\}.\coqdoceol
\coqdocemptyline
\end{coqdoccode}
 \paragraph{Casting rationals.}

   \noindent The property \coqref{Casts.records.Rat bottom cond}{\coqdocprojection{Rat\_bottom\_cond}} is obviously decidable.
   It is less clear for the property \coqref{Casts.records.Rat irred cond}{\coqdocprojection{Rat\_irred\_cond}}, which uses
   universal quantification. Indeed, in general, there is no decision
   procedure for a universally-quantified decidable property over
   natural numbers, because the set of natural numbers is infinite. So
   it seems we cannot use the cast framework to create rationals
   without having to provide proofs of their associated properties.

Interestingly, it {\em is} possible to use casts for rationals despite the fact that \coqref{Casts.records.Rat irred cond}{\coqdocprojection{Rat\_irred\_cond}} cannot be directly declared to be decidable. We review three different approaches in this section. They all exploit the fact that if we can prove that a decidable property is equivalent to \coqref{Casts.records.Rat irred cond}{\coqdocprojection{Rat\_irred\_cond}}, then \coqref{Casts.records.Rat irred cond}{\coqdocprojection{Rat\_irred\_cond}} is decidable (Section~\ref{sec:decidable}).

We define a cast operator for \coqref{Casts.records.Rat}{\coqdocrecord{Rat}}, which
   takes the three values for \coqref{Casts.records.sign}{\coqdocprojection{sign}}, \coqref{Casts.records.top}{\coqdocprojection{top}} and \coqref{Casts.records.bottom}{\coqdocprojection{bottom}}, two (implicitly-passed) decision
   procedures \coqdocvar{dec\_rat\_bottom} and \coqdocvariable{dec\_rat\_irred}, and checks the two
   properties: \begin{coqdoccode}
\coqdocemptyline
\coqdocemptyline
\coqdocnoindent
\coqdockw{Definition} \coqdef{Casts.records.cast Rat}{cast\_Rat}{\coqdocdefinition{cast\_Rat}} (\coqdocvar{s}:\coqexternalref{bool}{http://coq.inria.fr/stdlib/Coq.Init.Datatypes}{\coqdocinductive{bool}}) (\coqdocvar{t} \coqdocvar{b}: \coqexternalref{nat}{http://coq.inria.fr/stdlib/Coq.Init.Datatypes}{\coqdocinductive{nat}})\coqdoceol
\coqdocindent{1.00em}
\{\coqdocvar{dec\_rat\_bottom} : \coqdocclass{$\mathsf{Decidable}$} \coqdocvar{\_}\}\coqdoceol
\coqdocindent{1.00em}
\{\coqdocvar{dec\_rat\_irred}  : \coqexternalref{:type scope:x '->' x}{http://coq.inria.fr/stdlib/Coq.Init.Logic}{\coqdocnotation{(}}0 \coqexternalref{:type scope:x '<>' x}{http://coq.inria.fr/stdlib/Coq.Init.Logic}{\coqdocnotation{\ensuremath{\not=}}} \coqdocvariable{b}\coqexternalref{:type scope:x '->' x}{http://coq.inria.fr/stdlib/Coq.Init.Logic}{\coqdocnotation{)}} \coqexternalref{:type scope:x '->' x}{http://coq.inria.fr/stdlib/Coq.Init.Logic}{\coqdocnotation{\ensuremath{\rightarrow}}} \coqdocclass{$\mathsf{Decidable}$} \coqdocvar{\_}\} : \coqref{Casts.records.Rat}{\coqdocrecord{Rat}} :=\coqdoceol
\coqdocindent{1.00em}
\coqdockw{match} \coqdocdefinition{dec} \coqdocvar{\_}  \coqdockw{with}\coqdoceol
\coqdocindent{2.00em}
\ensuremath{|} \coqexternalref{inl}{http://coq.inria.fr/stdlib/Coq.Init.Datatypes}{\coqdocconstructor{inl}} \coqdocvar{Hb} \ensuremath{\Rightarrow}\coqdoceol
\coqdocindent{2.50em}
\coqdockw{match} \coqdocdefinition{dec} (\coqdocvar{$\mathsf{Decidable}$} := \coqdocvariable{dec\_rat\_irred} \coqdocvar{Hb}) \coqdocvar{\_}  \coqdockw{with} \coqdoceol
\coqdocindent{3.50em}
\ensuremath{|} \coqexternalref{inl}{http://coq.inria.fr/stdlib/Coq.Init.Datatypes}{\coqdocconstructor{inl}} \coqdocvar{Hi} \ensuremath{\Rightarrow} \coqref{Casts.records.mkRat}{\coqdocconstructor{mkRat}} \coqdocvariable{s} \coqdocvariable{t} \coqdocvariable{b} \coqdocvar{Hb} \coqdocvar{Hi}\coqdoceol
\coqdocindent{3.50em}
\ensuremath{|} \coqdocvar{\_} \ensuremath{\Rightarrow} \coqref{Casts.records.failed cast Rat}{\coqdocaxiom{failed\_cast\_Rat}} \coqdocvariable{s} \coqdocvariable{t} \coqdocvariable{b}\coqdoceol
\coqdocindent{3.00em}
\coqdockw{end}\coqdoceol
\coqdocindent{2.00em}
\ensuremath{|} \coqdocvar{\_} \ensuremath{\Rightarrow} \coqref{Casts.records.failed cast Rat}{\coqdocaxiom{failed\_cast\_Rat}} \coqdocvariable{s} \coqdocvariable{t} \coqdocvariable{b}\coqdoceol
\coqdocindent{1.00em}
\coqdockw{end}.\coqdoceol
\coqdocemptyline
\coqdocemptyline
\end{coqdoccode}
\noindent As before, the definition of the cast operator appeals to an inconsistent axiom in the case a property is violated. The \coqref{Casts.records.failed cast Rat}{\coqdocaxiom{failed\_cast\_Rat}} axiom states that any three values are adequate to make up a \coqref{Casts.records.Rat}{\coqdocrecord{Rat}}:\footnote{It is necessary to define custom axioms and cast operators for
each new record type. This limitation was not apparent with subset
types, because they are a general purpose structure, while records are
specific. To limit the burden of adoption, it would be interesting to
define a Coq plugin that automatically generates the axioms and cast
operators (whose definitions are quite straightforward).}\\

\coqdockw{Axiom} \coqref{Casts.records.failed cast Rat}{\coqdocaxiom{failed\_cast\_Rat}} : \coqdockw{\ensuremath{\forall}} (\coqdocvariable{s}:\coqexternalref{bool}{http://coq.inria.fr/stdlib/Coq.Init.Datatypes}{\coqdocinductive{bool}}) (\coqdocvariable{t} \coqdocvariable{b}: \coqexternalref{nat}{http://coq.inria.fr/stdlib/Coq.Init.Datatypes}{\coqdocinductive{nat}}), \coqref{Casts.records.Rat}{\coqdocrecord{Rat}}.
\vspace{2mm}

  \noindent Note that we use a type dependency in \coqref{Casts.records.cast Rat}{\coqdocdefinition{cast\_Rat}} to allow the decision procedure of \coqdocvariable{dec\_rat\_irred} to use the fact that \coqref{Casts.records.Rat bottom cond}{\coqdocprojection{Rat\_bottom\_cond}} holds in the branch where it is used.
 \begin{coqdoccode}
\coqdocemptyline
\coqdocemptyline
\end{coqdoccode}

 \paragraph{A decision procedure based on bounded quantification.}

A first approach to establish a decision procedure for irreducibility is to exploit that it is equivalent to the same property that quantifies over {\em bounded} natural numbers. We first define the type of bounded naturals (and we introduce an implicit coercion from \coqref{Casts.records.bnat}{\coqdocinductive{bnat}} to \coqexternalref{nat}{http://coq.inria.fr/stdlib/Coq.Init.Datatypes}{\coqdocinductive{nat}}):
\begin{coqdoccode}
\coqdocemptyline
\coqdocnoindent
\coqdockw{Definition} \coqdef{Casts.records.bnat}{bnat}{\coqdocinductive{bnat}} (\coqdocvar{n}:\coqexternalref{nat}{http://coq.inria.fr/stdlib/Coq.Init.Datatypes}{\coqdocinductive{nat}}) := \coqexternalref{:type scope:'x7B' x ':' x '|' x 'x7D'}{http://coq.inria.fr/stdlib/Coq.Init.Specif}{\coqdocnotation{\{}}\coqdocvar{m} \coqexternalref{:type scope:'x7B' x ':' x '|' x 'x7D'}{http://coq.inria.fr/stdlib/Coq.Init.Specif}{\coqdocnotation{:}} \coqexternalref{nat}{http://coq.inria.fr/stdlib/Coq.Init.Datatypes}{\coqdocinductive{nat}} \coqexternalref{:type scope:'x7B' x ':' x '|' x 'x7D'}{http://coq.inria.fr/stdlib/Coq.Init.Specif}{\coqdocnotation{\ensuremath{|}}} \coqdocvar{m} \coqexternalref{:nat scope:x '<=' x}{http://coq.inria.fr/stdlib/Coq.Init.Peano}{\coqdocnotation{\ensuremath{\le}}} \coqdocvariable{n}\coqexternalref{:type scope:'x7B' x ':' x '|' x 'x7D'}{http://coq.inria.fr/stdlib/Coq.Init.Specif}{\coqdocnotation{\}}}.\coqdoceol
\coqdocemptyline
\end{coqdoccode}
\noindent and define a general instance of \coqdocclass{$\mathsf{Decidable}$}, which allows building a decision procedure for any universally-quantified property over bounded naturals:
 \begin{coqdoccode}
\coqdocemptyline
\coqdocemptyline
\coqdocemptyline
\coqdocnoindent
\coqdockw{Instance} \coqdef{Casts.records.Decidable forall bounded}{Decidable\_forall\_bounded}{\coqdocinstance{Decidable\_forall\_bounded}} \coqdocvar{k}\coqdoceol
\coqdocindent{1.00em}
(\coqdocvar{P}:\coqref{Casts.records.bnat}{\coqdocinductive{bnat}} \coqdocvariable{k}\coqexternalref{:type scope:x '->' x}{http://coq.inria.fr/stdlib/Coq.Init.Logic}{\coqdocnotation{\ensuremath{\rightarrow}}}\coqdockw{Prop}) (\coqdocvar{HP} : \coqdockw{\ensuremath{\forall}} \coqdocvar{n}, \coqdocclass{$\mathsf{Decidable}$} (\coqdocvariable{P} \coqdocvariable{n})) :\coqdoceol
\coqdocindent{1.00em}
\coqdocclass{$\mathsf{Decidable}$} (\coqdockw{\ensuremath{\forall}} \coqdocvar{n}, \coqdocvariable{P} \coqdocvariable{n}).\coqdoceol
\coqdocemptyline
\coqdocemptyline
\end{coqdoccode}

We can then establish how to synthesize a decision procedure for \coqref{Casts.records.Rat irred cond}{\coqdocprojection{Rat\_irred\_cond}} by establishing that it is equivalent to a similar property, where the quantification is bounded by the max of \coqref{Casts.records.top}{\coqdocprojection{top}} and \coqref{Casts.records.bottom}{\coqdocprojection{bottom}}:
 \begin{coqdoccode}
\coqdocemptyline
\coqdocnoindent
\coqdockw{Definition} \coqdef{Casts.records.Rat irred cond bounded}{Rat\_irred\_cond\_bounded}{\coqdocdefinition{Rat\_irred\_cond\_bounded}} \coqdocvar{top} \coqdocvar{bottom} `(0 \coqexternalref{:type scope:x '<>' x}{http://coq.inria.fr/stdlib/Coq.Init.Logic}{\coqdocnotation{\ensuremath{\not=}}} \coqdocvariable{bottom}):\coqdoceol
\coqdocindent{1.00em}
\coqexternalref{:type scope:x '<->' x}{http://coq.inria.fr/stdlib/Coq.Init.Logic}{\coqdocnotation{(}}\coqdockw{\ensuremath{\forall}} \coqdocvar{x} \coqdocvar{y} \coqdocvar{z}: \coqref{Casts.records.bnat}{\coqdocinductive{bnat}} (\coqexternalref{max}{http://coq.inria.fr/stdlib/Coq.Init.Nat}{\coqdocdefinition{max}} \coqdocvariable{top} \coqdocvariable{bottom}),\coqdoceol
\coqdocindent{2.50em}
\coqdocvariable{y} \coqexternalref{:nat scope:x '*' x}{http://coq.inria.fr/stdlib/Coq.Init.Nat}{\coqdocnotation{\ensuremath{\times}}} \coqdocvariable{x} \coqexternalref{:type scope:x '=' x}{http://coq.inria.fr/stdlib/Coq.Init.Logic}{\coqdocnotation{=}} \coqdocvariable{top} \coqexternalref{:type scope:x '/x5C' x}{http://coq.inria.fr/stdlib/Coq.Init.Logic}{\coqdocnotation{\ensuremath{\land}}} \coqdocvariable{z} \coqexternalref{:nat scope:x '*' x}{http://coq.inria.fr/stdlib/Coq.Init.Nat}{\coqdocnotation{\ensuremath{\times}}} \coqdocvariable{x} \coqexternalref{:type scope:x '=' x}{http://coq.inria.fr/stdlib/Coq.Init.Logic}{\coqdocnotation{=}} \coqdocvariable{bottom} \coqexternalref{:type scope:x '->' x}{http://coq.inria.fr/stdlib/Coq.Init.Logic}{\coqdocnotation{\ensuremath{\rightarrow}}} 1 \coqexternalref{:type scope:x '=' x}{http://coq.inria.fr/stdlib/Coq.Init.Logic}{\coqdocnotation{=}} \coqdocvariable{x}\coqexternalref{:type scope:x '<->' x}{http://coq.inria.fr/stdlib/Coq.Init.Logic}{\coqdocnotation{)}} \coqexternalref{:type scope:x '<->' x}{http://coq.inria.fr/stdlib/Coq.Init.Logic}{\coqdocnotation{\ensuremath{\leftrightarrow}}}\coqdoceol
\coqdocindent{1.00em}
\coqexternalref{:type scope:x '<->' x}{http://coq.inria.fr/stdlib/Coq.Init.Logic}{\coqdocnotation{(}}\coqdockw{\ensuremath{\forall}} \coqdocvar{x} \coqdocvar{y} \coqdocvar{z}: \coqexternalref{nat}{http://coq.inria.fr/stdlib/Coq.Init.Datatypes}{\coqdocinductive{nat}},  \coqdocvariable{y} \coqexternalref{:nat scope:x '*' x}{http://coq.inria.fr/stdlib/Coq.Init.Nat}{\coqdocnotation{\ensuremath{\times}}} \coqdocvariable{x} \coqexternalref{:type scope:x '=' x}{http://coq.inria.fr/stdlib/Coq.Init.Logic}{\coqdocnotation{=}} \coqdocvariable{top} \coqexternalref{:type scope:x '/x5C' x}{http://coq.inria.fr/stdlib/Coq.Init.Logic}{\coqdocnotation{\ensuremath{\land}}} \coqdocvariable{z} \coqexternalref{:nat scope:x '*' x}{http://coq.inria.fr/stdlib/Coq.Init.Nat}{\coqdocnotation{\ensuremath{\times}}} \coqdocvariable{x} \coqexternalref{:type scope:x '=' x}{http://coq.inria.fr/stdlib/Coq.Init.Logic}{\coqdocnotation{=}} \coqdocvariable{bottom} \coqexternalref{:type scope:x '->' x}{http://coq.inria.fr/stdlib/Coq.Init.Logic}{\coqdocnotation{\ensuremath{\rightarrow}}} 1 \coqexternalref{:type scope:x '=' x}{http://coq.inria.fr/stdlib/Coq.Init.Logic}{\coqdocnotation{=}} \coqdocvariable{x}\coqexternalref{:type scope:x '<->' x}{http://coq.inria.fr/stdlib/Coq.Init.Logic}{\coqdocnotation{)}}.\coqdoceol
\coqdocemptyline
\end{coqdoccode}
\noindent Note that it is crucial to be able to use the fact that 0 \ensuremath{\not=} \coqref{Casts.records.bottom}{\coqdocprojection{bottom}} holds in the proof of equivalence, as it simply does not hold when \coqref{Casts.records.bottom}{\coqdocprojection{bottom}} = 0.

Then, the \coqdocclass{$\mathsf{Decidable}$} instance for \coqref{Casts.records.Rat irred cond}{\coqdocprojection{Rat\_irred\_cond}} is simply defined by connecting it to the bounded property through the \coqref{Casts.records.Decidable equivalent}{\coqdocdefinition{Decidable\_equivalent}} instance:
\begin{coqdoccode}
\coqdocemptyline
\coqdocnoindent
\coqdockw{Instance} \coqdef{Casts.records.Rat irred cond dec bounded}{Rat\_irred\_cond\_dec\_bounded}{\coqdocinstance{Rat\_irred\_cond\_dec\_bounded}} \coqdocvar{top} \coqdocvar{bottom}\coqdoceol
\coqdocindent{4.50em}
`(0 \coqexternalref{:type scope:x '<>' x}{http://coq.inria.fr/stdlib/Coq.Init.Logic}{\coqdocnotation{\ensuremath{\not=}}} \coqdocvariable{bottom})  : \coqdocclass{$\mathsf{Decidable}$} \coqdocvar{\_} :=\coqdoceol
\coqdocindent{1.00em}
\coqref{Casts.records.Decidable equivalent}{\coqdocdefinition{Decidable\_equivalent}}\coqdoceol
\coqdocindent{2.00em}
(\coqref{Casts.records.Rat irred cond bounded}{\coqdocdefinition{Rat\_irred\_cond\_bounded}} \coqdocvar{top} \coqdocvar{bottom} \coqdocvar{H}).\coqdoceol
\coqdocemptyline
\end{coqdoccode}
    \paragraph{Example.} It is now possible to define a rational number without having to prove the 
    two side conditions.  
 \begin{coqdoccode}
\coqdocemptyline
\coqdocnoindent
\coqdockw{Definition} \coqdef{Casts.records.Rat good}{Rat\_good}{\coqdocdefinition{Rat\_good}} := \coqref{Casts.records.cast Rat}{\coqdocdefinition{cast\_Rat}} \coqexternalref{true}{http://coq.inria.fr/stdlib/Coq.Init.Datatypes}{\coqdocconstructor{true}} 5 6.\coqdoceol
\coqdocemptyline
\coqdocnoindent
\coqdockw{Eval} \coqdoctac{compute} \coqdoctac{in} \coqref{Casts.records.top}{\coqdocprojection{top}} \coqref{Casts.records.Rat good}{\coqdocdefinition{Rat\_good}}.\coqdoceol
\coqdocemptyline
\end{coqdoccode}
\begin{verbatim}
= 5
: nat
\end{verbatim}

 \noindent
    Exactly in the same way as the first projection of a dependent pair cannot be recovered 
    if the cast fails, \coqref{Casts.records.sign}{\coqdocprojection{sign}}, \coqref{Casts.records.top}{\coqdocprojection{top}} or \coqref{Casts.records.bottom}{\coqdocprojection{bottom}} can not be recovered if \coqref{Casts.records.cast Rat}{\coqdocdefinition{cast\_Rat}} fails. 
 \begin{coqdoccode}
\coqdocemptyline
\coqdocnoindent
\coqdockw{Definition} \coqdef{Casts.records.Rat bad}{Rat\_bad}{\coqdocdefinition{Rat\_bad}} := \coqref{Casts.records.cast Rat}{\coqdocdefinition{cast\_Rat}} \coqexternalref{true}{http://coq.inria.fr/stdlib/Coq.Init.Datatypes}{\coqdocconstructor{true}} 5 10.\coqdoceol
\coqdocemptyline
\coqdocnoindent
\coqdockw{Eval} \coqdoctac{compute} \coqdoctac{in} \coqref{Casts.records.top}{\coqdocprojection{top}} \coqref{Casts.records.Rat bad}{\coqdocdefinition{Rat\_bad}}.\coqdoceol
\end{coqdoccode}
\begin{verbatim}
= let (_, top, bottom, _, _) :=
      failed_cast_Rat true 5 10  in top
: nat
\end{verbatim}

 \noindent Note that the evaluation of \coqref{Casts.records.top}{\coqdocprojection{top}} \coqref{Casts.records.Rat bad}{\coqdocdefinition{Rat\_bad}} takes a
    significant amount of time, because the decision procedure
    involves checking every possible \coqdocvariable{x} \coqdocvariable{y} \coqdocvariable{z}: \coqref{Casts.records.bnat}{\coqdocinductive{bnat}} 10, which amounts
    to checking more than $1000$ properties.  Indeed, a simple cast as
    above takes around $2$ seconds on a recent computer.

  We now show that we can improve the cast on rational numbers by using more efficient decision procedures over equivalent properties.

 \paragraph{A decision procedure using binary natural numbers.}

  In the Coq standard library, there is a {\em binary} representation
of integers, \coqexternalref{Z}{http://coq.inria.fr/stdlib/Coq.Numbers.BinNums}{\coqdocinductive{Z}}, which is much more efficient but less easy to reason
about.  We can exploit this representation by showing that the property
\coqref{Casts.records.Rat irred cond}{\coqdocprojection{Rat\_irred\_cond}} in \coqexternalref{Z}{http://coq.inria.fr/stdlib/Coq.Numbers.BinNums}{\coqdocinductive{Z}} implies the property in \coqexternalref{nat}{http://coq.inria.fr/stdlib/Coq.Init.Datatypes}{\coqdocinductive{nat}}: \begin{coqdoccode}
\coqdocemptyline
\coqdocemptyline
\coqdocnoindent
\coqdockw{Definition} \coqdef{Casts.records.Rat irred cond Z}{Rat\_irred\_cond\_Z}{\coqdocdefinition{Rat\_irred\_cond\_Z}} \coqdocvar{top} \coqdocvar{bottom} `(0 \coqexternalref{:type scope:x '<>' x}{http://coq.inria.fr/stdlib/Coq.Init.Logic}{\coqdocnotation{\ensuremath{\not=}}} \coqdocvariable{bottom}):\coqdoceol
\coqdocindent{1.00em}
\coqexternalref{:type scope:x '<->' x}{http://coq.inria.fr/stdlib/Coq.Init.Logic}{\coqdocnotation{(}}\coqdockw{\ensuremath{\forall}} \coqdocvar{x} \coqdocvar{y} \coqdocvar{z}: \coqref{Casts.records.bnat}{\coqdocinductive{bnat}} (\coqexternalref{max}{http://coq.inria.fr/stdlib/Coq.Init.Nat}{\coqdocdefinition{max}} \coqdocvariable{top} \coqdocvariable{bottom}),\coqdoceol
\coqdocindent{2.00em}
\coqexternalref{Z.mul}{http://coq.inria.fr/stdlib/Coq.ZArith.BinInt}{\coqdocdefinition{Z.mul}} \coqdocvariable{y} \coqdocvariable{x} \coqexternalref{:type scope:x '=' x}{http://coq.inria.fr/stdlib/Coq.Init.Logic}{\coqdocnotation{=}} \coqdocvariable{top} \coqexternalref{:type scope:x '/x5C' x}{http://coq.inria.fr/stdlib/Coq.Init.Logic}{\coqdocnotation{\ensuremath{\land}}} \coqexternalref{Z.mul}{http://coq.inria.fr/stdlib/Coq.ZArith.BinInt}{\coqdocdefinition{Z.mul}} \coqdocvariable{z} \coqdocvariable{x} \coqexternalref{:type scope:x '=' x}{http://coq.inria.fr/stdlib/Coq.Init.Logic}{\coqdocnotation{=}} \coqdocvariable{bottom} \coqexternalref{:type scope:x '->' x}{http://coq.inria.fr/stdlib/Coq.Init.Logic}{\coqdocnotation{\ensuremath{\rightarrow}}} 1 \coqexternalref{:type scope:x '=' x}{http://coq.inria.fr/stdlib/Coq.Init.Logic}{\coqdocnotation{=}} \coqdocvariable{x}\coqexternalref{:type scope:x '<->' x}{http://coq.inria.fr/stdlib/Coq.Init.Logic}{\coqdocnotation{)}} \coqexternalref{:type scope:x '<->' x}{http://coq.inria.fr/stdlib/Coq.Init.Logic}{\coqdocnotation{\ensuremath{\leftrightarrow}}}\coqdoceol
\coqdocindent{1.00em}
\coqexternalref{:type scope:x '<->' x}{http://coq.inria.fr/stdlib/Coq.Init.Logic}{\coqdocnotation{(}}\coqdockw{\ensuremath{\forall}} \coqdocvar{x} \coqdocvar{y} \coqdocvar{z}: \coqexternalref{nat}{http://coq.inria.fr/stdlib/Coq.Init.Datatypes}{\coqdocinductive{nat}}, \coqdocvariable{y} \coqexternalref{:nat scope:x '*' x}{http://coq.inria.fr/stdlib/Coq.Init.Nat}{\coqdocnotation{\ensuremath{\times}}} \coqdocvariable{x} \coqexternalref{:type scope:x '=' x}{http://coq.inria.fr/stdlib/Coq.Init.Logic}{\coqdocnotation{=}} \coqdocvariable{top} \coqexternalref{:type scope:x '/x5C' x}{http://coq.inria.fr/stdlib/Coq.Init.Logic}{\coqdocnotation{\ensuremath{\land}}} \coqdocvariable{z} \coqexternalref{:nat scope:x '*' x}{http://coq.inria.fr/stdlib/Coq.Init.Nat}{\coqdocnotation{\ensuremath{\times}}} \coqdocvariable{x} \coqexternalref{:type scope:x '=' x}{http://coq.inria.fr/stdlib/Coq.Init.Logic}{\coqdocnotation{=}} \coqdocvariable{bottom} \coqexternalref{:type scope:x '->' x}{http://coq.inria.fr/stdlib/Coq.Init.Logic}{\coqdocnotation{\ensuremath{\rightarrow}}} 1 \coqexternalref{:type scope:x '=' x}{http://coq.inria.fr/stdlib/Coq.Init.Logic}{\coqdocnotation{=}} \coqdocvariable{x}\coqexternalref{:type scope:x '<->' x}{http://coq.inria.fr/stdlib/Coq.Init.Logic}{\coqdocnotation{)}}.\coqdoceol
\coqdocemptyline
\coqdocnoindent
\coqdockw{Instance} \coqdef{Casts.records.Rat irred cond dec}{Rat\_irred\_cond\_dec}{\coqdocinstance{Rat\_irred\_cond\_dec}} \coqdocvar{top} \coqdocvar{bottom} `(0 \coqexternalref{:type scope:x '<>' x}{http://coq.inria.fr/stdlib/Coq.Init.Logic}{\coqdocnotation{\ensuremath{\not=}}} \coqdocvariable{bottom}):\coqdoceol
\coqdocindent{1.00em}
\coqdocclass{$\mathsf{Decidable}$} \coqdocvar{\_} :=\coqdoceol
\coqdocindent{1.00em}
\coqref{Casts.records.Decidable equivalent}{\coqdocdefinition{Decidable\_equivalent}}\coqdoceol
\coqdocindent{2.00em}
(\coqref{Casts.records.Rat irred cond Z}{\coqdocdefinition{Rat\_irred\_cond\_Z}} \coqdocvar{top} \coqdocvar{bottom} \coqdocvar{H}).\coqdoceol
\coqdocemptyline
\end{coqdoccode}
In this manner, the time for evaluating the same ``bad'' rational number cast as 
\coqref{Casts.records.Rat bad}{\coqdocdefinition{Rat\_bad}} decreases by a factor of 10!\begin{coqdoccode}
\coqdocemptyline
\end{coqdoccode}
\paragraph{A decision procedure based on gcd.}

We can go even one step further and avoid doing an exhaustive (even if finite) check: the property \coqref{Casts.records.Rat irred cond}{\coqdocprojection{Rat\_irred\_cond}} is actually equivalent to the gcd of \coqref{Casts.records.top}{\coqdocprojection{top}} and \coqref{Casts.records.bottom}{\coqdocprojection{bottom}} being equal to 1:
 \begin{coqdoccode}
\coqdocemptyline
\coqdocnoindent
\coqdockw{Definition} \coqdef{Casts.records.Rat irred cond gcd}{Rat\_irred\_cond\_gcd}{\coqdocdefinition{Rat\_irred\_cond\_gcd}} \coqdocvar{top} \coqdocvar{bottom} `(0 \coqexternalref{:type scope:x '<>' x}{http://coq.inria.fr/stdlib/Coq.Init.Logic}{\coqdocnotation{\ensuremath{\not=}}} \coqdocvariable{bottom}) :\coqdoceol
\coqdocindent{1.00em}
\coqexternalref{:type scope:x '<->' x}{http://coq.inria.fr/stdlib/Coq.Init.Logic}{\coqdocnotation{(}}\coqexternalref{Z.gcd}{http://coq.inria.fr/stdlib/Coq.ZArith.BinInt}{\coqdocdefinition{Z.gcd}} (\coqdocvariable{top}:\coqexternalref{nat}{http://coq.inria.fr/stdlib/Coq.Init.Datatypes}{\coqdocinductive{nat}}) \coqdocvariable{bottom} \coqexternalref{:type scope:x '=' x}{http://coq.inria.fr/stdlib/Coq.Init.Logic}{\coqdocnotation{=}} 1\coqexternalref{:type scope:x '<->' x}{http://coq.inria.fr/stdlib/Coq.Init.Logic}{\coqdocnotation{)}} \coqexternalref{:type scope:x '<->' x}{http://coq.inria.fr/stdlib/Coq.Init.Logic}{\coqdocnotation{\ensuremath{\leftrightarrow}}}\coqdoceol
\coqdocindent{1.00em}
\coqexternalref{:type scope:x '<->' x}{http://coq.inria.fr/stdlib/Coq.Init.Logic}{\coqdocnotation{(}}\coqdockw{\ensuremath{\forall}} \coqdocvar{x} \coqdocvar{y} \coqdocvar{z}, \coqdocvariable{y} \coqexternalref{:nat scope:x '*' x}{http://coq.inria.fr/stdlib/Coq.Init.Nat}{\coqdocnotation{\ensuremath{\times}}} \coqdocvariable{x} \coqexternalref{:type scope:x '=' x}{http://coq.inria.fr/stdlib/Coq.Init.Logic}{\coqdocnotation{=}} \coqdocvariable{top} \coqexternalref{:type scope:x '/x5C' x}{http://coq.inria.fr/stdlib/Coq.Init.Logic}{\coqdocnotation{\ensuremath{\land}}} \coqdocvariable{z} \coqexternalref{:nat scope:x '*' x}{http://coq.inria.fr/stdlib/Coq.Init.Nat}{\coqdocnotation{\ensuremath{\times}}} \coqdocvariable{x} \coqexternalref{:type scope:x '=' x}{http://coq.inria.fr/stdlib/Coq.Init.Logic}{\coqdocnotation{=}} \coqdocvariable{bottom} \coqexternalref{:type scope:x '->' x}{http://coq.inria.fr/stdlib/Coq.Init.Logic}{\coqdocnotation{\ensuremath{\rightarrow}}} 1 \coqexternalref{:type scope:x '=' x}{http://coq.inria.fr/stdlib/Coq.Init.Logic}{\coqdocnotation{=}} \coqdocvariable{x}\coqexternalref{:type scope:x '<->' x}{http://coq.inria.fr/stdlib/Coq.Init.Logic}{\coqdocnotation{)}}.\coqdoceol
\coqdocemptyline
\coqdocnoindent
\coqdockw{Instance} \coqdef{Casts.records.Rat irred cond gcd dec}{Rat\_irred\_cond\_gcd\_dec}{\coqdocinstance{Rat\_irred\_cond\_gcd\_dec}} \coqdocvar{top} \coqdocvar{bottom}\coqdoceol
\coqdocindent{4.50em}
(\coqdocvar{Hbot} : 0 \coqexternalref{:type scope:x '<>' x}{http://coq.inria.fr/stdlib/Coq.Init.Logic}{\coqdocnotation{\ensuremath{\not=}}} \coqdocvariable{bottom}) : \coqdocclass{$\mathsf{Decidable}$} \coqdocvar{\_} :=\coqdoceol
\coqdocindent{1.00em}
\coqref{Casts.records.Decidable equivalent}{\coqdocdefinition{Decidable\_equivalent}}\coqdoceol
\coqdocindent{2.00em}
(\coqref{Casts.records.Rat irred cond gcd}{\coqdocdefinition{Rat\_irred\_cond\_gcd}} \coqdocvar{top} \coqdocvar{bottom} \coqdocvar{Hbot}).\coqdoceol
\coqdocemptyline
\end{coqdoccode}
\noindent Computing the same bad cast is now instantaneous.
 \begin{coqdoccode}
\coqdocemptyline
\coqdocemptyline
\end{coqdoccode}

\section{Related Work}
\label{sec:related-work}

There is plenty of work on rich types like refinement types~\cite{freemanPfenning:pldi91,xiPfenning:pldi98,liquid:popl2008,bengstonAl:toplas2011} (which
roughly correspond to the subset types of
Coq~\cite{sozeau:types2006}), focusing mostly on how to maintain
statically decidable checking (eg. through SMT solvers) while offering
a refinement logic as expressive as possible. Liquid
types~\cite{liquid:popl2008}, and their subsequent
improvements~\cite{chughAl:popl2012,vazouAl:esop2013}, are one of the
most salient example of this line of work. Of course, to remain
statically decidable, the refinement logics are necessarily less
expressive than higher-order logics such as Coq and Agda. In this work
we focus on Coq, giving up fully automatic verification. This being
said, Coq allows a mix of automatic and manual theorem proving, and we
extend this combination with the possibility to lift proofs of
decidable properties to delayed checks with casts. Notably, the set
(and shape) of decidable properties is not hardwired in the language,
but is derived from an extensible library. We believe our approach is
applicable to Agda as well, since the main elements (axioms and type
classes) are also supported in Agda. However, the devil is certainly in the details.

Interestingly, Seidel {\em et al.} recently developed an approach
called type targeted testing to exploit refinement type annotations
not for static checking, but for randomized property-based
testing~\cite{seidelAl:esop2015}. This supports a progressive approach
by which programmers can first enjoy some benefits of (unchecked)
refinement type annotations for testing, and then eventually turn to
full static checking when they desire. While the authors informally
qualify the methodology as ``gradual'', it is quite different from other
gradual checking work, which focuses on mixing static and dynamic
checking~\cite{siekTaha:sfp2006}. 
Gradual typing has been extended to
a whole range of rich type disciplines:
typestates~\cite{wolffAl:ecoop2011,garciaAl:toplas2014}, information
flow typing and security
types~\cite{disneyFlanagan:stop2011,fennellThiemann:csf2013},
ownership types~\cite{sergeyClarke:esop2012}, annotated type
systems~\cite{thiemannFennell:esop2014}, and
effects~\cite{banadosAl:icfp2014}, but not to a full-blown
dependently-typed language.

This work is directly related to the work of Ou et
al. on combining dependent types and simple types~\cite{ouAl:tcs2004},
as well as the work on hybrid type
checking~\cite{knowlesFlanagan:toplas2010}, as supported in
Sage~\cite{sage}.  Ou et al. develop a core language with dependent
function types and subset types augmented
with three special commands: \textbf{simple}\{$e$\}, to denote that
expression $e$ is simply well-typed, \textbf{dependent}\{$e$\}, to
denote that the type checker should statically check all dependent
constraints in $e$, and \textbf{assert}$(e, \tau)$ to check at runtime
that $e$ produces a value of (possibly-dependent) type $\tau$. The
semantics of the source language is given by translation to an
internal language, which uses a type coercion judgment that inserts
runtime checks when needed. In hybrid type checking, the
language includes arbitrary refinements on base types, and
the type system tries to statically decide implications between
predicates using an external theorem prover. If it is not statically
possible to either verify or refute an implication, a cast is inserted
to defer checking to runtime. 

In both approaches, refinements are directly expressed in the base
language, as boolean expressions; therefore it suffices to evaluate
the refinement expression itself at runtime to dynamically determine
whether the refinement holds. (In hybrid type checking, refinements
are not guaranteed to terminate, while in Ou et al., refinements are
drawn from a pure subset of expressions.) In both cases,
arbitrary logical properties cannot be expressed: the refinements
directly correspond to boolean decision procedures, without the
possibility to specify their logical meaning (see also
Appendix~\ref{sec:refl} for a discussion of boolean reflection). In
particular, there are no ways for programmers to give proof terms
explicitly, which means that it is impossible to marry non-decidable
(explicitly proven) properties with decidable ones (which may
voluntarily be proven or deferred).

\section{Conclusion}
\label{sec:conclusion}

We exposed an approach to support gradual certified programming in
Coq. When initially engaging in this project, we anticipated a painful
extension to the theory and implementation of Coq. Much to our
surprise, it was possible to achieve our objectives in a simple and
elegant (albeit slightly heretical) manner, exploiting axioms and type
classes. The cast framework is barely over 50 lines of Coq, to which
we have to add the expansion of the Coq/HoTT \cind{Decidable} library,
which is useful beyond this work, and could be replaced by a different
decidability framework. A limitation of the internalized
approach is that it does not support blame
assignment~\cite{findlerFelleisen:icfp2002}, because it would be 
necessary to modify reduction to track blame labels transparently.

An interesting track to explore is to make the axiomatic approach to
casts less heretical, by requiring the claimed property to be
inhabited (this would rule out direct claims of \cind{False}, for
instance). The counterpart is that it requires some additional effort
from the programmer---it may be possible to automatically find
witnesses in certain cases. Also, the monadic version seems perfectly
reasonable if extraction is the main objective, because upon
extraction we can eliminate the success case of the error monad, and
turn the failure case into a runtime exception. Additionally, the
decidability constraint could be relaxed by only requiring a sound
approximation of the property to be decidable, not necessarily the
property itself. Finally, we can optimize the cast procedure so that
it does not execute the decision procedure if the property has been
statically proven.

\acks

We thank Jonathan Aldrich, R{\'e}mi Douence, St{\'e}phane Glondu,
Ronald Garcia, Fran\c{c}ois Pottier, Ilya Sergey and Matthieu Sozeau
for providing helpful feedback on this work and article. Ilya Sergey
also integrated the cast framework with Ssreflect as a decidability
framework. We also thank the anonymous DLS reviewers, and the
participants of the Coq workshop 2015 participants for their feedback,
especially Georges Gonthier and Gabriel Scherer who suggested very
interesting venues for future work.


 \bibliographystyle{abbrvnat}



\clearpage
\appendix

\coqlibrary{Casts.notes}{Library }{Casts.notes}

\begin{coqdoccode}
\coqdocemptyline
\coqdocemptyline
\end{coqdoccode}
\section{A Note on Boolean Reflection}

\label{sec:refl}

An alternative approach for the definition of decision procedures is
to use {\em boolean reflection}, i.e. considering that the decision
procedure {\em is} the property. \begin{coqdoccode}
\coqdocemptyline
\coqdocemptyline
\coqdocnoindent
\coqdockw{Instance} \coqdef{Casts.notes.Decidable bool}{Decidable\_bool}{\coqdocinstance{Decidable\_bool}} (\coqdocvar{b} : \coqexternalref{bool}{http://coq.inria.fr/stdlib/Coq.Init.Datatypes}{\coqdocinductive{bool}}) : \coqdoceol
\coqdocindent{2.00em}
\coqdocclass{$\mathsf{Decidable}$} (\coqdockw{if} \coqdocvariable{b} \coqdockw{then} \coqexternalref{True}{http://coq.inria.fr/stdlib/Coq.Init.Logic}{\coqdocinductive{True}} \coqdockw{else} \coqexternalref{False}{http://coq.inria.fr/stdlib/Coq.Init.Logic}{\coqdocinductive{False}}).\coqdoceol
 \coqdocemptyline
\end{coqdoccode}

However, while using boolean reflection can be convenient, there is no ``safeguard'' that the procedure is correctly implemented: the implementation {\em is} the specification. Another limitation is that the information reported to the programmer is unhelpful: if the cast succeeds, the proof term is \coqdocvar{I}; if the cast fails, the failed property is \coqexternalref{False}{http://coq.inria.fr/stdlib/Coq.Init.Logic}{\coqdocinductive{False}}. While the proof term is arguably irrelevant, the information about the failed property can be very helpful for debugging.

Both issues can nevertheless been solved by having both the boolean and the property, and 
formally establishing the relation between both, similarly to what is
done in the Ssreflect~\cite{ssreflect} library or the \cind{reflect} inductive in
Coq. This boolean/proposition relation mechanism is also provided in
the \href{http://ssg.ustcsz.edu.cn/~zz/doc/coq/stdlib/Coq.Classes.DecidableClass.html}{DecidableClass} library of Coq. To avoid name conflicts (the
class is also named \coqdocclass{$\mathsf{Decidable}$}), we provide the same class under the
name \coqref{Casts.notes.Decidable relate}{\coqdocrecord{Decidable\_relate}}:\begin{coqdoccode}
\coqdocemptyline
\coqdocnoindent
\coqdockw{Class} \coqdef{Casts.notes.Decidable relate}{Decidable\_relate}{\coqdocrecord{Decidable\_relate}} (\coqdocvar{P} : \coqdockw{Prop}) := \{\coqdoceol
\coqdocindent{1.50em}
\coqdef{Casts.notes.Decidable witness}{Decidable\_witness}{\coqdocprojection{Decidable\_witness}}: \coqexternalref{bool}{http://coq.inria.fr/stdlib/Coq.Init.Datatypes}{\coqdocinductive{bool}};\coqdoceol
\coqdocindent{1.50em}
\coqdef{Casts.notes.Decidable spec}{Decidable\_spec}{\coqdocprojection{Decidable\_spec}}: \coqref{Casts.notes.Decidable witness}{\coqdocmethod{Decidable\_witness}} \coqexternalref{:type scope:x '=' x}{http://coq.inria.fr/stdlib/Coq.Init.Logic}{\coqdocnotation{=}} \coqexternalref{true}{http://coq.inria.fr/stdlib/Coq.Init.Datatypes}{\coqdocconstructor{true}} \coqexternalref{:type scope:x '<->' x}{http://coq.inria.fr/stdlib/Coq.Init.Logic}{\coqdocnotation{\ensuremath{\leftrightarrow}}} \coqdocvariable{P}\coqdoceol
\coqdocnoindent
\}.\coqdoceol
\coqdocemptyline
\end{coqdoccode}
\noindent 
Actually the two presentations of decidability are equivalent. Indeed, the same development has been done in Ssreflect\footnote{The Ssreflect implementation was done by Ilya Sergey.}, using canonical structures~\cite{saibi:popl97} instead of type classes to automatically infer complex decision procedures from simpler ones~\cite{gonthierAl:jfp2013}.
This shows that  the decidability mechanism is orthogonal to the cast operators we propose. \footnote{The \coqdocclass{$\mathsf{Decidable}$} library is currently much less furnished than the Ssreflect library using boolean reflection, but its extension with instances similar to the ones implemented in Ssreflect is straightforward.} 
 \begin{coqdoccode}
\coqdocemptyline
\coqdocemptyline
\coqdocemptyline
\end{coqdoccode}
\section{A Note on Evaluation Regimes}

\label{sec:eval}

Recall that Coq does not impose any fixed reduction strategy. Instead, \ckw{Eval} is parameterized by a reduction strategy, called a conversion tactic, such as \coqdoctac{cbv} (aka. \coqdoctac{compute}), \coqdoctac{lazy}, \coqdoctac{hnf}, \coqdoctac{simpl}, etc.

In addition to understanding the impact of reduction strategies on the results of computations with casts, it is crucial to understand the impact of representing cast failures through an axiom. Consider a function \coqref{Casts.notes.g}{\coqdocdefinition{g}} that expects a \{\coqdocvar{n}:\coqexternalref{nat}{http://coq.inria.fr/stdlib/Coq.Init.Datatypes}{\coqdocinductive{nat}} \ensuremath{|} \coqdocvar{n} $>$ 0\}, but actually never uses its argument: \begin{coqdoccode}
\coqdocemptyline
\coqdocnoindent
\coqdockw{Definition} \coqdef{Casts.notes.g}{g}{\coqdocdefinition{g}} (\coqdocvar{x}: \coqexternalref{:type scope:'x7B' x ':' x '|' x 'x7D'}{http://coq.inria.fr/stdlib/Coq.Init.Specif}{\coqdocnotation{\{}}\coqdocvar{n}\coqexternalref{:type scope:'x7B' x ':' x '|' x 'x7D'}{http://coq.inria.fr/stdlib/Coq.Init.Specif}{\coqdocnotation{:}}\coqexternalref{nat}{http://coq.inria.fr/stdlib/Coq.Init.Datatypes}{\coqdocinductive{nat}} \coqexternalref{:type scope:'x7B' x ':' x '|' x 'x7D'}{http://coq.inria.fr/stdlib/Coq.Init.Specif}{\coqdocnotation{\ensuremath{|}}} \coqdocvar{n} \coqexternalref{:nat scope:x '>' x}{http://coq.inria.fr/stdlib/Coq.Init.Peano}{\coqdocnotation{$>$}} 0\coqexternalref{:type scope:'x7B' x ':' x '|' x 'x7D'}{http://coq.inria.fr/stdlib/Coq.Init.Specif}{\coqdocnotation{\}}}) := 1.\coqdoceol
\coqdocemptyline
\end{coqdoccode}
Typically, one would expect that evaluating \coqref{Casts.notes.g}{\coqdocdefinition{g}} (? 0) with a \coqdoctac{lazy} reduction would produce 1, while using an eager strategy like \coqdoctac{compute} would reveal the failed cast. However: \begin{coqdoccode}
\coqdocemptyline
\coqdocnoindent
\coqdockw{Eval} \coqdoctac{compute} \coqdoctac{in} \coqref{Casts.notes.g}{\coqdocdefinition{g}} (? 0).\coqdoceol
\coqdocemptyline
\end{coqdoccode}

\begin{verbatim}
= 1
: nat \end{verbatim}

 The reason is that a cast error in Coq is not an error per se (Coq has no such mechanism): it is just a non-canonical normal form. Therefore, even with an eager strategy, \coqref{Casts.notes.g}{\coqdocdefinition{g}} (? 0) simply returns 1. The cast is eagerly evaluated, and fails; but this only means that \coqref{Casts.notes.g}{\coqdocdefinition{g}} is called with {\tt failed\_cast} as a fully-evaluated argument. Because \coqref{Casts.notes.g}{\coqdocdefinition{g}} does not touch its argument, the cast failure goes unnoticed.
\begin{coqdoccode}
\coqdocemptyline
\coqdocemptyline
\end{coqdoccode}

On the contrary, if we extract the code to Ocaml (recall Section~\ref{sec:extraction}), the cast violation is reported immediately as an exception:
\begin{coqdoccode}
\coqdocemptyline
\coqdocnoindent
\coqdockw{Definition} \coqdef{Casts.notes.client}{client}{\coqdocdefinition{client}} (\coqdocvar{x}: \coqexternalref{nat}{http://coq.inria.fr/stdlib/Coq.Init.Datatypes}{\coqdocinductive{nat}}) := \coqref{Casts.notes.g}{\coqdocdefinition{g}} (? \coqdocvariable{x}).\coqdoceol
\coqdocemptyline
\coqdocnoindent
\coqdockw{Extraction} \coqdocvar{Language} \coqdocvar{Ocaml}.\coqdoceol
\coqdocemptyline
\coqdocnoindent
\coqdockw{Extraction} "test.ml" \coqdocvar{client}.\coqdoceol
\coqdocemptyline
\end{coqdoccode}

\begin{verbatim}
# client 1;;
- : int = 1
# client 0;;
Exception: Failure "Cast has failed".
\end{verbatim}

While, as expected, the error goes unnoticed in Haskell, because of its lazy evaluation regime.
\begin{coqdoccode}
\coqdocemptyline
\coqdocnoindent
\coqdockw{Extraction} \coqdocvar{Language} \coqdocvar{Haskell}.\coqdoceol
\coqdocnoindent
\coqdockw{Extraction} "test.hs" \coqdocvar{client}.\coqdoceol
\coqdocemptyline
\end{coqdoccode}
\begin{verbatim}
*Test> client 1
1
*Test> client 0
1
\end{verbatim}

 \paragraph{Extraction of axioms in eager languages.\label{sec:eager}} There is one last detail to discuss when considering extraction to eager languages. As defined, \coqdocvar{failed\_cast} and \coqdocvar{cast} are extracted as follows in Ocaml:

\begin{verbatim}
let failed_cast = 
  failwith "Cast has failed"

let cast dec a =
  match dec a with
  | Inl _ -> a
  | Inr _ -> failed_cast
\end{verbatim}

While these definitions are perfectly fine for a lazy language like
 Haskell, in an eager language like Ocaml or Scheme they imply that
 loading the definition of {\tt failed\_cast} fails directly. The
 solution is to enforce the inlining of \coqdocvar{failed\_cast}: \begin{coqdoccode}
\coqdocemptyline
\coqdocnoindent
\coqdockw{Extraction} \coqdockw{Inline} \coqdocvar{failed\_cast}.\coqdoceol
\coqdocemptyline
\end{coqdoccode}
As a result, \coqdocvar{failed\_cast} is not extracted as a separate definition, and \coqdocvar{cast} uses the Ocaml {\tt failwith} function directly. \begin{coqdoccode}
\end{coqdoccode}




\end{document}